\documentclass[reprint,aps,prd,superscriptaddress,showpacs,nofootinbib,floatfix]{revtex4-1} 

\usepackage{times}
\usepackage{mathptmx}
\usepackage{amsmath}
\usepackage{amsfonts}
\usepackage{amssymb}
\usepackage{graphicx}
\usepackage{color}
\usepackage[colorlinks]{hyperref}

\allowdisplaybreaks[4]

\definecolor{CiteColor}{rgb}{0,0.5,0}
\hypersetup{citecolor=CiteColor}
\definecolor{RefColor}{rgb}{0.55,0,0}
\hypersetup{linkcolor=RefColor}
\definecolor{darkgreen}{rgb}{0.2,0.7,0.2}

\DeclareMathOperator{\sgn}{sgn}

\newcommand{\ud}{\mathrm{d}}
\newcommand{\calO}{\mathcal{O}}
\newcommand{\beq}{\begin{equation}}
\newcommand{\eeq}{\end{equation}}
\newcommand{\ra}[1]{\renewcommand{\arraystretch}{#1}}

\newcommand{\UMD}{\affiliation{Maryland Center for Fundamental Physics \& Joint Space-Science Institute, Department of Physics, University of Maryland, College Park, MD 20742, USA}}
\newcommand{\CITA}{\affiliation{Canadian Institute for Theoretical Astrophysics, University of Toronto, Toronto, Ontario M5S 3H8, Canada}}
\newcommand{\CIFAR}{\affiliation{Canadian Institute for Advanced Research, 180 Dundas St.~West, Toronto, Ontario M5G 1Z8, Canada}}
\newcommand{\Cornell}{\affiliation{Center for Radiophysics and Space Research, Cornell University, Ithaca, NY 14853, USA}}
\newcommand{\CSUF}{\affiliation{Gravitational Wave Physics and Astronomy Center, California State University Fullerton, Fullerton, CA 92831, USA}}
\newcommand{\Caltech}{\affiliation{Theoretical Astrophysics 350-17, California Institute of Technology, Pasadena, CA 91125, USA}}

\begin{document}

\title{Periastron Advance in Spinning Black Hole Binaries: \\ Gravitational Self-Force from Numerical Relativity}

\author{Alexandre Le Tiec}\UMD
\author{Alessandra Buonanno}\UMD
\author{Abdul H. Mrou\'e}\CITA
\author{Harald P. Pfeiffer}\CITA\CIFAR
\author{Daniel A. Hemberger}\Caltech\Cornell
\author{Geoffrey Lovelace}\CSUF\Caltech
\author{Lawrence E. Kidder}\Cornell
\author{Mark A. Scheel}\Caltech
\author{Bela Szil\'agyi}\Caltech
\author{Nicholas W. Taylor}\Caltech
\author{Saul A. Teukolsky}\Cornell

\begin{abstract}
We study the general relativistic periastron advance in spinning black hole binaries on quasi-circular orbits, with spins aligned or anti-aligned with the orbital angular momentum, using numerical-relativity simulations, the post-Newtonian approximation, and black hole perturbation theory. By imposing a symmetry by exchange of the bodies' labels, we devise an improved version of the perturbative result, and use it as the leading term of a new type of expansion in powers of the symmetric mass ratio. This allows us to measure, for the first time, the gravitational self-force effect on the periastron advance of a non-spinning particle orbiting a Kerr black hole of mass $M$ and spin $S = -0.5M^2$, down to separations of order $9M$. Comparing the predictions of our improved perturbative expansion with the exact results from numerical simulations of equal-mass and equal-spin binaries, we find a remarkable agreement over a wide range of spins and orbital separations.
\end{abstract}

\date{\today}


\maketitle

\section{Introduction}\label{sec:intro}

Accounting for the observed anomalous advance of Mercury's perihelion was the first successful test of Einstein's theory of general relativity \cite{Ei.15}. More recently, the same effect---but with a much larger amplitude, of the order a few degrees per year---has been observed in the orbital motion of binary pulsars \cite{St.03,Lo.08}. Today, the prospect of observing gravitational radiation from binary systems of compact objects (black holes and neutron stars) is triggering further interest in the relativistic periastron advance. A worldwide effort is currently underway to achieve the first direct detection of gravitational waves by using kilometer-scale, ground-based laser interferometers such as advanced LIGO \cite{Sh.10} and advanced Virgo \cite{Ac.al.08}, as well as future space-based antennas, such as the eLISA mission \cite{eLISA}. The detection and analysis of these signals require very accurate theoretical predictions, for use as template waveforms to be cross-correlated against the output of the detectors. Hence, an accurate modeling of the relativistic orbital dynamics of compact-object binary systems is crucially needed.

For binaries with small orbital velocities/large separations, but otherwise arbitrary mass ratios, the periastron advance has been computed to increasingly high orders using the post-Newtonian (PN) approximation to general relativity \cite{Bl.06}. For non-spinning binaries moving on generic (bound) orbits, the 1PN, 2PN and 3PN results were derived in Refs.~\cite{Ro.38,DaSc.88,Da.al.00}. Spin-orbit and spin-spin effects were computed up to 3.5PN order for aligned or anti-aligned spins \cite{Te.al.10,Te.al.13}, as well as for generic spin orientations in special binary configurations \cite{KoGo.05,Te.09}; see Ref.~\cite{DaSc.88} for earlier references. For binaries with extreme mass ratios, the orbital motion can be studied using black-hole perturbation theory \cite{SaTa.03,Ba.09,Po.al.11}. In the test-mass approximation, the periastron advance of a non-spinning particle on a generic (bound) geodesic orbit around a Schwarzschild or Kerr black hole has been computed in Refs.~\cite{Cu.al.94,Sc.02}. The corrections linear and quadratic in the spin of the small body were computed in the companion paper \cite{Hi.al.13}, for nearly circular orbits. The first-order mass-ratio correction to the geodesic result was obtained in Ref.~\cite{BaSa.11} for a Schwarzschild background, but the result is still unknown in the Kerr case. Using the effective-one-body (EOB) formalism \cite{BuDa.99,Da.al3.00,Da.al2.08,BaBu.10}, the periastron advance has been computed for non-spinning \cite{Da.10} as well as for spinning compact binaries \cite{Hi.al.13} on quasi-circular orbits.

Following the breakthrough in the numerical simulation of the late inspiral and merger of binary black hole (BBH) systems \cite{Pr.05,Ba.al.06,Ca.al.06} (see Ref.~\cite{Pf.12} for a recent review), it has recently become possible to study the periastron advance using fully non-linear numerical relativity (NR) simulations. The first NR results for the periastron advance were presented in Ref.~\cite{Mr.al.10} and an improved analysis using longer and more accurate numerical simulations was done in Ref.~\cite{Le.al.11}. More recently, the periastron advance has also been measured in a mixed neutron star/black hole binary \cite{Fo.al.13}. In this paper we extend the earlier works \cite{Le.al.11,Mr.al.10} for non-spinning black hole binaries to spinning systems. We make use of accurate NR simulations of the late inspiral of spinning BBHs on quasi-circular orbits, with spins aligned or anti-aligned with the orbital angular momentum. The simulations we analyze have two different origins: (i) the series of equal-mass, equal-spin binaries presented in Refs.~\cite{He.al.13,Lo.al.12}, with a focus on the properties of binaries with nearly extremal spins, and (ii) the unequal-mass spinning simulations presented in Ref.~\cite{MrPf.12}.

After deriving explicit expressions for the periastron advance at the highest PN order currently known, we compare those predictions to the NR data. We then use the mathematical structure of the PN expansion for the periastron advance, together with explicit formulas for the periastron advance of a non-spinning and spinning particle in Kerr spacetime, to derive an improved version of the perturbative result that is fully symmetrized by exchange of the bodies' labels. Indeed, earlier works \cite{Sm.79,FiDe.84,Fa.al.04,Le.al.11,Sp.al2.11,Le.al2.12,Na.13} suggested that working with a ``symmetrized background'' can successfully extend the domain of validity of perturbative calculations. Finally, we show how to employ the improved, perturbative result to extract the gravitational self-force (GSF) correction to the periastron advance from NR simulations. As a proof of principle, we first use the NR simulations of non-spinning BBH systems with mass ratios $1-8$, extract the GSF correction to the periastron advance and compare it with the known, exact result from perturbative calculations \cite{BaSa.11}. Then, we consider NR simulations of single-spin BBH systems with mass ratios $1.5-8$ and {\it predict} the GSF correction to the periastron advance for a non-spinning particle moving on a circular equatorial orbit around a Kerr black hole of mass $M$ and spin $S = -0.5 M^2$. These results are summarized in Fig.~\ref{fig:self-force} below.

This paper is organized as follows. Section \ref{sec:NR} explains how the periastron advance is extracted from NR simulations of binary black holes, and how the error estimates are computed. In Sec.~\ref{sec:PN} we establish the 3.5PN-accurate expression of the periastron advance for quasi-circular orbits, including all spin-orbit and spin-spin effects. The perturbative result for a point mass orbiting a Kerr black hole on a circular equatorial orbit is obtained in Sec.~\ref{sec:pert}. In Sec.~\ref{sec:sym} we impose known symmetries on the perturbative result, and make use of this expression as a background to extract GSF information by using NR results in Sec.~\ref{sec:SF}. We summarize our main findings and discuss future prospects in Sec.~\ref{sec:sum}. Throughout this paper we set $G = c = 1$.

\section{Numerical Relativity}\label{sec:NR}

In this section we provide an in-depth discussion of the techniques used in Ref.~\cite{Le.al.11} to extract the periastron advance from BBH simulations, and further refine these techniques. Henceforth, we use the sum $m = m_1 + m_2$ of the irreducible masses of the black holes to define dimensionless frequencies.

\subsection{Basic procedure}

The analysis of the periastron advance is based on the coordinate trajectories of the centers of the apparent horizons, as computed during BBH
evolutions~\cite{Mr.al.13,MrPf.12,Bu.al.12,Ma.al2.13,He.al.13}
using the Spectral Einstein Code (SpEC)~\cite{SpECwebsite}.  Let
$\mathbf {c}_{i}(t)$ denote the coordinates of the center of each black
hole, and define their relative separation $\mathbf{r}(t)=
\mathbf{c}_1(t)-\mathbf{c}_2(t)$.  The instantaneous orbital
frequency $\Omega(t)$ is computed by
\begin{equation}\label{eq:omega-inst}
	\Omega(t) \equiv \frac{\vert \mathbf{r}(t) \times \dot{\mathbf{r}}(t) \vert}{r^2(t)}\,, 
\end{equation}
where the Euclidean cross product and norm are used, and an overdot stands for $\ud / \ud t$. The orbital frequency $\Omega(t)$ is the sum of a secular
quasi-circular piece [given by the average frequency
$\Omega_\varphi(t)$] and a small oscillatory remainder containing
information about the eccentricity and the radial frequency.  Both
components drift slowly in time due to the radiation-reaction driven
inspiral of the black holes. To separate $\Omega(t)$ into these two
components, we perform a fit to the model
\begin{align}\label{eq:fitting-function}
	\Omega(t) &= p_0 \left[p_1-(t-T)\right]^{p_2} \nonumber \\ &+ p_3 \, \cos{\big[p_4 + p_5 {(t-T)}+ p_6(t-T)^2\big]} \, .
\end{align}
The $p_i$'s are parameters to be determined by the fit.  The first
term in Eq.~(\ref{eq:fitting-function}), with fitting parameters $(p_0,
p_1, p_2)$, is intended to capture the monotonic, non-oscillatory
inspiral behavior of a non-eccentric binary. Writing this as a single
power-law term ensures monotonic behavior which would not be
guaranteed if this term were a polynomial of order $2$ or higher.  The
second term is designed to capture oscillations in $\Omega(t)$ that
arise from orbital eccentricity.  The amplitude $p_3$ will be
proportional to the eccentricity.  Because $\Omega$ is linked to the
radius through angular momentum conservation, the phase of the
oscillations (parameters $p_4, p_5, p_6$) will give the phase of the
radial motion of the binary.

The model \eqref{eq:fitting-function} is fitted over an interval $t \in [ T - \frac{\Delta T}{2}, T + \frac{\Delta T}{2}]$ centered around the time $T$, with width $\Delta T = {\varpi} \times 2\pi / \Omega(T)$ parametrized by the number {$\varpi$} of orbits within this interval. The instantaneous orbital frequency $\Omega_\varphi(T)$ and the radial frequency $\Omega_r(T)$ at time $T$ are computed by evaluating the monotonic and oscillatory parts of the fit at $t=T$:
\begin{subequations}\label{eq:Omegas}
	\begin{align}
		\Omega_\varphi(T) &= p_0 \, p_1^{p_2} \, , \label{eq:Omega_varphi} \\
		\Omega_r(T)& = p_5 \, . \label{eq:Omega_r}
	\end{align}
\end{subequations}
Finally, the periastron advance is given by the ratio
\begin{equation}\label{eq:K_NR}
	K_{\rm NR}(T)=\frac{\Omega_\varphi(T)}{\Omega_r(T)}.
\end{equation}
Repeating this procedure for many different times $T$ results in the
periastron advance $K_{\rm NR}(\Omega_\varphi)$ as a function of the
average quasi-circular orbital frequency $\Omega_\varphi$.

\begin{figure}
	\includegraphics[scale=0.49]{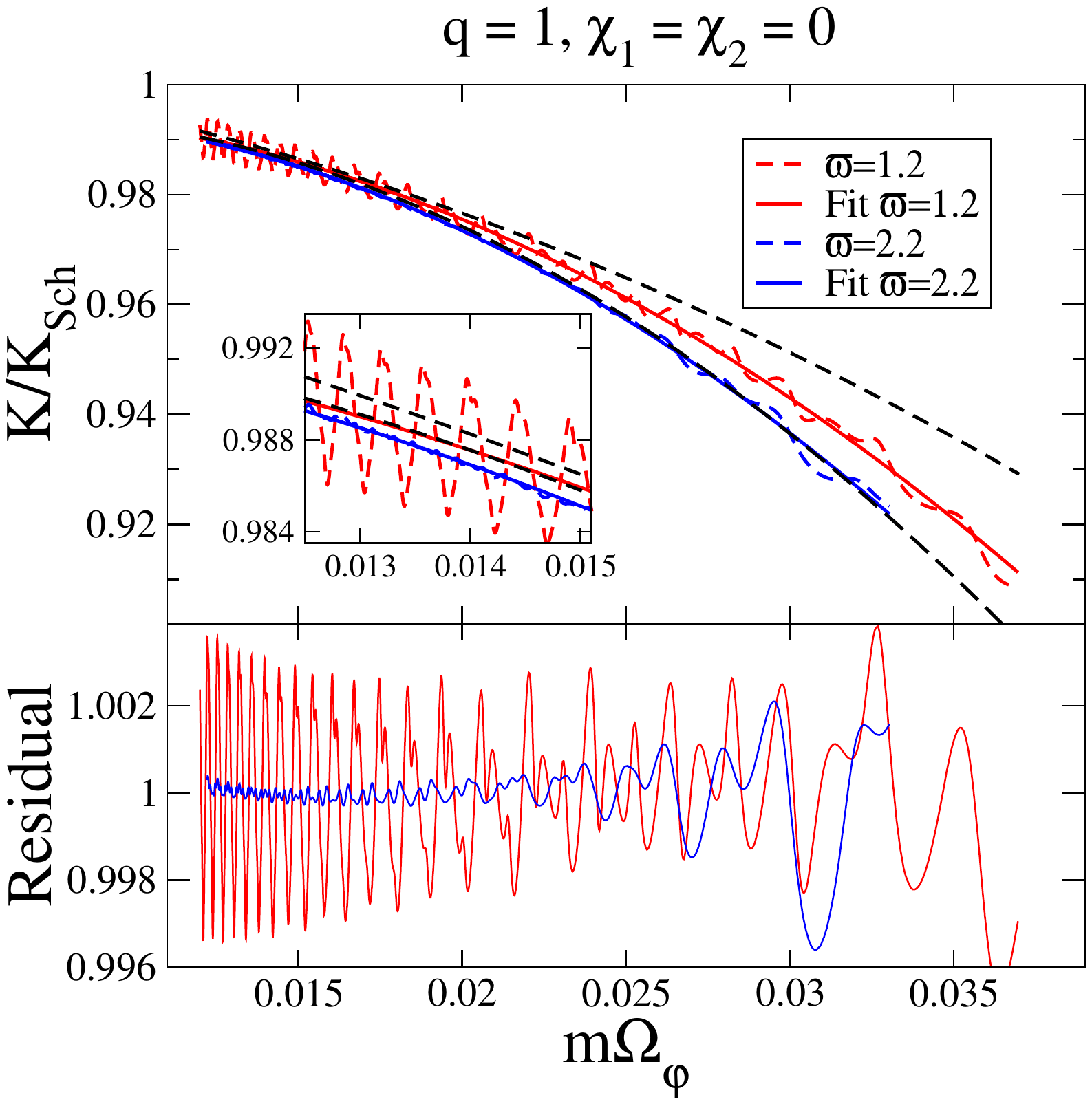}
	\caption{Periastron advance extracted from numerical simulations. \textit{Upper panel:} The dashed curves show $K_{\rm NR}(\Omega_\varphi)/K_{\rm Sch}(\Omega_\varphi)$ as computed from Eqs.~(\ref{eq:Omegas}) and (\ref{eq:K_NR}) using fitting intervals with two different widths {$\varpi$}. The solid lines show polynomial fits to $K_{\rm NR}/K_{\rm Sch}$. \textit{Lower panel:} Residuals of the polynomial fits.}
	\label{fig:pa-q1-ecc}
\end{figure}

Figure~\ref{fig:pa-q1-ecc} shows an example of this procedure, applied to an equal-mass, non-spinning BBH system. The red-dashed and blue-dashed curves are the output of Eq.~\eqref{eq:K_NR} for two different values of {$\varpi$}, normalized by the periastron advance $K_\text{Sch} = [1 - 6(m\Omega_\varphi)^{2/3}]^{-1/2}$ of a test mass orbiting a Schwarzschild black hole (\textit{cf.} Sec.~\ref{sec:pert} below) to reduce the dynamical range; note that the $y$-scale of Fig.~\ref{fig:pa-q1-ecc} represents only a relative variation of $8\%$ of $K_{\rm NR}$. The solid lines represent power-law fits to the dashed data, with error regions indicated by the dashed black lines. This is the procedure that was used in the analysis in Le Tiec, Mrou{\'e} \textit{et al.} \cite{Le.al.11}.

\begin{figure}
	\includegraphics[scale=0.48]{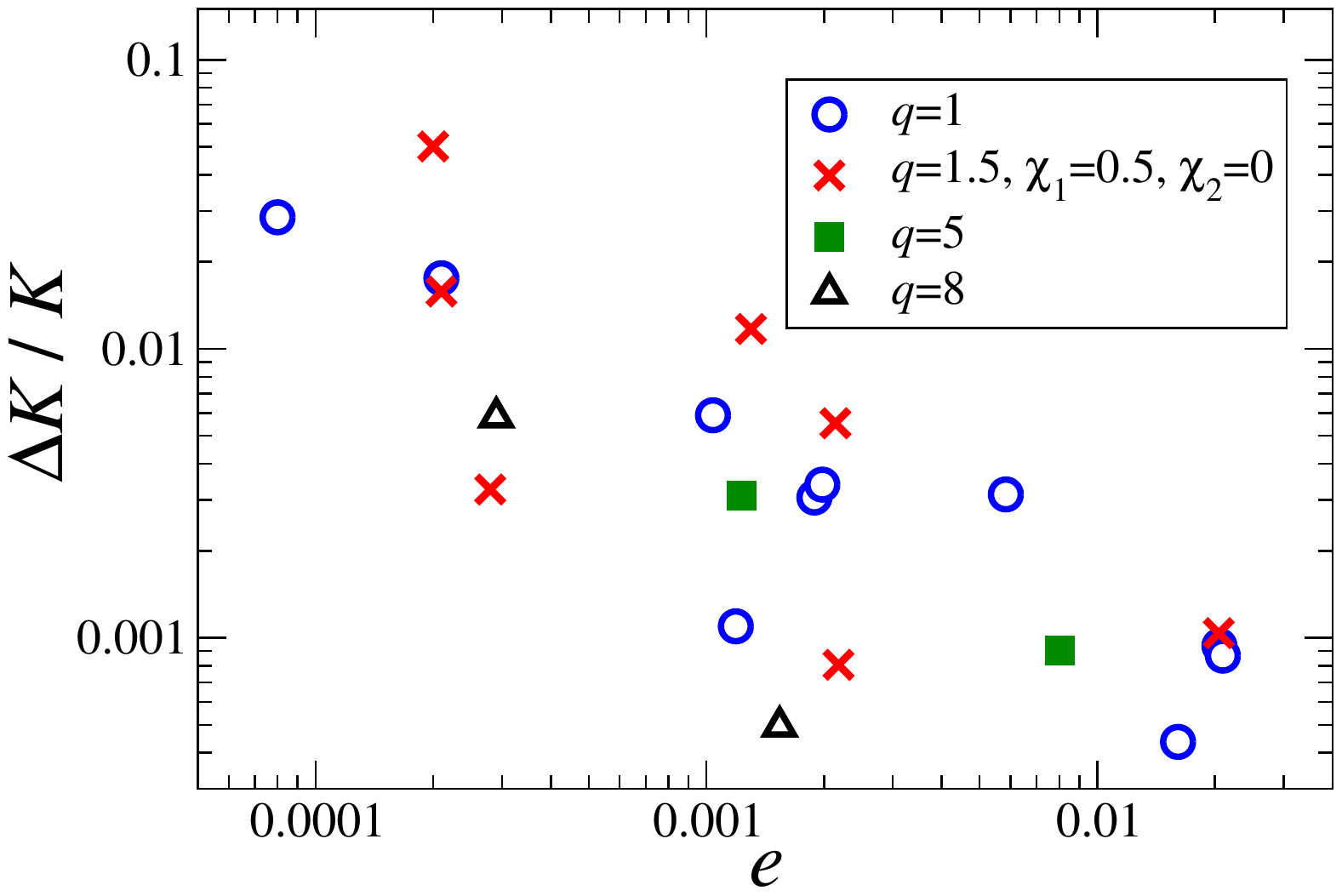}
	\caption{Relative uncertainty {$\Delta K / K$} in the numerical-relativity periastron advance as a function of the eccentricity $e$ of the configuration. Shown are data for four black-hole binaries with different mass ratios $q = m_1 / m_2$, one of them with a non-zero spin. Each symbol represents a separate numerical binary black hole evolution. The results shown here were computed at the orbital frequency $m\Omega_\varphi=0.033$.}
	\label{fig:pa-many-sim-all-ecc}
\end{figure}
\begin{figure}
	\includegraphics[width=\columnwidth]{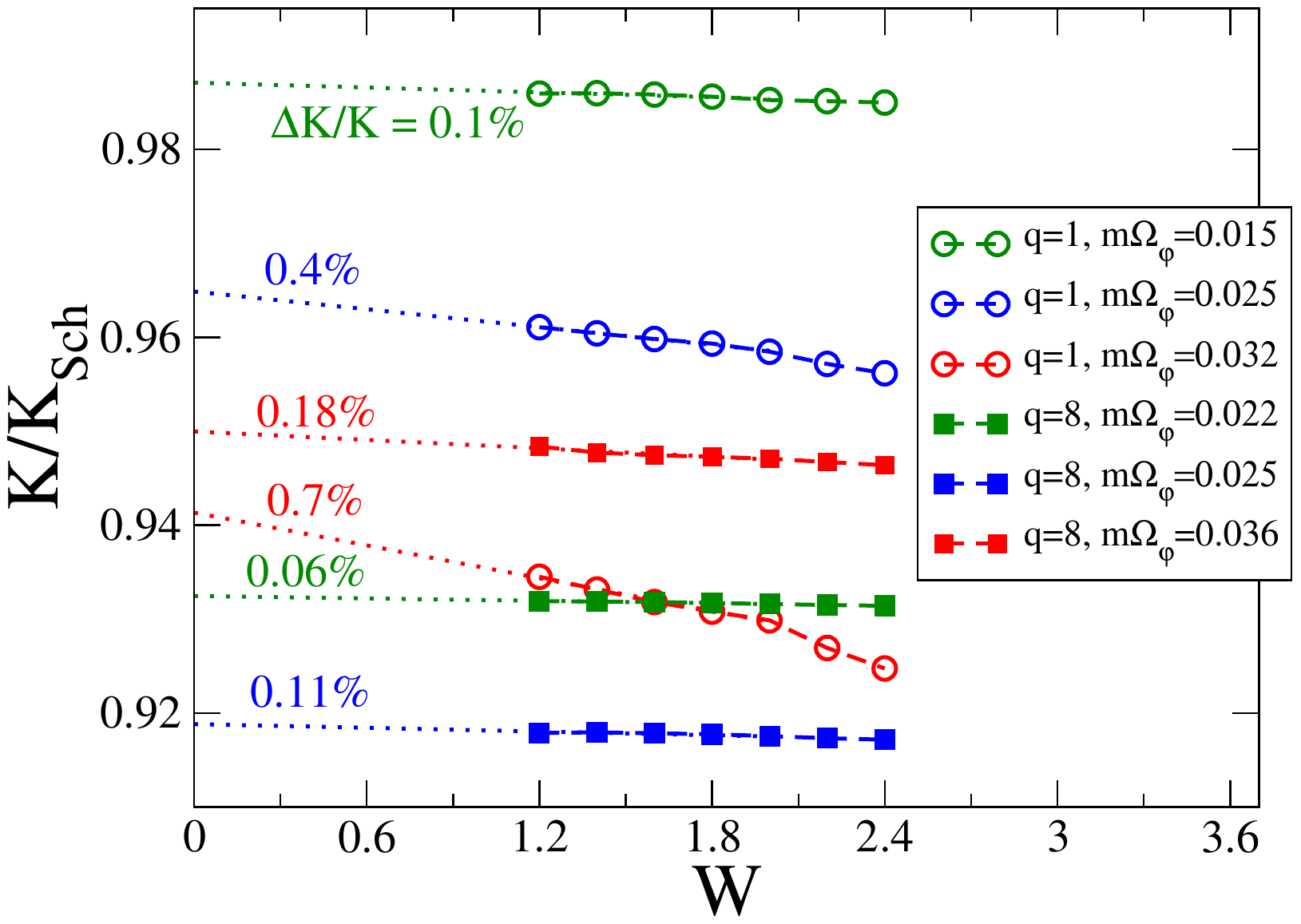}
	\caption{Effect of the choice of width {$\varpi$} on the measured periastron advance. Shown are data for the three reference frequencies $\Omega_{e/m/l}$ and for two exemplary runs: $(q,\chi_1,\chi_2) = (1,0,0)$ and $(q,\chi_1,\chi_2) = (8,0.5,0)$. The symbols denote $K_{\rm NR}/K_{\rm Sch}$ as measured with width {$\varpi$} indicated on the $x$-axis. The dotted lines denote fits indicating the extrapolation to zero width, ${\varpi}\to 0$. The number next to each dotted line indicates the fractional change in $K_{\rm NR}/K_{\rm Sch}$ between ${\varpi}=1.2$ and ${\varpi}\to 0$. For ease of plotting, the data for $q=8$ and $m\Omega_\varphi=0.036$ has been shifted up by $0.1$.}
	\label{fig:W-extrapolation}
\end{figure}

\subsection{Systematic effects}\label{sec:NR:systematics}

The procedure just outlined is subject to three effects which impact
$K_{\rm NR}$ at the $0.1-1\%$ level. The first of these
effects is already clearly visible in Fig.~\ref{fig:pa-q1-ecc}:
$K_{\rm NR}(\Omega_\varphi)$ as obtained by Eq.~(\ref{eq:K_NR})
oscillates around its mean.  These oscillations arise because the
fitting function \eqref{eq:fitting-function} does not perfectly
capture the features of $\Omega(t)$: eccentricity-related effects and the
radiation-reaction driven inspiral are more complicated than the rather
simple fitting formula \eqref{eq:fitting-function} used.  Early in the inspiral, these oscillations are
typically of order $0.1-0.2\%$, and they grow during the inspiral.  The
amplitude of these oscillations is furthermore strongly dependent on
the width {$\varpi$} of the fitting window.  This dependence arises because a
longer fitting interval includes a larger number of the
eccentricity-induced oscillations in $\Omega(t)$ that the fitting
function \eqref{eq:fitting-function} is designed to capture, and
therefore reduces the uncertainty of the fit.

A second important effect enters through the magnitude of the
eccentricity.  The oscillatory term in Eq.~(\ref{eq:fitting-function})
will be proportional to the eccentricity of the orbit.  With
decreasing eccentricity, this oscillatory term will be increasingly
hard to isolate and $\Omega_r$ will be increasingly difficult to
measure.  This effect is illustrated in
Fig.~\ref{fig:pa-many-sim-all-ecc} which provides a survey of NR
simulations at different eccentricities.  An eccentricity $e\sim
0.01$ typically allows one to measure $K$ with a relative accuracy of order $0.1\%$.
For smaller eccentricities, the uncertainty in $K_{\rm NR}$
increases roughly inversely proportionally to $e$. For larger
eccentricities, eventually the eccentricity-dependent corrections to
the periastron advance will become noticeable; the leading relative correction
is proportional to $e^2$, and hence still negligible for $e\sim 0.01$.
Figure~\ref{fig:pa-many-sim-all-ecc} shows data obtained at the orbital
frequency $m\Omega_\varphi=0.033$.  As one moves closer to the merger, the uncertainty
{$\Delta K$} increases. 

A third systematic effect arises from the choice of the width {$\varpi$} of
the fitting interval.  Larger {$\varpi$} systematically underestimate $K_{\rm NR}$
because the average radial frequency over the fitting interval is biased
toward larger values, as already visible in Fig.~\ref{fig:pa-q1-ecc}.
Figure~\ref{fig:W-extrapolation} demonstrates this drift more
clearly.  As can be seen, $K_{\rm NR}$ drifts by an amount of order $0.1\%$ to $1$\%;
the drift is generally smaller at large separations (where the
inspiral motion is very ``small''), and more pronounced at small
separations. This systematic error also gets smaller as 
the mass ratio of the binary increases (more unequal masses).

\subsection{Refined procedure}

The three effects described in Sec.~\ref{sec:NR:systematics} depend
strongly on the eccentricity $e$ of the run being analyzed, on the
width {$\varpi$} of the fitting interval, and on the orbital frequency
$\Omega_\varphi$ under consideration for each binary configuration.  
All three effects couple non-linearly, and have a large impact on how accurately $K_{\rm NR}$
can be measured at a given combination of $(e, {\varpi}, \Omega_\varphi)$. 
Furthermore, we generally do {\em not} have control over the
eccentricity $e$.  Numerical-relativity simulations are
computationally costly.  To maximize the scientific returns of these
simulations, we extract the periastron advance from simulations
originally performed for other purposes, even if the eccentricity is
smaller than desired for optimal extraction of $K_{\rm NR}$.  (The
data shown in Fig.~\ref{fig:pa-many-sim-all-ecc}, based
on Ref.~\cite{MrPf.12} is exceptional, as the goal of these
simulations was precisely the study of eccentricity).
Therefore, we proceed as follows for each BBH
configuration (specified by mass ratio and spins):
\begin{enumerate}
\item Pick three tentative target frequencies $\Omega_e$, $\Omega_m$,
  and $\Omega_l$.  These are chosen to fall into the early inspiral,
  into the middle of the inspiral, and late in the inspiral, but such
  that for all three frequencies we can still obtain good periastron
  advance measurements.
\item \label{step:2} If simulations with different orbital eccentricities are available
  for the considered configuration, perform fits similar to
  those shown in Fig.~\ref{fig:pa-q1-ecc} for each available eccentricity.  Manually
  assess which eccentricity gives the most reliable fits (these can be
  different runs at the various frequencies $\Omega_{e/m/l}$).
  Determine an error bar on $K_{\rm NR}$ from manual inspection.
\item Consider the dependence on {$\varpi$} by using plots similar to
  Fig.~\ref{fig:W-extrapolation}.  Take the periastron advance
  extrapolated to ${\varpi}\to 0$ as the final value reported.  If the change
  in $K_{\rm NR}$ between ${\varpi}=1.2$ and ${\varpi}\to 0$ is larger than the
  error bar determined in step~\ref{step:2}, then increase the
  error bar to this difference. 
\item To obtain convenient analytical approximations of the behavior
  of $K_{\rm NR}/K_{\rm Sch}$, fit the values for $K_{\rm
    NR}/K_{\rm Sch}$ at the three frequencies $\Omega_{e/m/l}$ with a 
  quadratic polynomial in $m \Omega_\varphi$, 
  \begin{equation}
    \frac{K_{\rm NR}}{K_{\rm Sch}} = a_0 + a_1 \, (m\Omega_\varphi) + a_2 \, (m\Omega_\varphi)^2.
  \end{equation}
\end{enumerate}
Because of the variety of simulations to be analyzed, manual
inspection as indicated in the procedure above was crucial to improve
the accuracy of $K_{\rm NR}$ over the earlier, more automatic procedure
used in Ref.~\cite{Le.al.11}. Table~\ref{tab:NR-fits} lists the numerical
results for the periastron advance obtained for the simulations considered here.

\begin{table*}
	\ra{1.2}
	\begin{tabular}{@{}rrrrcrrrcrrrcrrrcrr@{}}
		\toprule
		& & & & \phantom{a} & \multicolumn{3}{c}{$K$} & \phantom{a} & \multicolumn{3}{c}{$K + \Delta K$} & \phantom{a} & \multicolumn{3}{c}{$K - \Delta K$} \\
		\cline{6-8} \cline{10-12} \cline{14-16}
		$q$ & $\chi_1$ & $\chi_2$ & $10^4e$ && $a_0$ & $a_1$ & $a_2$ && $a_0$ & $a_1$ & $a_2$ && $a_0$ & $a_1$ & $a_2$ && $ m\Omega_i$ & $ m\Omega_f$ \\
		\hline
		$1$ & $0.97$& $0.97$ & 6 && $1.00764$ & $-3.9948$ & $-70.807$ && $1.0065$ & $-3.9405$ & $-67.121$ && $0.99417$ & $-2.7579$ & $-101.543$ && $0.0169$ & $0.0344$ \\
		$1$ & $0.95$ & $0.95$ & 1 && $0.98829$ & $-2.2363$ & $-107.11$ && $0.99952$ & $-3.2597$ & $-79.724$ && $0.98340$ & $-1.7802$ & $-122.724$ && $0.0184$ & $0.0318$ \\
		$1$ & $0.9$ & $0.9$ & 5 && $0.96487$ & $-0.3254$ & $-138.67$ && $0.96828$ & $-0.5883$ & $-130.568$ && $0.99319$ & $-2.6814$ & $-94.833$ && $0.020$ & $0.031$ \\
		$1$ & $0.8$ & $0.8$ & 5 && $0.98881$ & $-1.8427$ & $-104.636$ && $1.00304$ & $-3.1415$ & $-73.025$ && $0.97868$ & $-0.9218$ & $-127.882$ && $0.0177$ & $0.0317$ \\
		$1$ & $0.6$ & $0.6$ & 4 &&  $0.99922$ & $-2.0355$ & $-86.060$ && $1.01226$ & $-3.1796$ & $-56.734$ && $0.97886$ & $-0.2337$ & $-128.612$ && $0.019$ & $0.031$ \\
		$1$ & $-0.9$ & $-0.9$ & 7 && $0.96721$ & $6.4391$ & $-34.411$ && $1.3842$ & $-38.2326$ & $1175.23$ && $0.68088$ & $37.9372$ & $-908.291$ && $0.0177$ & $0.024$ \\
		$1$ & $-0.95$ & $-0.95$ & 10 && $1.09949$ & $-7.4342$ & $346.477$ && $1.32874$ & $-33.6076$ & $1099.42$ && $0.78659$ & $26.3466$ & $-570.337$ && $0.0177$ & $0.026$ \\
		$1$ & $0.5$ & $0$ & 3 && $0.98950$ & $0.2892$ & $-106.77$ && $1.01884$ & $-3.0265$ & $-8.075$ && $0.957$ & $3.8257$ & $-210.184$ && $0.0155$ & $0.025$ \\
		$1$ & $0$ & $0$ & 282 && $0.99554$ & $0.5048$ & $-76.340$ && $0.99678$ & $0.2800$ & $-62.419$ && $0.99430$ & $0.7296$ & $-90.261$ && $0.012$ & $0.032$ \\
		$1$ & $-0.5$ & $0$ & 4 && $0.93781$ & $6.5574$ & $-171.793$ && $1.2331$ & $-23.1674$ & $588.235$ && $0.84533$ & $17.1947$ & $-486.223$ && $0.0195$ & $0.0259$ \\
		$1.5$ & $0.5$ & $0$ & 0.6 && $0.97522$ & $1.4334$ & $-139.448$ && $1.03313$ & $-4.6662$ & $30.686$ && $0.92706$ & $6.5006$ & $-281.776$ && $0.0158$ & $0.0259$ \\
		$1.5$ & $0$ & $0$ & 228 && $0.99849$ & $0.1745$ & $-66.444$ && $1.00508$ & $-0.6835$ & $-36.986$ && $0.99190$ & $1.0326$ & $-95.902$ && $0.013$ & $0.032$ \\
		$1.5$ & $-0.5$ & $0$ & 25 && $0.99987$ & $1.0477$ & $-30.021$ && $1.00286$ & $0.6444$ & $-15.295$ && $0.99588$ & $1.5908$ & $-49.195$ && $0.0123$ & $0.0215$ \\
		$3$ & $0.5$ & $0$ & 3 && $1.00301$ & $-1.7335$ & $-65.616$ && $1.02202$ & $-3.7817$ & $-7.465$ && $0.99159$ & $-0.4448$ & $-105.151$ && $0.0164$ & $0.0287$ \\
		$3$ & $0$ & $0$ & 21 && $1.00277$ & $-0.0865$ & $-50.201$ && $1.0178$ & $-1.5553$ & $-11.582$ && $0.98773$ & $1.3822$ & $-88.819$ && $0.019$ & $0.029$ \\
		$3$ & $-0.5$ & $0$ & 229 && $1.00559$ & $0.7584$ & $17.064$ && $1.01162$ & $0.0920$ & $38.352$ && $0.99854$ & $1.5502$ & $-8.129$ && $0.013$ & $0.027$ \\
		$5$ & $0.5$ & $0$ & 356 && $0.99812$ & $-1.2904$ & $-76.358$ && $0.99779$ & $-1.1426$ & $-79.708$ && $0.99845$ & $-1.4382$ & $-73.008$ && $0.0169$ & $0.0280$ \\
		$5$ & $0$ & $0$ & 367 && $0.99279$ & $0.7364$ & $-54.033$ && $1.00428$ & $-0.1182$ & $-36.789$ && $0.98130$ & $1.5911$ & $-71.276$ && $0.020$ & $0.041$ \\
		$5$ & $-0.5$ & $0$ & 229 && $1.02734$ & $-1.3157$ & $101.025$ && $1.03345$ & $-1.9244$ & $117.851$ && $1.02648$ & $-1.2086$ & $95.785$ && $0.0179$ & $0.036$ \\
		$8$ & $0.5$ & $0$ & 37 && $0.97198$ & $0.7118$ & $-114.923$ && $0.98182$ & $0.0285$ & $-102.411$ && $0.96137$ & $1.4528$ & $-128.537$ && $0.021$ & $0.042$ \\ 
		$8$ & $0$ & $0$ & 84 && $0.99868$ & $0.2793$ & $35.300$ && $1.0045$ & $-0.2028$ & $-24.723$ && $0.98878$ & $1.0538$ & $50.982$ && $0.021$ & $0.036$ \\
		$8$ & $-0.5$ & $0$ & 17 && $1.02556$ & $-1.2577$ & $130.85$ && $1.05938$ & $-4.3455$ & $203.072$ && $0.99952$ & $1.2217$ & $69.698$ && $0.020$ & $0.030$ \\
		\botrule
	\end{tabular}
	\caption{\textit{Fitting parameters for the NR data.} Here $q = m_1 / m_2$ is the mass ratio, $m = m_1 + m_2$ the total mass, $\chi_i = S_i / m_i^2$ (with $i=1,2$) the dimensionless spins, and $e$ the eccentricity. The fits are of the form $K = [a_0 + a_1 (m\Omega_\varphi) + a_2(m\Omega_\varphi)^2] / [1 - 6(m\Omega_\varphi)^{2/3}]^{1/2}$. The estimated uncertainties $K\pm \Delta K$ have a similar format. The fitting parameters $(a_0,a_1,a_2)$ are computed for the restricted frequency range $\Omega_i \leqslant \Omega_\varphi \leqslant \Omega_f$.} 
	\label{tab:NR-fits}
\end{table*}

\section{Post-Newtonian Approximation}\label{sec:PN}

\subsection{Post-Newtonian calculation to 3.5PN order}

In the context of the post-Newtonian approximation to general relativity, we consider a binary system of spinning point particles (modeling two rotating black holes) with constant masses $m_i$ ($i=1,2$) and canonical spins $\mathbf{S}_i = S_i \, \hat{\mathbf{L}}$ aligned or anti-aligned with the orbital angular momentum $\mathbf{L} = L \, \hat{\mathbf{L}}$, with $\hat{\mathbf{L}}$ the unit vector pointing in the direction of $\mathbf{L}$, such that $L > 0$ and $\vert S_i \vert < m_i^2$. In this section, using the results of Ref.~\cite{Te.al.13} we explicitly write down the PN expression of the periastron advance for circular orbits, including all spin-independent, spin-orbit (SO), and spin-spin (SS) contributions up to 3.5PN order included. Higher-order interactions in the spins \cite{HeSc.08,HeSc2.08} will be neglected; hence we do not include the leading-order 3.5PN terms cubic in the spins. We restrict to the conservative part of the dynamics, neglecting the dissipative effects related to gravitational-wave emission.

Reference \cite{Te.al.13} provides an explicit, 3.5PN-accurate solution of the orbital equations of motion of a binary system of spinning point particles (at quadratic order in the spins $S_i$), for a generic bound orbit and aligned or anti-aligned spins, in the form of a quasi-Keplerian parametrization of the motion.\footnote{The expressions for the mean motion $n$ and periastron advance per radial period $\Phi$ as functions of $\vert E \vert$ and $L$ were not given in Ref.~\cite{Te.al.13}. We thank M. Tessmer and J. Hartung for making these results available to us.} The orbital elements are expressed in terms of the two constants of the motion: the reduced binding energy $\varepsilon \equiv \vert E \vert / (m \nu)$ (recall that $E < 0$ for bound orbits) and the dimensionless angular momentum $h \equiv L / (m^2 \nu)$, where $m = m_1 + m_2$ is the total mass and $\nu = m_1 m_2 / m^2$ the symmetric mass ratio, such that $\nu = 1/4$ for equal masses and $\nu \to 0$ in the extreme mass-ratio limit. The 3.5PN expression of the (reduced) periastron advance per radial period, $K \equiv \Phi / (2 \pi)$, reads\footnote{We use the black-hole value $C_Q = 1$ for the constant parameter characterizing the quadrupolar deformation of a compact object under the effect of its intrinsic rotation \cite{Te.al.13}.}
\begin{widetext}
	\begin{align}\label{K}
		K &= 1 + \frac{3}{h^2} - \left( 2 + 2 \Delta - \nu \right) \frac{\chi_1}{h^3} + \left( - \frac{15}{2} + 3 \nu \right) \frac{\varepsilon}{h^2} + \left( \frac{105}{4} - \frac{15}{2} \nu \right) \frac{1}{h^4} + \left( \frac{3}{4} + \frac{3}{4} \Delta - \frac{3}{2} \nu \right) \frac{\chi_1^2}{h^4} \nonumber \\ &\qquad\! + \left( 12 + 12 \Delta - 16 \nu - 4 \Delta \, \nu + 2 \nu^2 \right) \chi_1 \, \frac{\varepsilon}{h^3} + \left( - 42 - 42 \Delta + \frac{147}{4} \nu + \frac{21}{4} \Delta \, \nu - \frac{3}{2} \nu^2 \right) \frac{\chi_1}{h^5} \nonumber \\ &\qquad\! + \left( \frac{1155}{4} - \left[ \frac{625}{2} - \frac{615}{128} \pi^2 \right] \nu + \frac{105}{8} \nu^2 \right) \frac{1}{h^6} + \left( - \frac{315}{2} + \left[ 218 - \frac{123}{64} \pi^2 \right] \nu - \frac{45}{2} \nu^2 \right) \frac{\varepsilon}{h^4} \nonumber \\ &\qquad\! + \left( \frac{15}{4} - \frac{15}{4} \nu + 3 \nu^2 \right) \frac{\varepsilon^2}{h^2} + \left( \frac{105}{2} + \frac{105}{2} \Delta - 135 \nu - 30 \Delta \, \nu + \frac{45}{4} \nu^2 \right) \frac{\chi_1^2}{h^6} + \left( - \frac{33}{2} - \frac{33}{2} \Delta \right. \nonumber \\ &\left. \qquad\qquad + \frac{93}{2} \nu + \frac{27}{2} \Delta \, \nu - \frac{15}{2} \nu^2 \right) \frac{\varepsilon \, \chi_1^2}{h^4} + \left( - \frac{1485}{2} - \frac{1485}{2} \Delta + \frac{15165}{16} \nu + \frac{5265}{16} \Delta \, \nu - \frac{345}{2} \nu^2 \right. \nonumber \\ &\left. \qquad\qquad - \frac{75}{8} \Delta \, \nu^2 + \frac{15}{8} \nu^3 \right) \frac{\chi_1}{h^7} + \left( 420 + 420 \Delta - 717 \nu - 297 \Delta \, \nu + 207 \nu^2 + 21 \Delta \, \nu^2 - 6 \nu^3 \right) \frac{\chi_1 \, \varepsilon}{h^5} \nonumber \\ &\qquad\! + \left( - 15 - 15 \Delta + 42 \nu + \frac{39}{2} \Delta \, \nu - 27 \nu^2 - 6 \Delta \, \nu^2 + 3 \nu^3 \right) \frac{\chi_1 \, \varepsilon^2}{h^3} + 1 \leftrightarrow 2 + \calO(c^{-8}) \, , 
	\end{align}
\end{widetext}
where $\Delta \equiv (m_1 - m_2) / m = \sqrt{1-4\nu}$ is the reduced mass difference and $\chi_1 \equiv S_1 / m_1^2$ the dimensionless spin of particle $1$. (We assume, without any loss of generality, that $m_1 \geqslant m_2$.) The symbol $1 \leftrightarrow 2$ stands for all the \textit{spin-dependent} terms with the particle labels $1$ and $2$ exchanged ($\chi_1 \leftrightarrow \chi_2$ and $\Delta \to - \Delta$) that have to be added to the previous expression.

We now restrict to a circular orbit with constant azimuthal frequency $\Omega_\varphi$, and make use of the well-known expressions of $\varepsilon$ and $h$ as functions of the usual dimensionless, invariant PN parameter $x \equiv {(m \Omega_\varphi)}^{2/3}$. When including the leading-order 1.5PN and next-to-leading order 2.5PN spin-orbit couplings, as well as the leading-order 2PN spin-spin couplings, those expressions read~\cite{DaSc.88,Da.al3.00,Bl.al.06,Da.al.08}:
\begin{widetext}
	\begin{subequations}\label{E_h}
		\begin{align}
			\varepsilon &= \frac{x}{2} \, \biggl\{ 1 + \left( - \frac{3}{4} - \frac{\nu}{12} \right) x + \left( \frac{4}{3} + \frac{4}{3} \Delta - \frac{2}{3} \nu \right) \chi_1 \, x^{3/2} + \left( - \frac{27}{8} + \frac{19}{8} \nu - \frac{\nu^2}{24} \right) x^2 - \frac{1}{2} \left( 1 + \Delta - 2\nu \right) \chi_1^2 \, x^2 \nonumber \\ &\qquad\quad\;\, - \nu \, \chi_1 \chi_2 \; x^2 + \left( 4 + 4 \Delta - \frac{121}{18} \nu - \frac{31}{18} \Delta \, \nu + \frac{\nu^2}{9} \right) \chi_1 \, x^{5/2} + 1 \leftrightarrow 2 + \calO(x^3) \biggr\} \, , \\
			h &= \frac{1}{\sqrt{x}} \, \biggl\{ 1 + \left( \frac{3}{2} + \frac{\nu}{6} \right) x + \left( - \frac{5}{3} - \frac{5}{3} \Delta + \frac{5}{6} \nu \right) \chi_1 \, x^{3/2} + \left( \frac{27}{8} - \frac{19}{8} \nu + \frac{\nu^2}{24} \right) x^2 + \left( \frac{1}{2} + \frac{\Delta}{2} - \nu \right) \chi_1^2 \, x^2 \nonumber \\ &\qquad\qquad\, + \nu \, \chi_1 \chi_2 \; x^2 + \left( - \frac{7}{2} - \frac{7}{2} \Delta + \frac{847}{144} \nu + \frac{217}{144} \Delta \, \nu - \frac{7}{72} \nu^2 \right) \chi_1 \, x^{5/2} + 1 \leftrightarrow 2 + \calO(x^3) \biggr\} \, .
		\end{align}
	\end{subequations}
\end{widetext}
Note that to control the expansion for $K(x)$ up to 3.5PN order, we only need the expressions for $\varepsilon(x)$ and $h(x)$ at the relative 2.5PN accuracy. The expressions \eqref{E_h} can also be recovered from the quasi-Keplerian parametrization of Ref.~\cite{Te.al.13}, by imposing the zero-eccentricity condition $e_t = 0$ (or equivalently $e_r = 0$ or $e_\varphi = 0$) appropriate for a circular orbit.

Replacing the formulas \eqref{E_h} into Eq.~\eqref{K}, and expanding in powers of $1/c$, we obtain the 3.5PN result for the invariant relation $K(x;\nu,\chi_1,\chi_2)$, which can conveniently be split into non-spinning, spin-orbit, and spin-spin contributions:
\beq\label{K_PN}
	K = K_\text{NS} + K^\text{LO}_\text{SO} + K^\text{LO}_\text{SS} + K^\text{NLO}_\text{SO} + K^\text{NLO}_\text{SS} + K^\text{NNLO}_\text{SO} + \calO(c^{-8}) \, .
\eeq
The non-spinning (NS) contribution $K_\text{NS}$ is accurate to 3.5PN order. The leading-order (LO), next-to-leading order (NLO), and next-to-next-to-leading order (NNLO) spin-orbit (SO) terms $K^\text{LO}_\text{SO}$, $K^\text{NLO}_\text{SO}$ and $K^\text{NNLO}_\text{SO}$ contribute at 1.5PN, 2.5PN, and 3.5PN order, respectively. The leading-order 2PN and next-to-leading order 3PN spin-spin (SS) contributions can themselves be split into self-spin ($S_1^2$ and $S_2^2$) and cross-spin ($S_1S_2$) interactions: $K^\text{LO}_\text{SS} = K^\text{LO}_{S^2} + K^\text{LO}_{S_1S_2}$ and $K^\text{NLO}_\text{SS} = K^\text{NLO}_{S^2} + K^\text{NLO}_{S_1S_2}$. All these contributions explicitly read
\begin{subequations}\label{K_pieces}
	\begin{align}
		K_\text{NS} &= 1 + 3 x + \left( \frac{27}{2} - 7 \nu \right) x^2 \nonumber \\ &\quad + \left( \frac{135}{2} - \left[ \frac{649}{4} - \frac{123}{32} \pi^2 \right] \nu + 7 \nu^2 \right) x^3 \, , \label{K_NS} \\
		K^\text{LO}_\text{SO} &= \left( - 2 - 2 \Delta + \nu \right) \chi_1 \, x^{3/2} + 1 \leftrightarrow 2 \, , \label{K_SO^LO} \\
		K^\text{LO}_{S^2} &= \left( \frac{3}{4} + \frac{3}{4} \Delta - \frac{3}{2} \nu \right) \chi_1^2 \, x^2 + 1 \leftrightarrow 2 \, , \label{K_S2^LO} \\
		K^\text{LO}_{S_1S_2} &= 3 \nu \, \chi_1 \chi_2 \; x^2 \, , \label{K_S1S2^LO} \\
		K^\text{NLO}_\text{SO} &= \left( - 17 - 17 \Delta + \frac{81}{4} \nu + \frac{17}{4} \Delta \, \nu - \nu^2 \right) \chi_1 \, x^{5/2} \nonumber \\ &\quad+ 1 \leftrightarrow 2 \, , \label{K_SO^NLO} \\
		K^\text{NLO}_{S^2} &= \left( \frac{67}{4} + \frac{67}{4} \Delta - \frac{189}{4} \nu - \frac{55}{4} \Delta \, \nu + 6 \nu^2 \right) \chi_1^2 \, x^3 \nonumber \\ &\quad + 1 \leftrightarrow 2 \, , \label{K_S2^NLO} \\
		K^\text{NLO}_{S_1S_2} &= \left( 45 + 2 \nu \right) \nu \, \chi_1 \chi_2 \; x^3 \, , \label{K_S1S2^NLO} \\
		K^\text{NNLO}_\text{SO} &= \left( - 126 - 126 \Delta + \frac{11581}{48} \nu + \frac{5317}{48} \Delta \, \nu \right . \nonumber \\ & \quad \left. - \frac{733}{12} \nu^2 - \frac{11}{3} \Delta \, \nu^2 + \frac{\nu^3}{3} \right) \chi_1 \, x^{7/2} + 1 \leftrightarrow 2 \, . \label{K_SO^NNLO}
	\end{align}
\end{subequations}
The NS contribution \eqref{K_NS} is a strictly increasing function of frequency for all mass ratios ($0 \leqslant \nu \leqslant 1/4$). The 2PN and 3PN $S_1^2$ and $S_2^2$ contributions \eqref{K_S2^LO} and \eqref{K_S2^NLO} are positive for all spins and mass ratios, while the $S_1S_2$ contributions \eqref{K_S1S2^LO} and \eqref{K_S1S2^NLO} are positive if $\sgn{(S_1 S_2)} > 0$ and negative otherwise. The 1.5PN, 2.5PN and 3.5PN SO contributions \eqref{K_SO^LO}, \eqref{K_SO^NLO} and \eqref{K_SO^NNLO} are all negative (resp. positive) when both spins are aligned (resp. anti-aligned) with the angular momentum.

To ease the comparison with the perturbative result derived in Sec.~\ref{sec:pert} below, we also compute the quantity $W \equiv 1 / K^2$ introduced in Refs.~\cite{Da.10,Ba.al.10}. The 3.5PN-accurate expression for $W(x;\nu,\chi_1,\chi_2)$ is
\begin{widetext}
	\begin{align}\label{W_PN}
		W &= 1 - 6 x + \bigl[ \, \left( 4 + 4 \Delta - 2 \nu \right) \chi_1
  + \left( 4 - 4 \Delta - 2 \nu \right) \chi_2 \, \bigr] \, x^{3/2}
  + \biggl[ \, 14 \nu + \left( - \frac{3}{2} - \frac{3}{2} \Delta + 3 \nu \right) \chi_1^2
  - 6 \nu \, \chi_1 \chi_2 + \left( - \frac{3}{2} + \frac{3}{2} \Delta + 3 \nu \right) \chi_2^2
  \, \biggr] \, x^2 \nonumber \\ 
&\qquad\!\!\! - \biggl[ \left( 2 + 2
    \Delta + \frac{45}{2} \nu + \frac{17}{2} \Delta \, \nu - 2 \nu^2
  \right) \chi_1 + \left( 2 - 2 \Delta + \frac{45}{2} \nu -
    \frac{17}{2} \Delta \, \nu - 2 \nu^2 \right) \chi_2 \, \biggr] \,
  x^{5/2} + \biggl[ \left( \frac{397}{2} -
    \frac{123}{16} \pi^2 \right) \nu - 14 \nu^2 \nonumber \\ 
&\qquad\qquad\, + \left( 4 + 4 \Delta + \frac{15}{2} \nu + \frac{31}{2} \Delta \, \nu - 9
    \nu^2 \right) \chi_1^2 + \left( 36 + 2 \nu
  \right) \nu \, \chi_1 \chi_2 + \left( 4 - 4 \Delta + \frac{15}{2} \nu -
    \frac{31}{2} \Delta \, \nu - 9 \nu^2 \right) \chi_2^2 \, \biggr]
  \, x^3 \nonumber \\
&\qquad\!\!\! - \biggl[ \left( \frac{1465}{24} +
    \frac{1465}{24} \Delta - \frac{373}{6} \nu - \frac{22}{3} \Delta
    \, \nu + \frac{2}{3} \nu^2 \right) \nu \, \chi_1 + \left( \frac{1465}{24} - \frac{1465}{24} \Delta -
    \frac{373}{6} \nu + \frac{22}{3} \Delta \, \nu + \frac{2}{3} \nu^2
  \right) \nu \, \chi_2 \, \biggr] \, x^{7/2} + \calO(x^4) \, .
	\end{align}
\end{widetext}
Note that the 3.5PN spin-orbit terms in $W$ vanish in the test-particle limit $\nu \to 0$. Recall, however, that we are missing some contributions $\calO(S^3)$ at 3.5PN order, which may not vanish in that limit. Notice also that Eq.~\eqref{W_PN} is invariant by exchange $1 \leftrightarrow 2$ of the bodies' labels.

\begin{figure}
	\includegraphics[scale=0.37]{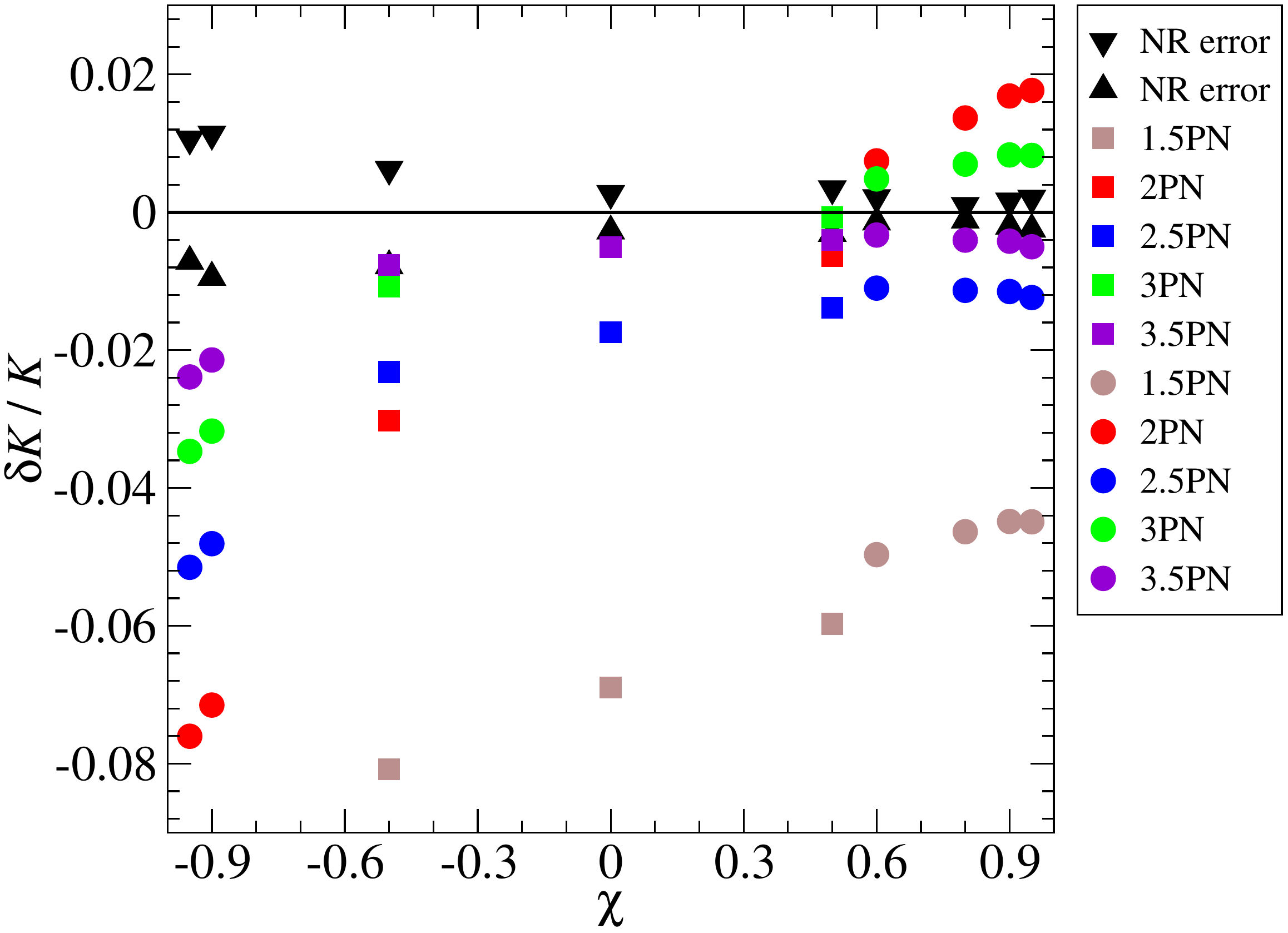}
	\caption{Fractional difference between the NR and PN predictions for the periastron advance $K$ as a function of spin, at different PN orders, for equal-mass black-hole binaries. We set $m\Omega_\varphi = 0.021$.}
	\label{fig:Fig1}
\end{figure}
\begin{figure}
	\includegraphics[scale=0.5]{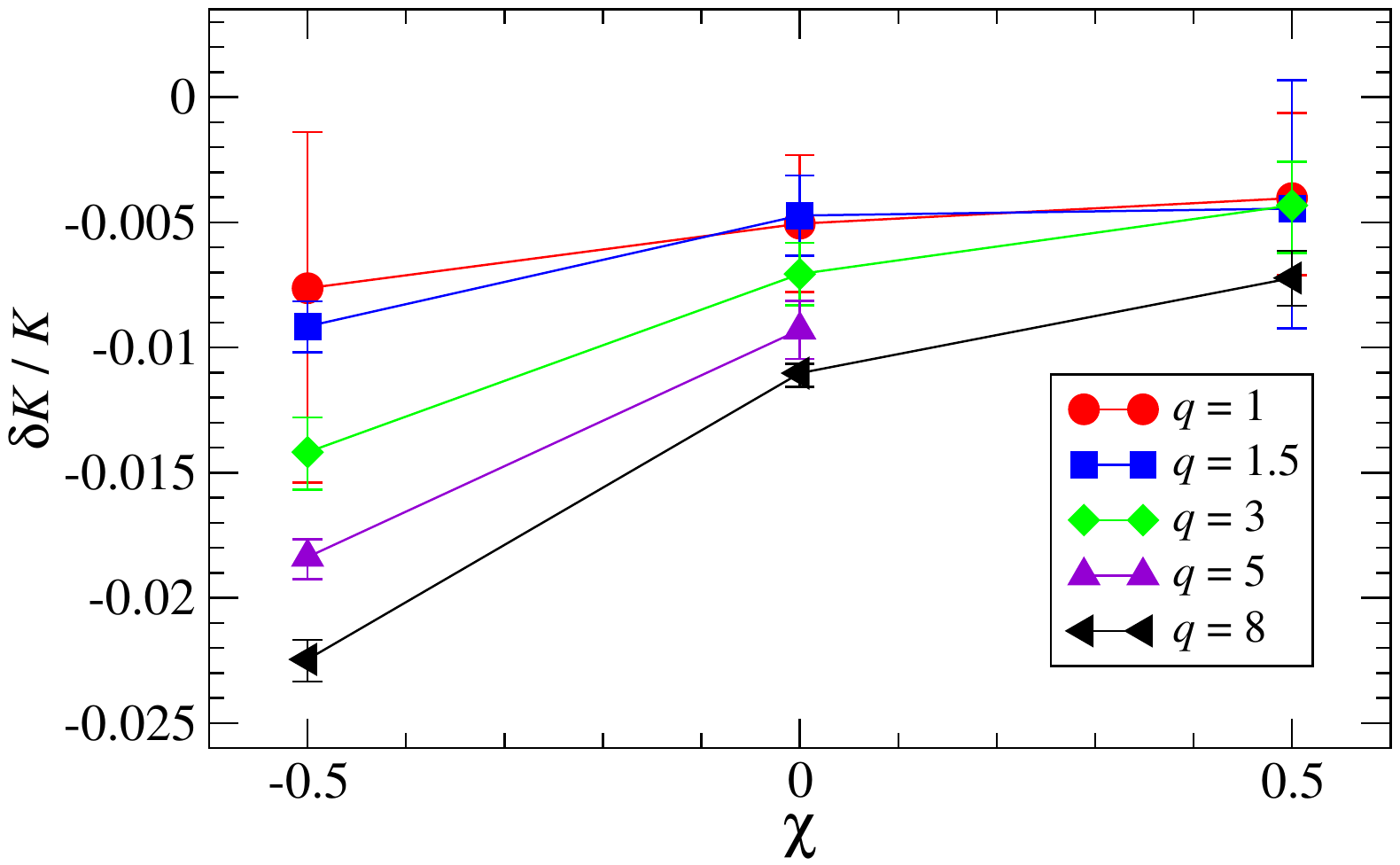}
	\caption{Fractional difference between the NR and PN predictions for $K$ for black-hole binaries with mass ratios $q \in \{1,1.5,3,5,8\}$ and spins $\chi \equiv \chi_1 \in \{-0.5,0,0.5\}$ and $\chi_2=0$. We set $m\Omega_\varphi = 0.021$.} 
	\label{fig:Fig2}
\end{figure}
\begin{figure}[t!]
	\includegraphics[scale=0.39]{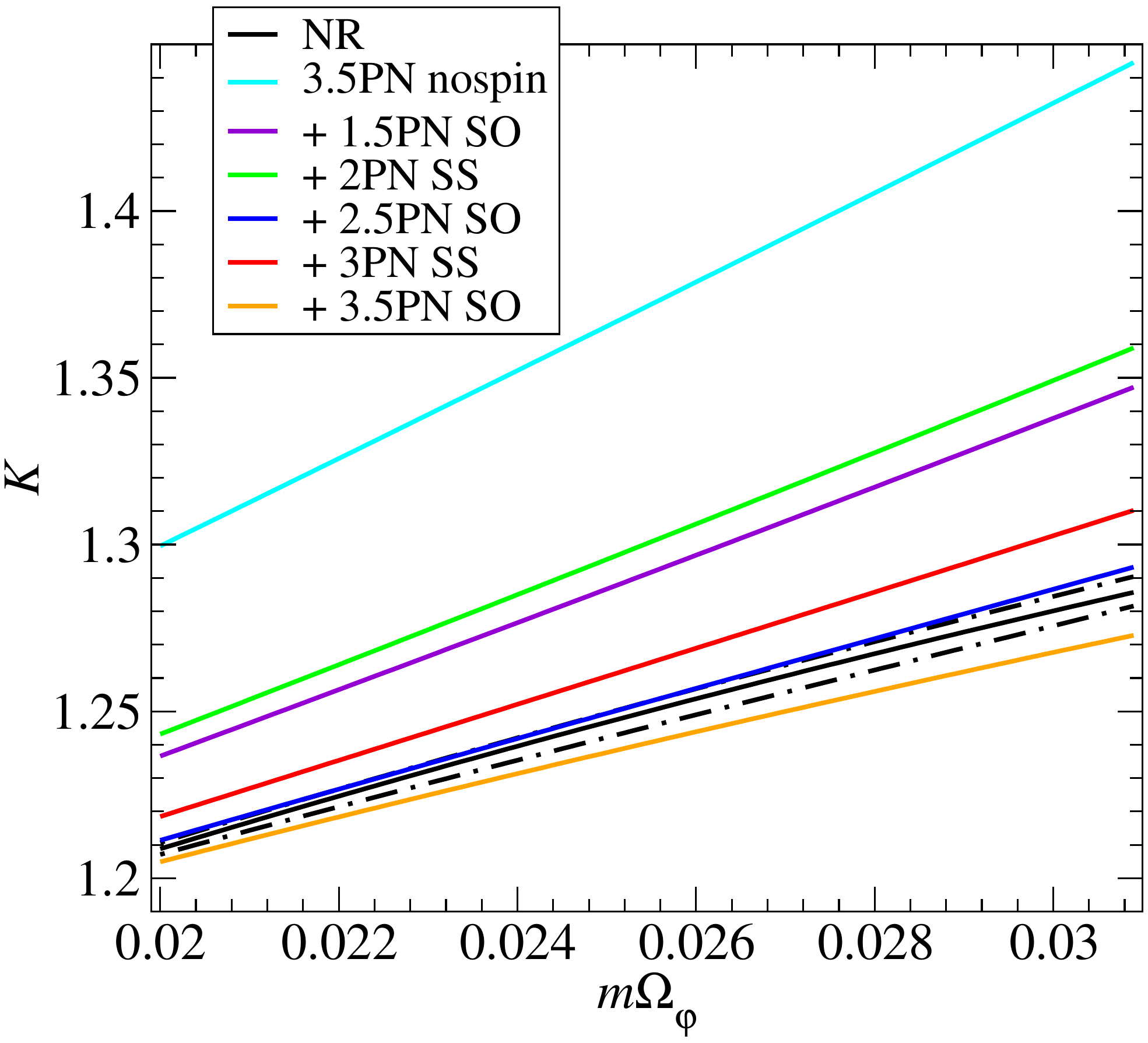}
	\includegraphics[scale=0.39]{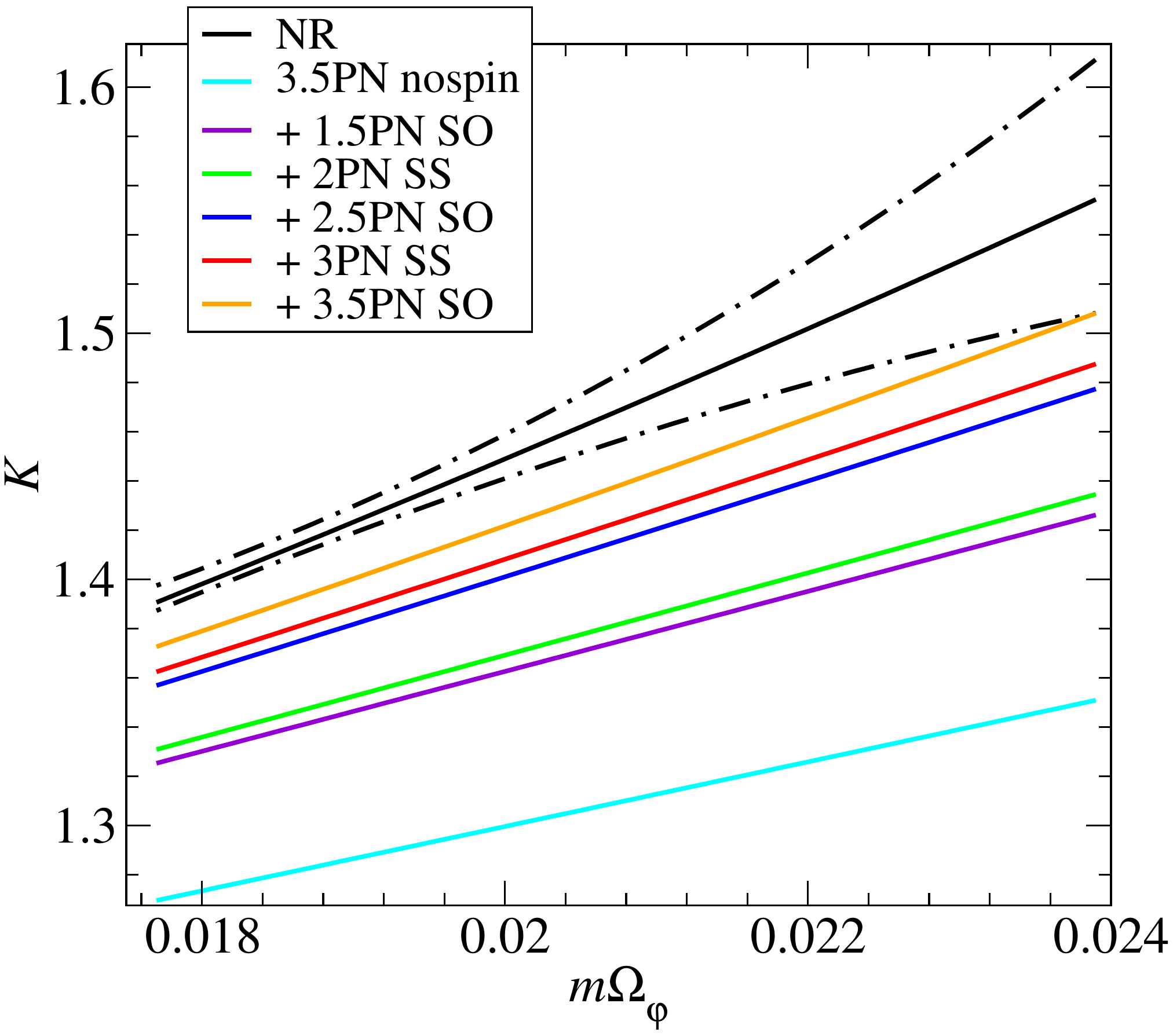}
	\caption{Periastron advance $K$ as a function of the orbital frequency $m\Omega_\varphi$, for equal-mass binaries with equal spins $\chi_1 = \chi_2=0.9$ (top) and $\chi_1 = \chi_2=-0.9$ (bottom). The black dashed lines show the estimated numerical-relativity uncertainties.}
	\label{fig:Figs3-4}
\end{figure}

\subsection{Comparison to numerical-relativity simulations}

We now compare the PN prediction \eqref{K_PN}--\eqref{K_pieces} with the NR results discussed in Sec.~\ref{sec:NR}. In Fig.~\ref{fig:Fig1} we show the fractional difference between the NR and PN predictions for $K$ as a function of spin, at different PN orders, for equal-mass black-hole binaries. We compute the periastron advance at the orbital frequency $m\Omega_\varphi = 0.021$, which is typically in the middle of the NR frequency range. We indicate with a dot the simulations in which both black holes are spinning and with a square the simulations in which only one black hole is spinning. For spins anti-aligned with the orbital angular momentum, the various contributions \eqref{K_pieces} are all positive, such that the successive PN approximations approach the NR results in a monotonic way. For spins aligned with the orbital angular momentum, the spin-squared contributions are still positive, but the spin-orbit ones are negative, such that the successive PN approximations approach the NR results in a non-monotonic way. At the moderate orbital frequency $m\Omega_\varphi = 0.021$, the 3.5PN results are almost within the numerical errors, with a relative difference of $1\%$ at most (except for large negative spins).

In Fig.~\ref{fig:Fig2} we plot the fractional difference between the NR and 3.5PN predictions for the periastron advance $K$, for black-hole binaries with mass ratios $q \in \{1,1.5,3,5,8\}$ and spins $\chi_1 \in \{-0.5,0,0.5\}$ and $\chi_2=0$, still at the orbital frequency $m\Omega_\varphi = 0.021$. The performance of the PN approximation deteriorates as the mass ratio increases (more unequal masses), consistent with previous findings \cite{Le.al.11, Ba.al.12}. This result is robust to changes in the orbital frequency. 

Figure \ref{fig:Figs3-4} shows the periastron advance $K$ as a function of the orbital frequency $m\Omega_\varphi$ for equal-mass binaries with equal spins $\chi_1 = \chi_2 = 0.9$ (top) and $\chi_1 = \chi_2 = -0.9$ (bottom). We show the NR results (black continuous curves) with their errors (black dashed curves) and the PN results at different PN orders. In particular, we plot the non-spinning 3.5PN result and show how the periastron advance varies when PN spin effects are successively added. The SO terms typically give larger contributions than the SS terms. Figure \ref{fig:Fig5} shows $K$ as a function of $m\Omega_\varphi$ for other equal-mass, equal-spins configurations. In all cases the 3.5PN approximation underestimates the exact result, typically by a few percent over our frequency ranges.

\section{Test-Particle Approximation}\label{sec:pert}

\subsection{Test mass in a Kerr background}

In this section we compute the periastron advance of a test particle on a circular orbit in the equatorial plane of a Kerr black hole; see also Refs.~\cite{Sc.02,AbFr.13,Hi.al.13} for alternative derivations. Our analysis closely follows that of Ref.~\cite{Da.10}, in which the circular-orbit limit of the periastron advance was recently computed within the (non-spinning) EOB framework. Although the properties of timelike geodesics of the Kerr geometry were explored in detail long ago \cite{Ba.al.72}, we recall some well-known formulae here for the sake of completeness, in order to make our perturbative analysis self-contained.

We consider a test particle of mass $\mu$ on a bound geodesic orbit in the equatorial plane of a Kerr black hole of mass $M$ and spin $S \equiv M a \equiv M^2 \chi$. We use Boyer-Lindquist coordinates $\{t,r,\theta,\phi\}$, defined such that the equatorial plane coincides with the plane $\theta = \pi / 2$. Using the proper time $\tau$ to parametrize the timelike geodesic followed by the particle, the orbital motion obeys
\begin{subequations}
	\begin{align}
		\left( \frac{\ud r}{\ud \tau} \right)^2 &= \left( e^2 - 1 \right) + \frac{2M}{r} - \frac{1}{r^2} \left[ j^2 + a^2(1-e^2) \right] \nonumber \\
& + \frac{2M}{r^3} \left( j - a e \right)^2 \, , \label{rdot} \\
		r^4 \left( \frac{\ud \varphi}{\ud \tau} \right)^2 &= j - a e + a \, \frac{e (r^2 + a^2) - a j}{r^2 - 2Mr + a^2} \, , \label{phidot} \\
		r^4 \left( \frac{\ud t}{\ud \tau} \right)^2 &= a \left( j - a e \right) + (r^2 + a^2) \, \frac{e (r^2 + a^2) - a j}{r^2 - 2Mr + a^2} \, , \label{tdot}
	\end{align}
\end{subequations}
where $e$ and $j$ are the conserved specific energy and angular momentum of the particle. Introducing the inverse separation $u \equiv 1 / r$, and parametrizing the orbital motion in terms of the Mino time parameter $\lambda$ \cite{Mi.03}, defined such that $\ud \tau / \ud \lambda = r^2$, the radial first integral of the motion, Eq.~\eqref{rdot}, can be rewritten in the simple form
\beq\label{radial}
	\dot{u}^2 + V(u) = 0 \, ,
\eeq
where the overdot stands for a derivative with respect to $\lambda$, and the radial potential $V$ is a third order polynomial in $u$:
\beq\label{V}
	V = 1 - e^2 - 2M \, u + \left[ j^2 + a^2(1-e^2) \right] u^2 - 2M \left( j - a e \right)^2 u^3 \, .
\eeq

To derive the expression of the periastron advance in the circular-orbit limit, we can restrict to a slightly eccentric orbit, treated as a linear perturbation of an exactly circular orbit with radius $r_0$. To first order in a parameter $\varepsilon$ measuring the deviation from perfect circularity, the radial motion can be written as
\beq
	u(\lambda) = u_0 + \varepsilon \, u_1(\lambda) + \calO(\varepsilon^2) \, ,
\eeq
where $u_0 = 1 / r_0$ satisfies the circular-orbit conditions $V(u_0) = V'(u_0) = 0$. The function $u_1(\lambda)$ encodes the effect of the eccentricity perturbation on the radial motion. To first order in $\varepsilon$, the differential equation \eqref{radial} reduces to
\beq\label{perturbation}
	\dot{u}_1^2 + \omega_r^2 \, u_1^2 = 0 \, ,
\eeq
where $\omega_r^2(u_0) \equiv \frac{1}{2} V''(u_0)$ is the radial frequency (squared) associated with the circular orbit of radius $r_0$. Using the explicit expression \eqref{V} of the radial potential $V(u)$, we have
\beq\label{Omega_r}
	\omega_r^2 = j^2 + a^2 (1 - e^2) - 6M \left( j - a e \right)^2 u_0 \, .
\eeq
The solution of the differential equation \eqref{perturbation} for the perturbation $u_1(\lambda)$ depends on the sign of the radial frequency squared: if $\omega_r^2 > 0$ then the perturbation is stable, as it obeys the harmonic evolution $u_1(\lambda) \propto \cos{\bigl( \omega_r \lambda + \varphi_0 \bigr)}$, where $\varphi_0$ is a constant; if $\omega_r^2 < 0$ then the perturbation is unstable, as it grows like $u_1(\lambda) \sim \exp{\bigl( \sqrt{-\omega_r^2} \, \lambda \bigr)}$ as $\lambda \to +\infty$. The boundary case $\omega_r^2 = 0$ corresponds to a marginally stable circular orbit, or innermost stable circular orbit (ISCO); its radius is given by
\beq\label{rISCO}
	r_\text{ISCO} = \frac{6M \, \left( j - a e \right)^2}{j^2 + a^2 (1 - e^2)} \, .
\eeq
In the limit $a \to 0$ of vanishing spin, the Boyer-Lindquist radial coordinate reduces to the usual Schwarzschild radial coordinate, and we recover the well-known location $r_\text{ISCO} = 6M$ of the Schwarzschild ISCO.

On the other hand, the instantaneous azimuthal frequency $\omega_\varphi \equiv \ud \varphi / \ud \lambda$ of the orbit is given, in Mino time, by Eq.~\eqref{phidot}. In the limit $\varepsilon \to 0$, it is constant and reads
\beq\label{Omega_phi}
	\omega_\varphi = \frac{j + 2M \left( a e - j \right) u_0}{1 - 2M \, u_0 + a^2 u_0^2} \, .
\eeq
In the circular-orbit limit, the periastron advance is given by the ratio $K \equiv \omega_\varphi / \omega_r$ of the two frequencies of the motion. Following Refs.~\cite{Da.10,Ba.al.10}, we find it more convenient to work with the quantity $W \equiv 1 / K^2$ instead. Using Eqs.~\eqref{Omega_r} and \eqref{Omega_phi}, we obtain
\begin{align}\label{W_u}
	W &= \bigl[ j^2 + a^2 (1 - e^2) - 6M \left( a e - j \right)^2 u_0 \bigr] \times \nonumber \\
&\qquad \biggl[ \frac{1 - 2M \, u_0 + a^2 u_0^2}{j + 2M \left( a e - j \right) u_0} \biggr]^2 .
\end{align}
Notice that the ratio of frequencies $W = \left( \omega_r / \omega_\varphi \right)^2$ does not depend on the time parametrization used to describe the motion; hence the result \eqref{W_u} is valid, e.g., in Mino time $\lambda$, in proper time $\tau$, and in Boyer-Lindquist coordinate time $t$.

Next, we use the conditions $V(u_0) = 0$ and $V'(u_0) = 0$ for a circular orbit to express the energy $e$ and angular momentum $j$ as functions of the orbital radius $r_0$. In terms of the coordinate ``velocity'' $v^2 \equiv M u_0 = M / r_0$, this yields \cite{Ba.al.72}
\begin{subequations}\label{E_L}
	\begin{align}
		e &= \frac{1 - 2 v^2 + \chi v^3}{\sqrt{1 - 3 v^2 + 2 \chi v^3}} \, , \\
		j &= \frac{M}{v} \frac{1 - 2 \chi v^3 + \chi^2 v^4}{\sqrt{1 - 3 v^2 + 2 \chi v^3}} \, .
	\end{align}
\end{subequations}
Replacing these formulas into Eq.~\eqref{W_u}, the algebra simplifies considerably, and we are left with the polynomial result
\beq\label{W_v}
	W = 1 - 6 v^2 + 8 \chi v^3 - 3 \chi^2 v^4 \, .
\eeq

This simple expression lends itself to a nice (but simplistic) physical interpretation: the first term in the right-hand side of Eq.~\eqref{W_v} corresponds to the Newtonian result (no periastron advance), the second term encodes the full general relativistic correction for a Schwarzschild black hole ($\chi$-independent), the third term is a spin-orbit coupling (linear in $\chi$), and the last term a spin-spin contribution (quadratic in $\chi$).

Notice that by substituting Eqs.~\eqref{E_L} into the expression \eqref{rISCO} previously derived for the coordinate location of the Kerr ISCO, we obtain an equation for $v$ that can easily be shown to be equivalent to the vanishing of the polynomial in the right-hand side of Eq.~\eqref{W_v}. This is expected because the condition $W = 0$ corresponds to a vanishing radial frequency (independently of the time parametrization used), which defines the ISCO \cite{Ba.al.72,Fa2.11}.

The test-particle result \eqref{W_v} being expressed in terms of the Boyer-Lindquist coordinate radius $r_0$ of the circular orbit, a meaningful comparison with the predictions from PN theory and NR simulations is not obvious. To ease such comparisons, we must first relate $r_0$ to the ``invariant'' circular-orbit frequency $\Omega_\varphi \equiv \ud \varphi / \ud t$, defined in terms of the coordinate time $t$ that coincides with the proper time of an asymptotic, inertial observer. By taking the ratio of the first integrals \eqref{phidot} and \eqref{tdot} for $\ud \varphi / \ud \tau$ and $\ud t / \ud \tau$, we find
\beq
	\Omega_\varphi = \frac{u_0^2 \left[ j + 2M \left( a e - j \right) u_0 \right]}{e + a \, u_0^2 \left[ a e + 2M \left( a e - j \right) u_0 \right]} = \left( a + \frac{M}{v^3} \right)^{-1} ,
\eeq
where we used Eqs.~\eqref{E_L} to substitute $e$ and $j$ in favor of $v$. Inverting this last result yields the expression of $v^2 = M u_0$ in terms of the dimensionless product $M \Omega_\varphi$ as \cite{Ba.al.72}
\beq
\label{vOmega}
v^3 = \frac{M \Omega_\varphi}{1 - \chi \, M \Omega_\varphi} \, .
\eeq
Substituting this expression into Eq.~\eqref{W_v}, we finally obtain the desired relationship $W(M\Omega_\varphi;\chi)$, valid in the test-mass limit. In the limit $\chi \to 0$ of vanishing spin, the result \eqref{W_v} reduces to the well-known expression $W = 1 - 6 (M \Omega_\varphi)^{2/3}$ for the periastron advance of a test particle on a circular orbit around a Schwarzschild black hole \cite{DaSc.88,Cu.al.94}.

A check of the validity of \eqref{W_v} is provided by the results of Schmidt \cite{Sc.02}, who performed a thorough analysis of the fundamental frequencies of the geodesic motion of a test particle on a \textit{generic} (bound) orbit around a Kerr black hole. Combining Eqs.~(40)--(42), (51), and (59)--(62) of Ref.~\cite{Sc.02} with Eqs.~\eqref{E_L} of this paper, the result \eqref{W_v} can easily be recovered. That expression was also established in Sec.~2.5 of Ref.~\cite{AbFr.13}.

\subsection{Test spin in a Kerr background}

Before ending this section, we consider the additional effects on the periastron advance $W$ if the particle has a spin. Using a pole-dipole-quadrupole model (gravitational skeleton approach) for the small black hole, the authors of the companion paper \cite{Hi.al.13} computed the periastron advance for a spinning particle of mass $\mu$ and spin $S_* \equiv \mu^2 \, \chi_*$ orbiting a Kerr black hole of mass $M$ and spin $S = M^2 \chi$, for circular equatorial orbits and spins aligned or anti-aligned with the orbital angular momentum. Thereafter, it will prove convenient to introduce the notation $\bar{q} \equiv 1 / q$ for the inverse mass ratio, such that $0 < \bar{q} \leqslant 1$ and the perturbative limit corresponds to $\bar{q} \to 0$. Discarding the terms quadratic in the spin variable $\bar{\chi}_* \equiv \bar{q} \, \chi_*$, the authors of Ref.~\cite{Hi.al.13} found
\begin{align}\label{Wspinpart}
	W &= 1 - 6 v^2 + \left( 8 \chi + 6 \bar{\chi}_* \right) v^3 - \left( 3 \chi^2 + 6 \chi \bar{\chi}_* \right) v^4 \nonumber \\ &\qquad\! - 18 \bar{\chi}_* \, v^5 + 30 \chi \bar{\chi}_* \, v^6 - 12 \chi^2 \bar{\chi}_* \, v^7 + \calO(\chi_*^2) \, .
\end{align}
Even when accounting for the terms linear in the spin $S_*$ of the small black hole, the result for the coordinate-invariant function $W(M\Omega_\varphi;\chi,\chi_*)$ takes the form of a polynomial in the ``velocity'' $v^2 = M / r_0$, given by Eq.~\eqref{vOmega} above. Note that higher powers in the spins appear at increasingly higher PN orders: 1.5PN, 2PN, and 3.5PN for linear (spin-orbit), quadratic (spin-spin), and cubic contributions. Since $0 \leqslant |\chi|,|\chi_*| < 1$, contributions of high order in the spins are further suppressed when $v \lesssim 1$.

To make contact with the PN result \eqref{W_PN}, valid for any mass ratio, we substitute \eqref{vOmega} in the expression \eqref{Wspinpart}, and expand the result in powers of the dimensionless PN parameter $y \equiv (M\Omega_\varphi)^{2/3}$ in the weak-field/small-velocity limit $M \Omega_\varphi \to 0$. At 3.5PN order, we obtain
\begin{align}\label{WspinPN}
	W = 1 &- 6 y + \left( 8 \chi + 6 \bar{q} \chi_* \right) y^{3/2} - \left( 3 \chi^2 + 6 \bar{q} \chi_* \chi \right) y^2 \nonumber \\ &- \left( 4 \chi + 18 \bar{q} \chi_* \right) y^{5/2} + \left( 8 \chi^2 + 36 \bar{q} \chi_* \chi \right) y^3 \nonumber \\ & - \left( 4 \chi^3 + 20 \bar{q} \chi_* \chi^2 \right) y^{7/2} + \calO(y^4,\chi_*^2) \, .
\end{align}
This expression is in complete agreement with the test-mass limit ($\nu \to 0$ and $\Delta \to 1$) of the PN result \eqref{W_PN}, as long as the mass $M$ and spin $\chi$ of the Kerr black hole, and the mass $\mu$ and spin $\chi_*$ of the particle, are identified with $(m_1$, $\chi_1)$ and $(m_2$, $\chi_2)$, respectively. In that limit the symmetric mass ratio reduces to $\nu = \bar{q} + \calO(\bar{q}^2)$. Note that we would need to control the (unknown) contribution $\calO(S^3)$ at 3.5PN order in the PN result to compare with the term $\calO(y^{7/2})$ in Eq.~\eqref{WspinPN}.

\section{Imposing a known symmetry on the perturbative result}\label{sec:sym}

\subsection{Motivation and guidance from post-Newtonian theory}

In the general relativistic two-body problem, most quantities of physical interest are \textit{symmetric} by exchange of the bodies' labels. For compact-object binaries on quasi-circular orbits, this property is satisfied, e.g., by the periastron advance, the binding energy, the total angular momentum, the fluxes of energy and angular momentum, and the gravitational-wave polarizations themselves, when expressed as functions of the circular-orbit frequency. This symmetry property can be seen in explicit PN expansions for these relations, such as Eq.~\eqref{W_PN} above, Eqs.~(3.13) and (3.15) of Ref.~\cite{Bo.al.13}, or Eqs.~(194), (231) and (237)--(241) of Ref.~\cite{Bl.06}. In the context of black hole perturbation theory, however, the central Kerr black hole and the small spinning compact object are, by design, not treated ``on equal footing.'' Any quantity of interest is usually computed as an expansion in powers of the usual mass ratio $\bar{q} = \mu / M$, and is therefore \textit{not} symmetric by exchange of the black hole and the particle.

One could hardly overstate the major role played by symmetries in physics. Symmetry considerations often drastically simplify the process of solving a given physics problem. References \cite{Bo.al2.08,BoKe.08} provide an example of the constraining power of symmetries in the context of the binary black-hole problem in general relativity. In the present context, enforcing the symmetry by exchange $1 \leftrightarrow 2$ on the perturbative expression \eqref{Wspinpart} could possibly enlarge the domain of validity of this relativistic formula. However, starting from Eq.~\eqref{Wspinpart}, one can devise many ways of imposing this symmetry property. We shall look for the \textit{simplest} such ``symmetrization,'' guided solely by well-established properties of the PN expansion.

Let us consider two spinning particles with masses $m_i$ and spins $S_i = m_i^2 \, \chi_i$, on a quasi-circular orbit with azimuthal frequency $\Omega_\varphi$. The PN expansion of any function $f$ that is symmetric under the exchange $1 \leftrightarrow 2$ of the particles' labels, and scales like $(v/c)^0$ at Newtonian order, takes the generic form\footnote{Because of gravitational tail effects, a logarithmic running appears starting at the relative 4PN order \cite{BlDa.88}. See, e.g., Ref.~\cite{Bl.al2.10} and references therein. We neglect those here to simplify the discussion.}
\begin{align}\label{f}
	f(\Omega_\varphi&;m_i,S_i) = \sum_{n=0}^N a_n(\nu) \, x^{n/2} \nonumber \\ &+ x^{3/2} \sum_{n=0}^{N-3} \left[ b_n(\nu) \, \chi_s +  c_n(\nu) \, \Delta \, \chi_a \right] x^{n/2} \nonumber \\ &+ x^2 \sum_{n=0}^{N-4} \left[ d_n(\nu) \, \chi_s^2 + e_n(\nu) \, \chi_s \, \Delta \, \chi_a + f_n(\nu) \, \chi_a^2 \right] x^{n/2} \nonumber \\ &+ x^{7/2} \sum_{n=0}^{N-7} \left[ g_n(\nu) \, \chi_s^3 + h_n(\nu) \, \chi_s^2 \, \Delta \, \chi_a + i_n(\nu) \, \chi_s \, \chi_a^2 \right. \nonumber \\ &\left. \qquad\qquad\quad + \, j_n(\nu) \, \Delta \, \chi_a^3 \right] x^{n/2} + o(x^{N/2}) \, ,
\end{align}
with $N \geqslant 7$ a fixed integer. The coefficients $a_n$, $b_n$, $c_n$, $\cdots$ are polynomials in the symmetric mass ratio $\nu$, and we introduced the half-sum and half-difference of the dimensionless spins,
\begin{subequations}\label{chi_s,a}
	\begin{align}
		\chi_s &\equiv \frac{1}{2} \left( \chi_1 + \chi_2 \right) , \\
		\chi_a &\equiv \frac{1}{2} \left( \chi_1 - \chi_2 \right) .
	\end{align}
\end{subequations}
Note that $\Delta \to - \Delta$ by exchange $1 \leftrightarrow 2$ of the particles' labels, such that the product $\Delta \, \chi_a$ appearing in Eq.~\eqref{f} is indeed symmetric. There is, of course, no unique way to write down the dependence on the spins $\chi_1$ and $\chi_2$ in the PN expansion \eqref{f}. However, given the present emphasis on symmetries, the variables $\chi_s$ and $\chi_a$ (or rather $\Delta \, \chi_a$) provide a natural choice, as Eq.~\eqref{W_PN} above suggests.

\subsection{Substitution rules for masses and spins}

While the perturbative result \eqref{Wspinpart}, or rather its PN expansion \eqref{WspinPN}, is most easily expressed in terms of the variables $(y,\bar{q},\chi,\chi_*)$, the generic PN formula \eqref{f} features the variables $(x,\nu,\chi_s,\chi_a)$. Therefore, to impose the symmetry by exchange $1 \leftrightarrow 2$ on the perturbative result \eqref{Wspinpart}, the mass $M$ of the Kerr black hole should be replaced by the sum $m = m_1 + m_2$ of the component masses, and the asymmetric mass ratio $\bar{q}$ by the symmetic mass ratio $\nu$:
\begin{subequations}\label{masses}
	\begin{align}
		y = (M \Omega_\varphi)^{2/3} &\longrightarrow x = (m \Omega_\varphi)^{2/3} \, , \label{y_x} \\
		\bar{q} = \mu / M &\longrightarrow \nu = m_1 m_2 / m^2 \, . \label{q_nu}
	\end{align}
\end{subequations}
The substitution \eqref{y_x} is commonly used while comparing results from perturbative calculations to those of numerical relativty simulations, the post-Newtonian approximation, or the EOB model \cite{Mr.al.10,Da.10,Ba.al.10,Le.al.11,Sp.al2.11,Ak.al.12}. As was pointed out earlier, the symmetric mass ratio $\nu = \bar{q} / (1+\bar{q})^2$ appears most naturally in PN calculations, and for small mass ratios we have $\nu = \bar{q} + \calO(\bar{q}^2)$, or equivalently $\bar{q} = \nu + \calO(\nu^2)$. These considerations motivated Refs.~\cite{Sm.79,Le.al.11,Sp.al2.11,Le.al2.12} to adopt the substitution \eqref{q_nu} while comparing the results of perturbative calculations to those of NR simulations.

Next, we note that in the test-mass limit $\nu \to 0$ the spin $\chi_2$ of the lightest body must disappear from Eq.~\eqref{f}, which can only depend on $m_2 \Omega_\varphi = M \Omega_\varphi$ and $\chi_1 = \chi$ in that limit; recall e.g. Eq.~\eqref{W_v} with \eqref{vOmega}. This implies that the polynomials $b_n(\nu)$, $c_n(\nu)$, $d_n(\nu)$, $f_n(\nu)$, $g_n(\nu)$, $j_n(\nu)$, $\cdots$ in Eq.~\eqref{f} must satisfy $b_n(0) = c_n(0)$, $d_n(0) = f_n(0)$, $g_n(0) = j_n(0)$, etc. This motivates substituting the spin $\chi$ of the Kerr black hole in Eq.~\eqref{Wspinpart} by the following \textit{symmetric linear combination} of the spin variables $\chi_s$ and $\chi_a$:
\beq\label{chi0}
	\chi \longrightarrow \chi_0 \equiv \chi_s + \Delta \, \chi_a \, .
\eeq
This replacement will indeed ensure that \textit{all} terms $\calO(\nu^0)$, including the terms $\calO(\Delta \, \nu^0)$, will be reproduced by the PN expansion of the symmetric version of the perturbative formula \eqref{Wspinpart}. An immediate consequence of the substitutions \eqref{y_x} and \eqref{chi0} is the following replacement:
\beq
	v^2 = \frac{y}{(1 - \chi \, y^{3/2})^{2/3}} \longrightarrow  u^2 \equiv \frac{x}{(1 - \chi_0 \, x^{3/2})^{2/3}} \, .
\eeq

Comparing the PN expansion \eqref{WspinPN} of the formula \eqref{Wspinpart}, valid in the test-particle limit, with the generic PN expansion \eqref{f}, valid for any mass ratio, it is clear that the numerical coefficients in front of the terms $\calO(\bar{q} \, \chi_*)$ in \eqref{WspinPN} come from the \textit{sum} of the numerical coefficients in front of the terms $\calO(\nu \, \chi_2)$ and $\calO(\Delta \, \nu \, \chi_2)$ in Eq.~\eqref{W_PN}, as $\Delta \to 1$ when $\nu \to 0$. Hence, following the substitution \eqref{chi0} of $\chi$ by a linear combination of $\chi_s$ and $\Delta \, \chi_a$, we make the following substitution for the spin $\chi_*$ of the small body:
\beq\label{chi*}
	\chi_* \longrightarrow c_s \, \chi_s + c_a \, (\Delta \, \chi_a) \, ,
\eeq
where $c_s$ and $c_a$ are \textit{a priori} unknown coefficients. The spin $\chi_*$ occurs at five different places in \eqref{Wspinpart}, each time multiplying a different power of the velocity $v$. Importantly, the coefficients $c_s$ and $c_a$ need not take the same numerical values in each of these five terms, contrary to the unique substitution \eqref{chi0} for $\chi$.

Finally, we point out that one could add in Eqs.~\eqref{chi0} or \eqref{chi*} any symmetric function of the masses and spins that vanish in the limit $\nu \to 0$. We refrain from doing so, making only the simplest substitutions compatible with the structure of the PN expansion, since we do not have any guiding principle motivating the introduction of additional mass-ratio corrections.

\subsection{Symmetric background}\label{subsec:sym}

We now need to determine the values of the coefficients $c_s$ and $c_a$ in each of the five occurrences of $\chi_*$. This is done by making the substitutions \eqref{masses}--\eqref{chi*} into Eq.~\eqref{Wspinpart}, expanding the result in powers of $x$ up to 3.5PN order, expanding again in powers of the mass ratio $\bar{q}$ to first order, and enforcing agreement with the PN expansion \eqref{WspinPN} of the perturbative result \eqref{Wspinpart}. Doing so and remembering that there can be no term $\calO(\Delta \, \nu)$ or $\calO(\nu^2)$ in the 1.5PN SO and 2PN SS contributions, we obtain the unique solutions $(c_s,c_a) = (-2/3,0)$ and $(c_s,c_a) = (0,0)$ for the terms $\calO(u^3)$ and $\calO(u^4)$. Furthermore, we find the relationships $c_s = c_a + 28$ for the term $\calO(u^5)$, $c_s = c_a + 44$ for the term $\calO(u^6)$, and $c_s = c_a - 16$ for the term $\calO(u^7)$. Our final formula for the ``symmetrized'' version of the perturbative result \eqref{Wspinpart} thus reads
\begin{align}\label{Wbar}
	W_\text{SB} &= 1 - 6 u^2 + \left( 8 \chi_0 - 4 \nu \chi_s \right) u^3 - 3 \chi_0^2 \, u^4 \nonumber \\ &\qquad\! - \nu \left[ (\alpha + 28) \, \chi_s + \alpha \,  \Delta \, \chi_a \right] u^5 \nonumber \\ &\qquad\! + \nu \left[ (\beta + 44) \, \chi_s + \beta \, \Delta \, \chi_a \right] \chi_0 \, u^6  \nonumber \\ &\qquad\! + \nu \left[ (\gamma - 16) \, \chi_s + \gamma \, \Delta \, \chi_a \right] \chi_0^2 \, u^7 \, .
\end{align}
By construction, Eq.~\eqref{Wbar} is symmetric by exchange $1 \leftrightarrow 2$ of the bodies' labels, and it reduces to the known result \eqref{Wspinpart} in the extreme mass-ratio limit $\nu \ll 1$. This expression effectively encodes some spin-dependent finite mass-ratio corrections through $\nu$, $\Delta$, and $\chi_0 = \chi_s + \Delta \, \chi_a$. Hereafter, we will refer to Eq.~\eqref{Wbar} as the \textit{symmetric background} (SB), and we will use it in Sec.~\ref{sec:SF} as the zeroth-order approximation, or background, for a new type of expansion in powers of the symmetric mass ratio $\nu$.

The numerical values of the coefficients $(\alpha,\beta,\gamma)$ are left unconstrained by our ``symmetrization.'' However, by considering the PN expansion of Eq.~\eqref{Wbar}, and using some information from the PN result \eqref{W_PN}, namely the coefficients $45/2$ and $15/2$ in front of the terms $\calO(\nu \, x^{5/2})$ and $\calO(\nu \, x^3)$, we readily fix the values of two of the coefficients as
\begin{subequations}\label{coeffs}
	\begin{align}
		\alpha = 17 \, , \\
		\beta = 11 \, .
	\end{align}
\end{subequations}
Unfortunately, we would need to know the contribution $\calO(S^3)$ at 3.5PN order in Eq.~\eqref{W_PN} to fix the value of $\gamma$. Nevertheless, we checked that for the range of frequencies, mass ratios and spins for which we have NR data, any value $\vert \gamma \vert \leqslant 100$ affects $W_\text{SB}$ at the relative $0.2\%$ level at most. This is because the term $\calO(u^7)$ in Eq.~\eqref{Wbar} is cubic in the spins and contributes at leading 3.5PN order. Henceforth, we shall thus use (simply out of convenience) the fiducial value $\gamma_\text{fid} = 0$ in Eq.~\eqref{Wbar}. A future PN calculation of the leading-order contribution $\calO(S^3)$ in the periastron advance would immediately provide the unique, correct value of the coefficient $\gamma$.

Hence, in the weak-field/small velocity limit $m \Omega_\varphi \to 0$, the 3PN expansion of the symmetric background \eqref{Wbar}--\eqref{coeffs} reads
\begin{widetext}
\begin{align}\label{Wbarexp}
  W_\text{SB} &= 1 - 6 x + \bigl[ \, \left( 4 + 4 \Delta - 2 \nu \right) \chi_1
  + \left( 4 - 4 \Delta - 2 \nu \right) \chi_2 \, \bigr] \, x^{3/2}
  + \biggl[ \left( - \frac{3}{2} - \frac{3}{2} \Delta + 3 \nu \right) \chi_1^2 - 6 \nu \, \chi_1 \chi_2
  + \left( - \frac{3}{2} + \frac{3}{2} \Delta + 3 \nu \right) \chi_2^2
  \, \biggr] \, x^2 \nonumber \\ 
&\qquad\! - \biggl[ \left( 2 + 2
    \Delta + \frac{45}{2} \nu + \frac{17}{2} \Delta \, \nu
  \right) \chi_1 + \left( 2 - 2 \Delta + \frac{45}{2} \nu -
    \frac{17}{2} \Delta \, \nu \right) \chi_2 \, \biggr] \,
  x^{5/2} + \biggl[ \left( 4 + 4 \Delta + \frac{15}{2} \nu + \frac{31}{2} \Delta \, \nu - 11 \nu^2 \right) \chi_1^2 \nonumber \\
&\qquad\qquad\!\! + \left( 36 + 22 \nu
  \right) \nu \, \chi_1 \chi_2 + \left( 4 - 4 \Delta + \frac{15}{2} \nu -
    \frac{31}{2} \Delta \, \nu - 11 \nu^2 \right) \chi_2^2 \, \biggr]
  \, x^3 + \calO(x^{7/2}) \, .
\end{align}
\end{widetext}
Comparing with the PN result \eqref{W_PN}, we find that the fully relativistic, symmetric background \eqref{Wbar}--\eqref{coeffs} reproduces the \textit{exact} leading-order 1.5PN spin-orbit and 2PN spin-spin terms, which are of course valid for any mass ratio.\footnote{The variable $S_0 = 4 m^2 \chi_0$ was previously introduced, in a PN context, as an effective spin that fully encodes the leading-order 2PN spin-spin terms in the Hamiltonian of two spinning particles \cite{Da.01}. Hence it is not surprising that the substitution \eqref{chi0} allows one to reproduce the exact 2PN spin-spin terms in the periastron advance.} It also reproduces the next-to-leading order 2.5PN spin-orbit and 3PN spin-spin terms, except for the contributions $\calO(\nu^2)$; these five quadratic terms could nonetheless be encoded in $W_\text{SB}$ by imposing the symmetry by exchange $1 \leftrightarrow 2$ to the known terms $\calO(\chi_*^2)$ \cite{Hi.al.13} in the perturbative result \eqref{Wspinpart}. Furthermore, because the test-spin expression \eqref{Wspinpart} does not include any spin-independent mass-ratio correction [$\bar{q}$ always appears in factors of $\chi_*$ in Eq.~\eqref{Wspinpart}], the formula \eqref{Wbar}--\eqref{coeffs} cannot reproduce the mass-type contributions $\calO(\nu)$ and $\calO(\nu^2)$ at 2PN and 3PN orders in Eq.~\eqref{W_PN}.

\section{Extracting self-force information from numerical-relativity simulations}\label{sec:SF}

Using the symmetric background \eqref{Wbar}--\eqref{coeffs}, we introduce a new type of perturbative expansion in Sec.~\ref{subsec:exp}. This allows us to use the results of NR simulations detailed in Sec.~\ref{sec:NR} to measure the GSF correction to the geodesic periastron advance of a particle orbiting a Schwarzschild (Kerr) black hole in Sec.~\ref{subsec:Schw} (Sec.~\ref{subsec:Kerr}). Finally, in Sec.~\ref{subsec:equal-spin} we compare the predictions of the new perturbative expansion to the NR results for equal-mass, equal-spin configurations.

\subsection{Expansion in the symmetric mass ratio}\label{subsec:exp}

In the PN approximation, one usually expands all quantities in powers of the small PN parameter $x = (m \Omega_\varphi)^{2/3}$, with coefficients depending on the symmetric mass ratio $\nu$ and the spins $\chi_i$ [see Eq.~\eqref{f}]; these coefficients encode all finite mass-ratio corrections at each PN order. By contrast, in black-hole perturbation theory, one usually expands all quantities in powers of the small (asymmetric) mass ratio $\bar{q}$, with coefficients depending on $y = (M\Omega_\varphi)^{2/3}$ and the spin $\chi$ of the central Kerr black hole; these coefficients encode all the relativistic corrections at each perturbative order.

Motivated by the generic form \eqref{f} of the PN expansion, as well as by the earlier works \cite{Sm.79,FiDe.84,Fa.al.04,Le.al.11,Sp.al2.11,Le.al2.12,Na.13} suggesting that the scaling $\bar{q} \to \nu = \bar{q} / (1+\bar{q})^2$ considerably extends the domain of validity of perturbative calculations, we introduce a new type of expansion in powers of the \textit{symmetric} mass ratio, with coefficients encoding all the relativistic corrections at each order, using the symmetric background \eqref{Wbar}--\eqref{coeffs} as the zeroth-order approximation. Therefore, we are considering a formal expansion of the type
\beq\label{Wexp}
	W = W_\text{SB} + \sum_{n=1}^\infty \nu^n \, W_n \, ,
\eeq
where the functions $W_n(\Omega_\varphi;m_i,S_i)$ encode the successive finite mass-ratio corrections to the background $W_\text{SB}$. The symmetry by exchange of the bodies' labels implies that these functions can always be written in the form
\beq\label{W_n}
	W_n(\Omega_\varphi;m_i,S_i) = f_n(x,\chi_s,\chi_a^2) + \Delta \, \chi_a \; g_n(x,\chi_s,\chi_a^2) \, ,
\eeq
where $f_n$ and $g_n$ are functions of the symmetric variables $x$, $\chi_s$ and $\chi_a^2$. The traditional PN and perturbative approximations are then recovered by expanding the formal series \eqref{Wexp}--\eqref{W_n} in powers of $x$ and $\bar{q}$, respectively.

Notice that the functions $W_n$ implicitly depend on the mass ratio $\bar{q}$ through the reduced mass difference $\Delta = \sqrt{1-4\nu}$ appearing in front of $g_n$ in Eq.~\eqref{W_n}. However, from the PN expansions \eqref{W_PN} and \eqref{Wbarexp} of $W$ and $W_\text{SB}$ we have the leading-order scalings $f_1 = \calO(x^2)$ and $g_1 = \calO(x^{7/2})$. Thus $g_1 \ll f_1$ in the frequency range $0.05 \lesssim  x \lesssim 0.1$ for which we have NR data, such that $W_1 \simeq f_1$ depends only weakly on the mass ratio. For non-spinning binaries, $\chi_s = \chi_a = 0$, we simply have $W_\text{SB} = 1 - 6x$ and $W_n = f_n(x)$ is independent of the mass ratio.

\subsection{Self-force in a Schwarzschild background}\label{subsec:Schw}

\begin{figure}
	\includegraphics[scale=0.4]{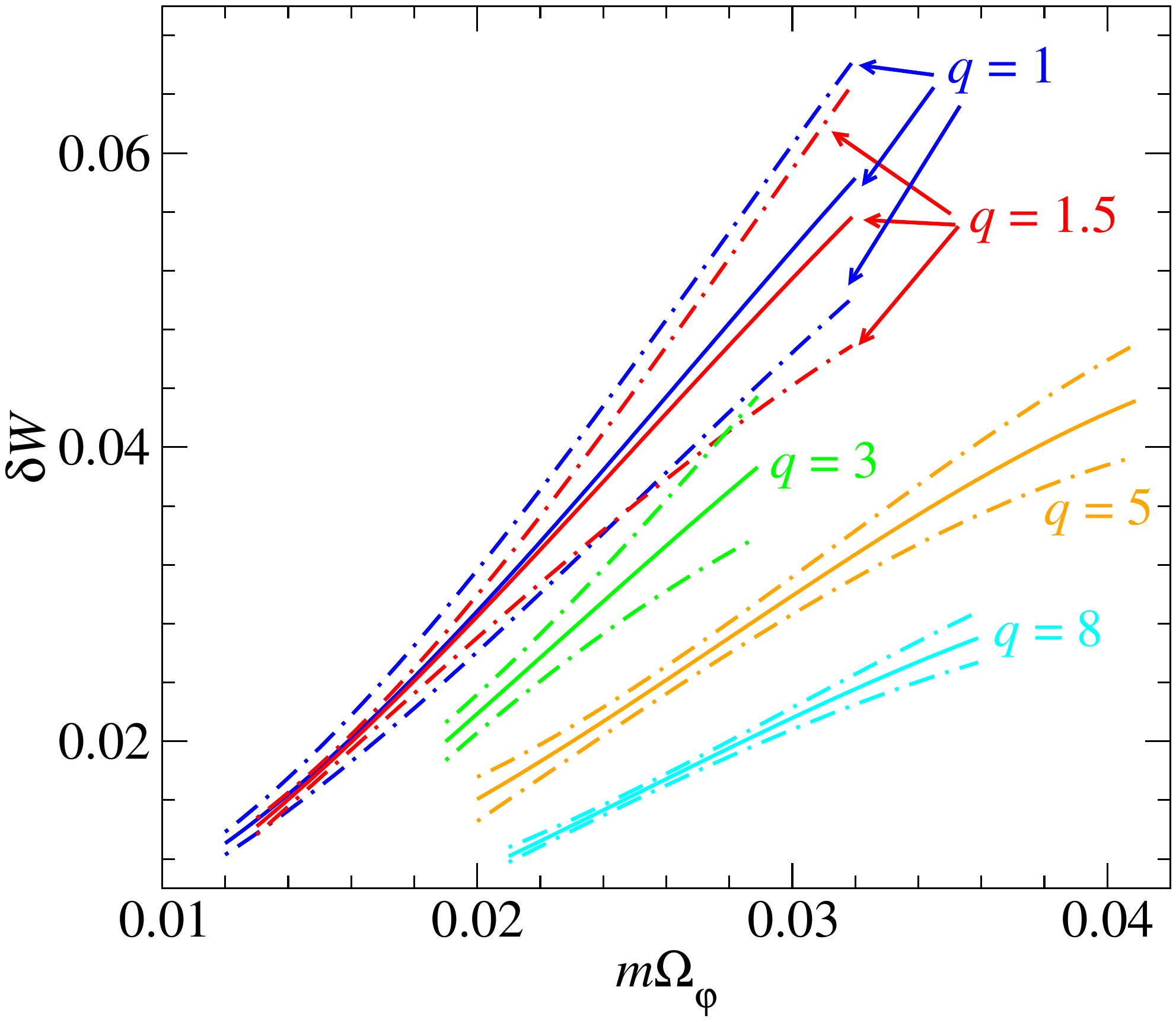}
	\caption{The difference $\delta W = W_\text{NR} - W_\text{SB}$ as a function of the orbital frequency $m \Omega_\varphi$, for non-spinning binaries with mass ratios $q = 1$ (blue), $1.5$ (red), $3$ (green), $5$ (orange), and $8$ (cyan). The dashed lines show the estimated NR uncertainties.}
	\label{fig:deltaW_non-spin}
\end{figure}

\begin{figure}
	\includegraphics[scale=0.4]{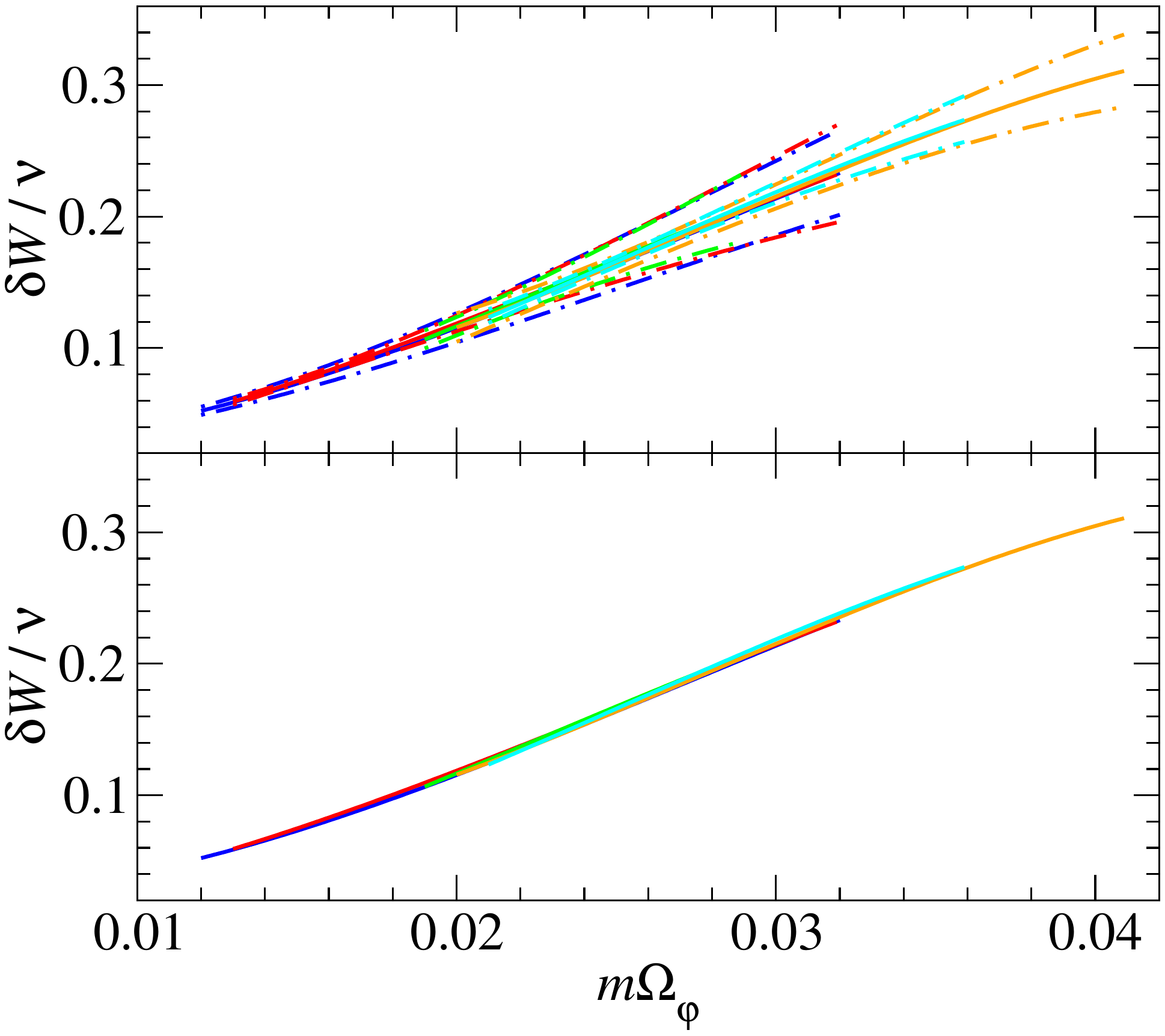}
	\caption{The rescaled difference $\delta W / \nu$ as a function of $m \Omega_\varphi$, for non-spinning binaries, including (top) and excluding (bottom) the uncertainties affecting the NR results.}
	\label{fig:SF_non-spin}
\end{figure}

Figure \ref{fig:deltaW_non-spin} shows the difference $\delta W \equiv W_\text{NR} - W_\text{SB}$ between the NR results for $W = 1/K^2$ and the symmetric background, as a function of the orbital frequency $m\Omega_\varphi$, for non-spinning black-hole binaries with mass ratios $q \in \{1,1.5,3,5,8\}$. The various differences $\delta W$ are of order $0.01$--$0.07$, showing that the background accounts for about $90\%$ of the exact result, for all mass ratios considered. Notice that $\delta W(\Omega_\varphi)$ depends sensitively on the mass ratio $q$. In Fig.~\ref{fig:SF_non-spin} the differences $\delta W$ are rescaled by the symmetric mass ratio $\nu$, still for mass ratios $q \in \{1,1.5,3,5,8\}$. The bottom panel shows that the five \textit{independent} curves for $\delta W / \nu$ overlap very well over a wide range of orbital frequencies. Their scatter is much smaller than the intrinsic NR error bars shown in the upper panel. The remarkable alignment of the various curves for $\delta W / \nu$ implies that (i) the fully relativistic numerical results for $W$ are well approximated by an expansion of the type \eqref{Wexp}, and that (ii) the finite mass-ratio corrections $\calO(\nu^2)$ or higher are significantly smaller than the sum of the contributions $\calO(\nu^0)$ and $\calO(\nu)$. Hence, the overlapping curves in the bottom panel of Fig.~\ref{fig:SF_non-spin} effectively measure the function $W_1(m\Omega_\varphi)$ appearing in Eq.~\eqref{Wexp} over the frequency range $0.012 < m\Omega_\varphi < 0.041$, which corresponds to a range of separations $8 m \lesssim r_\Omega \lesssim 19 m$, where $r_\Omega \equiv (m / \Omega_\varphi^2)^{1/3}$. We find that the numerical data can be captured by the compact analytic formula

\beq\label{fit}
	W_1^\text{fit} = 14 \, x^2 \, \frac{1 + c_1 x}{1 + c_2 x + c_3 x^2} \, ,
\eeq
where $c_1$, $c_2$, $c_3$ are fitting coefficients. The formula \eqref{fit} accounts for the leading-order (2PN) behavior of $W_1(x)$ when $x \to 0$ [see Eq.~\eqref{W_PN} above]. It was first introduced in Ref.~\cite{Ba.al.10} to model the GSF correction to the periastron advance of a particle orbiting a Schwarzschild black hole. We find for the best fit coefficients (the superscript stands for ``non-spinning'')
\begin{subequations}\label{ns}
	\begin{align}
		c_1^\text{ns} &= -5.4022 \, , \\
		c_2^\text{ns} &= -11.1172 \, , \\
		c_3^\text{ns} &= 38.8701 \, .
	\end{align}
\end{subequations}

As long as the dissipative radiation-reaction effects related to the emission of gravitational waves can be neglected, the first-order correction $W_1$ to $W_\text{SB}$ coincides with the conservative piece of the GSF contribution to the periastron advance, say $W_\text{GSF}$. This function  was computed in Ref.~\cite{Ba.al.10} with high numerical accuracy. The authors performed several fits of the GSF data for $W_\text{GSF}(x)$ in the range $6m < r_\Omega < 80m$. In particular, they found that these data can be accurately reproduced at the $2.4 \times 10^{-3}$ level by means of the fitting formula \eqref{fit}, with best fit coeffcients $c_1 = 13.3687$, $c_2 = 4.60958$, and $c_3 = -9.47696$. Figure \ref{fig:self-force} shows that the fit \eqref{fit}--\eqref{ns} of the NR results for $W_\text{GSF}(x)$ closely tracks the exact perturbative result \cite{Ba.al.10} (blue line) up to $m\Omega_\varphi \simeq 0.03$. The difference grows at larger frequencies, but remains within the NR uncertainty down to separations of order $r_\Omega \simeq 9m$, while the 3.5PN prediction (red line) overshoots over the entire frequency range.

\subsection{Self-force in a Kerr background}\label{subsec:Kerr}

\begin{figure}
	\includegraphics[scale=0.4]{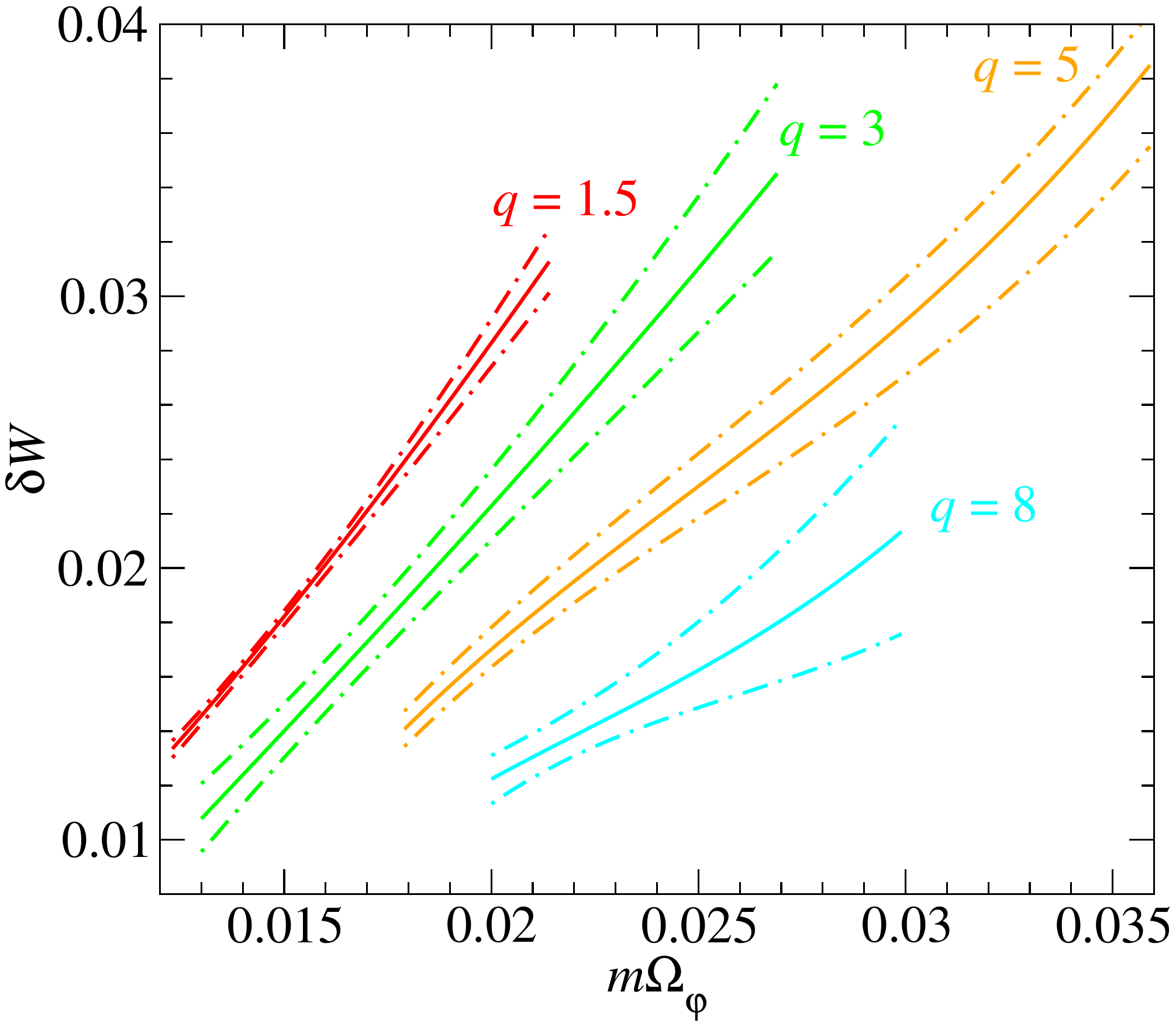}
	\caption{The difference $\delta W = W_\text{NR} - W_\text{SB}$ as a function of the orbital frequency $m \Omega_\varphi$, for spinning binaries with $(\chi_1,\chi_2) = (-0.5,0)$ and mass ratios $q = 1.5$ (red), $3$ (green), $5$ (orange), and $8$ (cyan). The dashed lines show the estimated NR uncertainties.}
	\label{fig:deltaW_spin-down}
\end{figure}

Next, we repeat the analysis of Sec.~\ref{subsec:Schw} in the case of spinning black-hole binaries with mass ratios $q \in \{1.5,3,5,8\}$ and spins $\chi_1 = - 0.5$ and $\chi_2 = 0$. (We do not use the NR data for $q = 1$ because it has much larger error bars than the other configurations; see the left panel of Fig.~\ref{fig:Fig2}.) In Fig.~\ref{fig:deltaW_spin-down} we plot the difference $\delta W = W_\text{NR} - W_\text{SB}$ for these configurations. As in the non-spinning case, the background accounts for more than $90\%$ of the full result and $\delta W$ depends strongly on $q$.

\begin{figure}
	\includegraphics[scale=0.4]{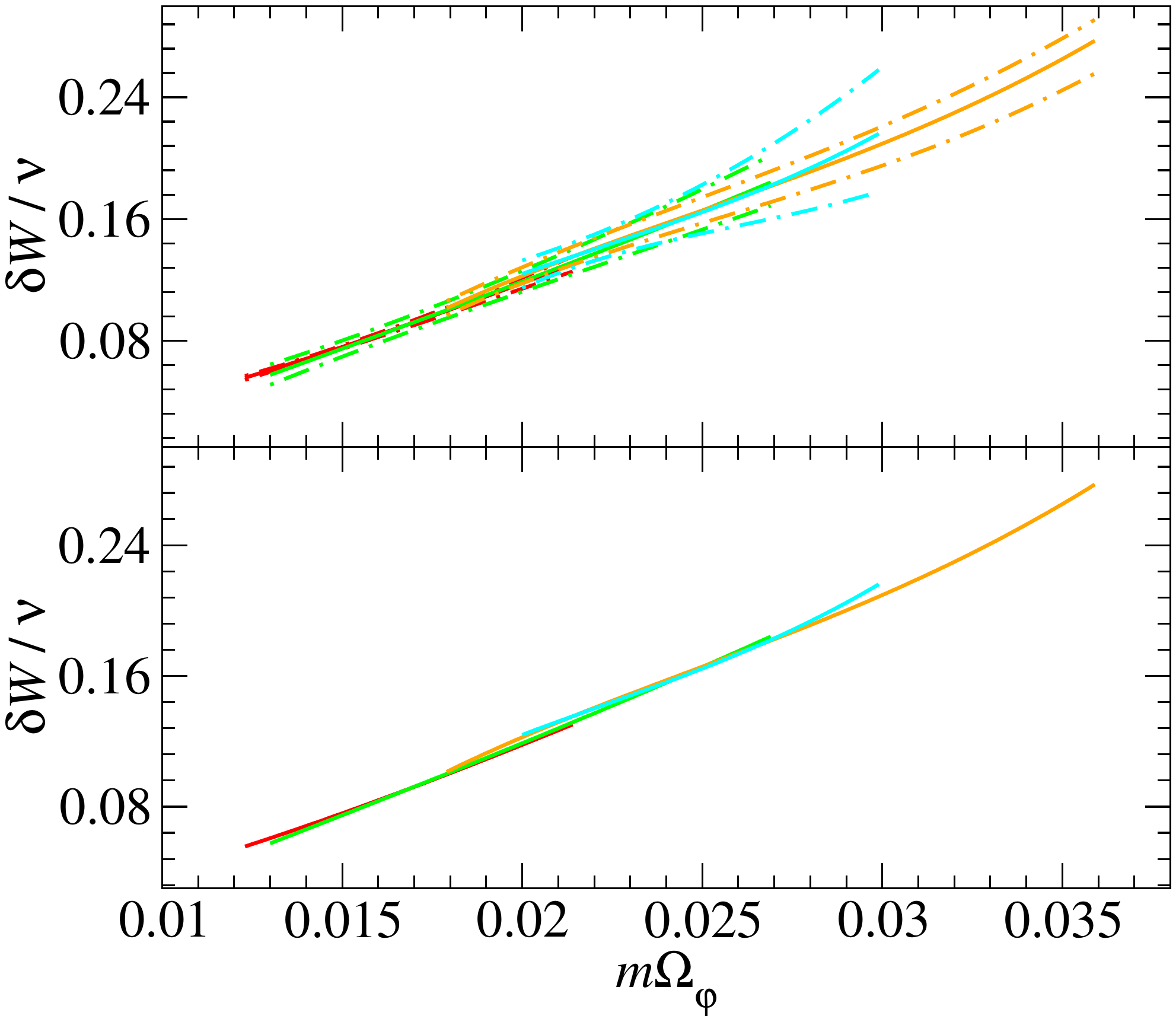}
	\caption{The rescaled difference $\delta W / \nu$ as a function of $m \Omega_\varphi$, for spinning binaries with $(\chi_1,\chi_2) = (-0.5,0)$, including (top) and excluding (bottom) the uncertainties affecting the NR results.}
	\label{fig:SF_spin-down}
\end{figure}

Figure \ref{fig:SF_spin-down} shows the rescaled difference $\delta W / \nu$, still for mass ratios $q \in \{1.5, 3, 5, 8\}$ and spins $\chi_1 = -0.5$ and $\chi_2 = 0$. Again the mean values align remarkably well, with little scatter. As discussed earlier, in our frequency range the first-order correction $W_1$ to $W_\text{SB}$ depends only weakly on the mass ratio $q$. The overlapping curves in the bottom panel of Fig.~\ref{fig:SF_spin-down} thus measure the function $W_1(\Omega_\varphi)$ over the frequency range $0.012 < m \Omega_\varphi < 0.036$, corresponding to separations $9m \lesssim r_\Omega \lesssim 19m$. Combining the NR results for the various mass ratios and performing a least-square fit to the model \eqref{fit}, we obtain the best fit values (the superscript stands for ``spin down'')
\begin{subequations}\label{down}
	\begin{align}
		c_1^\text{down} &= 1.1973 \, , \\
		c_2^\text{down} &= -6.88457 \, , \\
		c_3^\text{down} &= 37.3406 \, .
	\end{align}
\end{subequations}
Interestingly, the fits \eqref{fit}--\eqref{ns} and \eqref{fit}--\eqref{down} of the NR results for the non-spinning ($\chi_1 = 0$) and spinning ($\chi_1 = -0.5$) configurations agree to within $4 \%$ over their common frequency range $0.012 < m \Omega_\varphi < 0.036$. Therefore, the effects of the spin of the most massive black hole are almost entirely accounted for by the symmetric background $W_\text{SB}$.

\begin{figure}
	\includegraphics[scale=0.4]{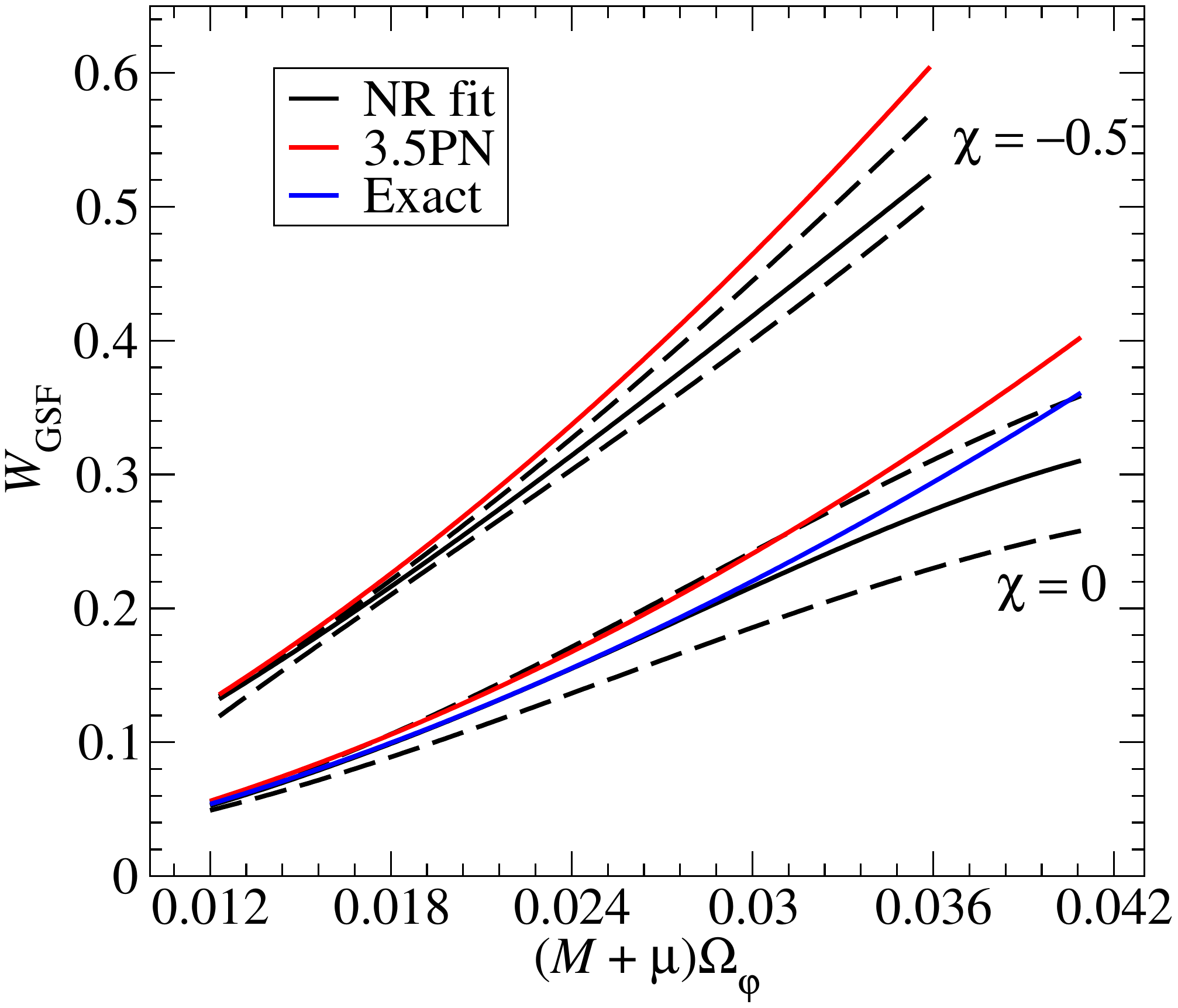}
	\caption{Gravitational self-force correction $W_\text{GSF}$ to the periastron advance of a non-spinning particle of mass $\mu$ orbiting a black hole of mass $M$ and spin $S \equiv \chi \, M^2$, as measured using NR simulations of black-hole binaries with mass ratios $q=1,1.5,3,5,8$. Also shown are the 3.5PN prediction (red) and the exact result for $\chi = 0$ (blue).}
	\label{fig:self-force}
\end{figure}

The prediction \eqref{fit}--\eqref{down} should be compared with a future calculation of the conservative part of the GSF correction to the periastron advance of a non-spinning particle on a circular, equatorial orbit around a Kerr black hole of mass $M$ and spin $S = -0.5 M^2$. Given the conventions usually adopted within the self-force community, such a future perturbative calculation would likely be formulated as an expansion in powers of the mass ratio $\bar{q} = \mu / M$ about a Kerr background. However, following Refs.~\cite{Ba.al.10,BaSa.11,Le.al.11} and keeping with the PN habit of using the total mass $M + \mu$ to adimensionalize frequencies (rather than the mass $M$ of the central black hole), we shall consider an expansion of the type
\beq\label{WexpKerr}
	W = W_\text{Kerr}(x;\chi) + \bar{q} \; W_\text{GSF}(x;\chi) + \calO(\bar{q}^2) \, ,
\eeq
where $W_\text{Kerr}$ is given by Eqs.~\eqref{W_v} and \eqref{vOmega} with $M \to M + \mu$, and $x = [(M+\mu)\Omega_\varphi]^{2/3}$. The expression \eqref{WexpKerr} should be compared to the expansion \eqref{Wexp}, in which the formula \eqref{Wbar} for the symmetric background $W_\text{SB}$ must be expanded in powers of $\bar{q}$ to first order, using the spins values $\chi_s = \chi_a = \chi / 2$. Comparing the two expressions, we obtain the following relationship between the GSF correction $W_\text{GSF}$ to the Kerr result and our first-order symmetric mass-ratio correction $W_1$:
\begin{align}\label{W_GSF}
	W_\text{GSF} &= W_1 - 10 \chi v^3 + 6 \chi^2 v^4 - 27 \chi v^5 \nonumber \\ &\qquad\;\; + 25 \chi^2 v^6 + (\gamma - 4) \, \chi^3 v^7 \, .
\end{align}
Here, the ``velocity'' $v$ is given by Eq.~\eqref{vOmega} with $M \to M + \mu$. (Recall that the numerical coefficient $\gamma$ will remain unknown until the terms $\calO(S^3)$ at 3.5PN order in Eq.~\eqref{W_PN} are computed, but that its precise numerical value is irrelevant for $x \lesssim 0.12$.) The additional spin-dependent terms in Eq.~\eqref{W_GSF} come from the mass-ratio expansion of the symmetric background $W_\text{SB}$. For a Schwarzschild black hole we simply have $W_\text{GSF} = W_1$; see the discussion at the end of Sec.~\ref{subsec:Schw}. The PN expansion of $W_\text{GSF} - W_1$ recovers all the spin-dependent terms $\calO(\bar{q})$ in Eq.~\eqref{W_PN} with $\chi_1 = \chi$ and $\chi_2 = 0$, except for the 3.5PN term linear in $\chi$ whose effect must be captured in $W_1(x;\chi)$.

For a Kerr black hole with spin $\chi = -0.5$, one should replace $W_1$ in Eq.~\eqref{W_GSF} by the fit \eqref{fit}--\eqref{down}. The GSF correction \eqref{W_GSF} for $\chi=-0.5$ (with $\gamma_\text{fid} = 0$) is plotted in Fig.~\ref{fig:self-force}. Clearly, the effect of the spin of the central black hole on the rate of periastron advance is significant: the GSF correction is more than doubled with respect to the non-spinning case. In particular we find that for retrograde orbits, the spin yields a \textit{decrease} in the self-force contribution to $K = 1 / \sqrt{W}$. However, given the error estimates on the NR results, our measurement of $W_\text{GSF}$ is only accurate at the $5$--$10\%$ level. The 3.5PN approximation for $W_\text{GSF}$ (red curve) clearly deviates from the NR-based prediction. It will be interesting to see how the exact GSF result compares with these predictions.

\begin{figure*}
	\includegraphics[scale=0.32]{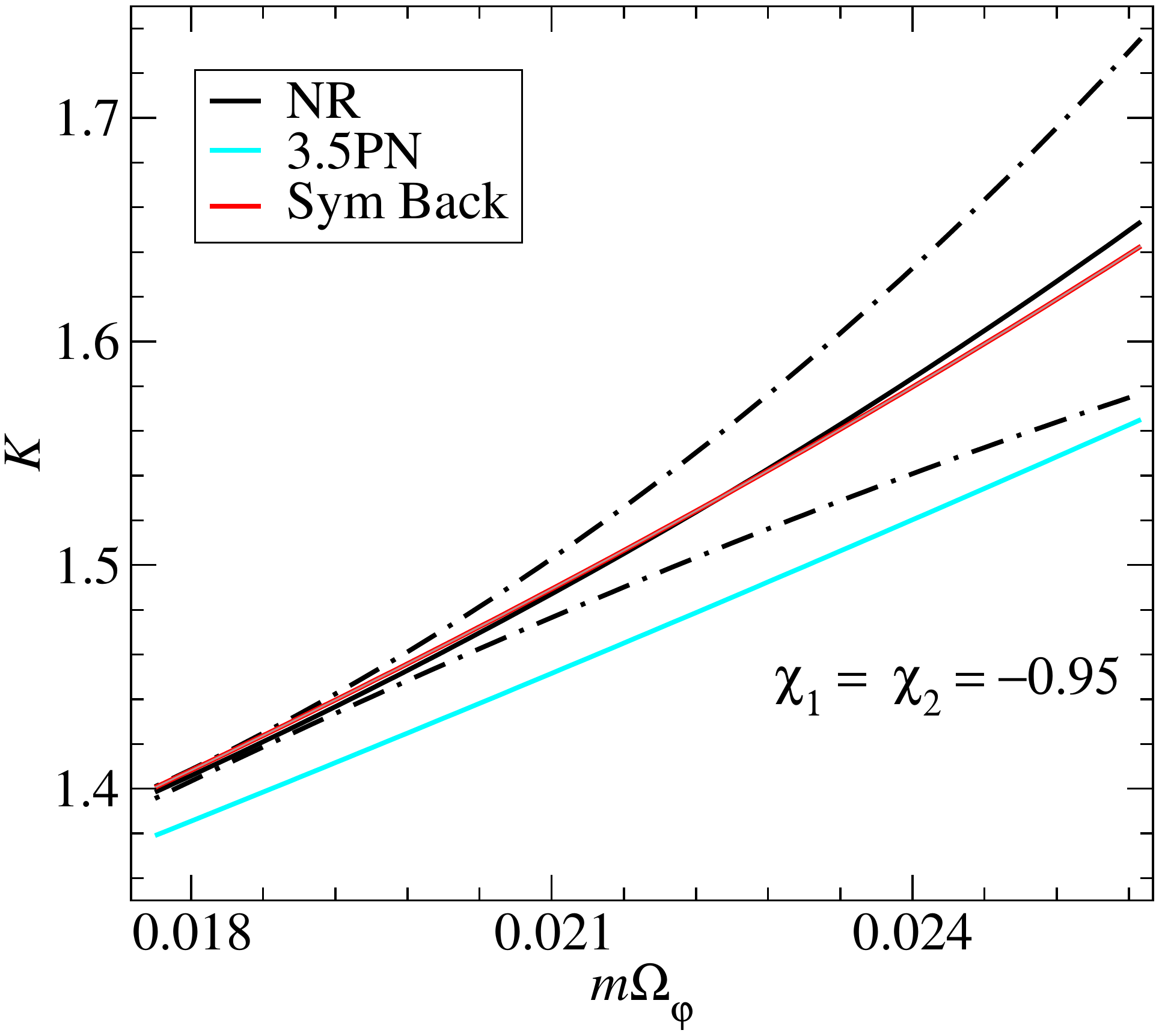}
	\hspace{1cm}
	\includegraphics[scale=0.32]{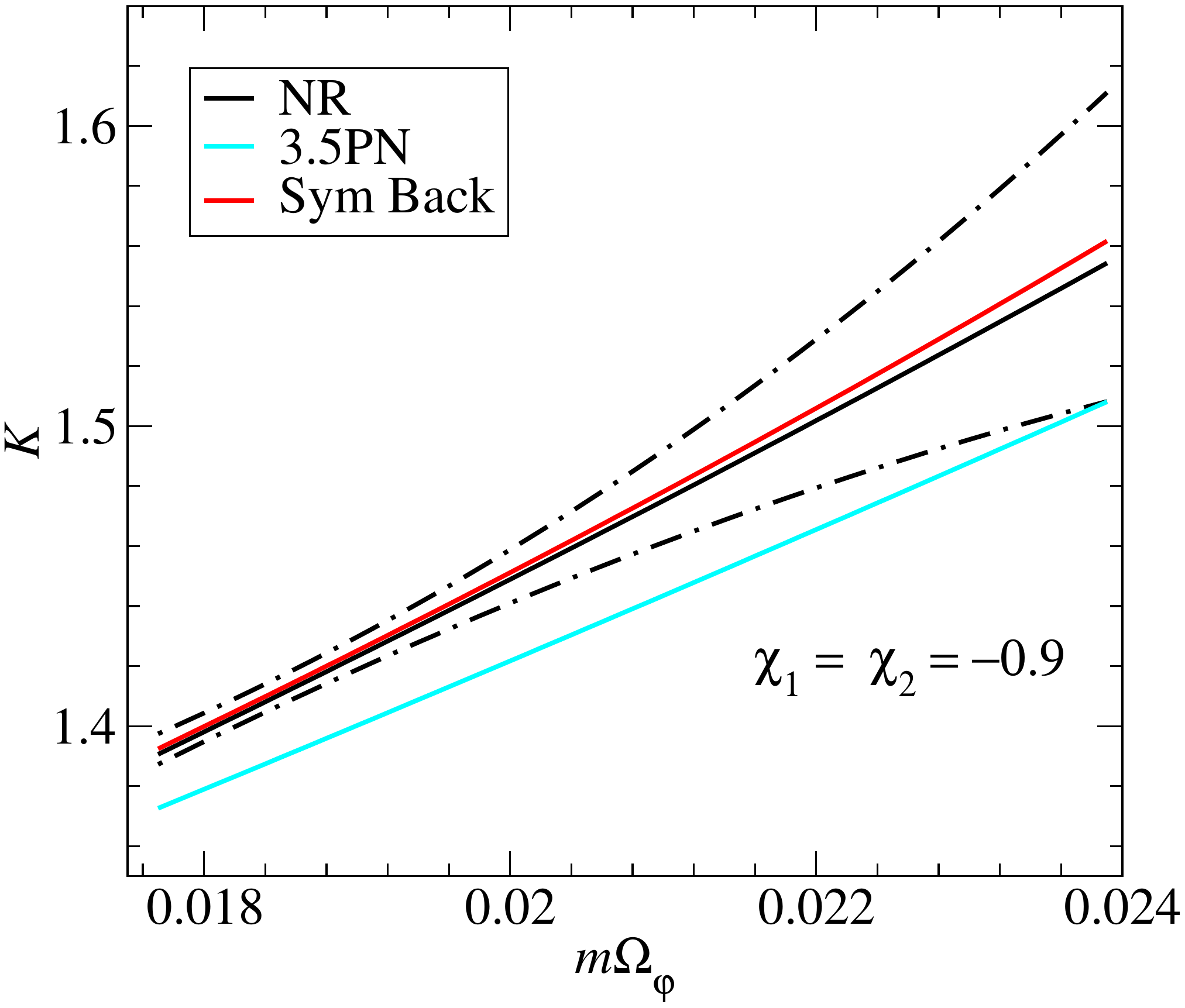}
	\hspace{1cm}
	\includegraphics[scale=0.32]{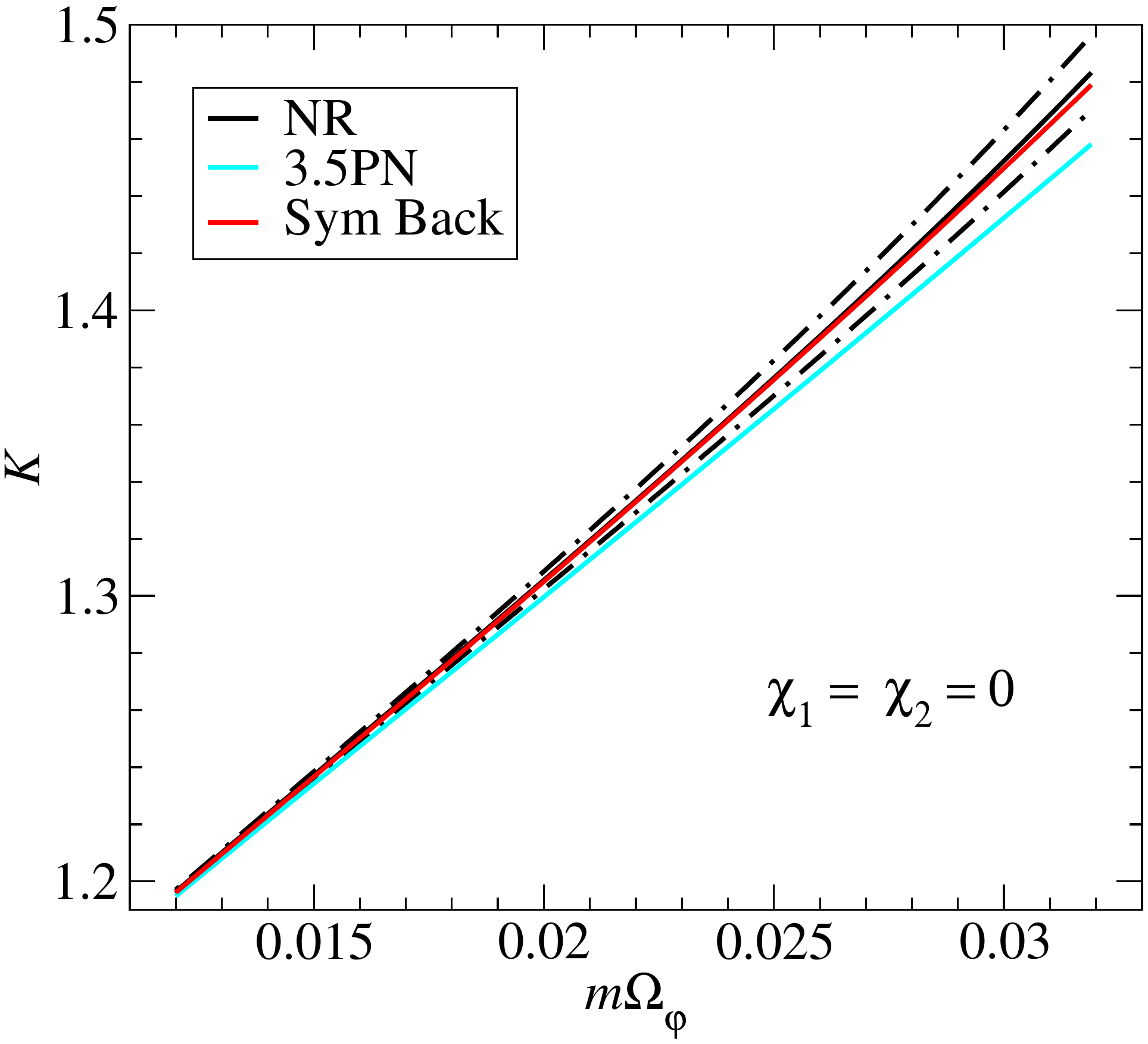}
	\hspace{1cm}
	\includegraphics[scale=0.32]{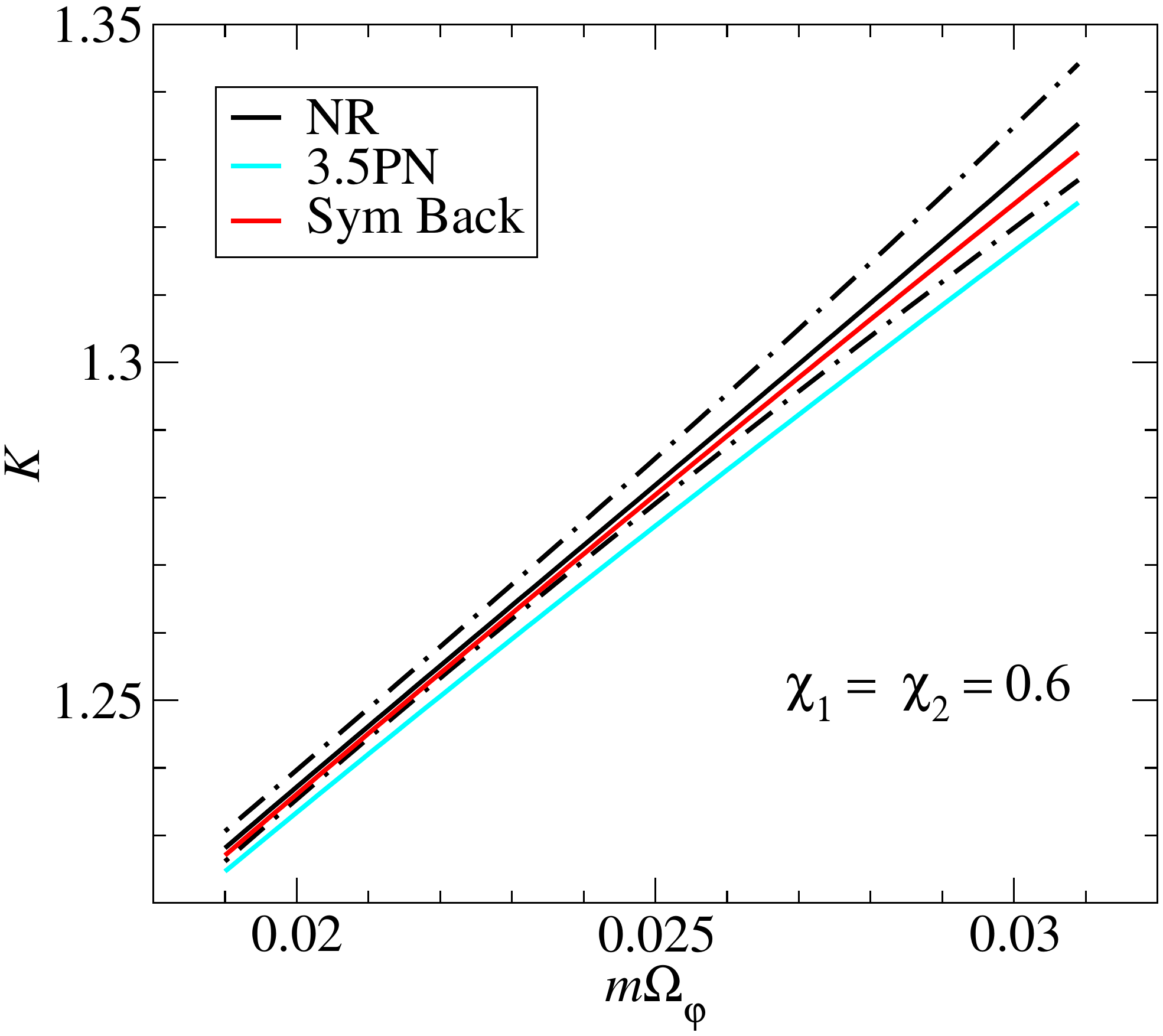}
	\hspace{1cm}
	\includegraphics[scale=0.32]{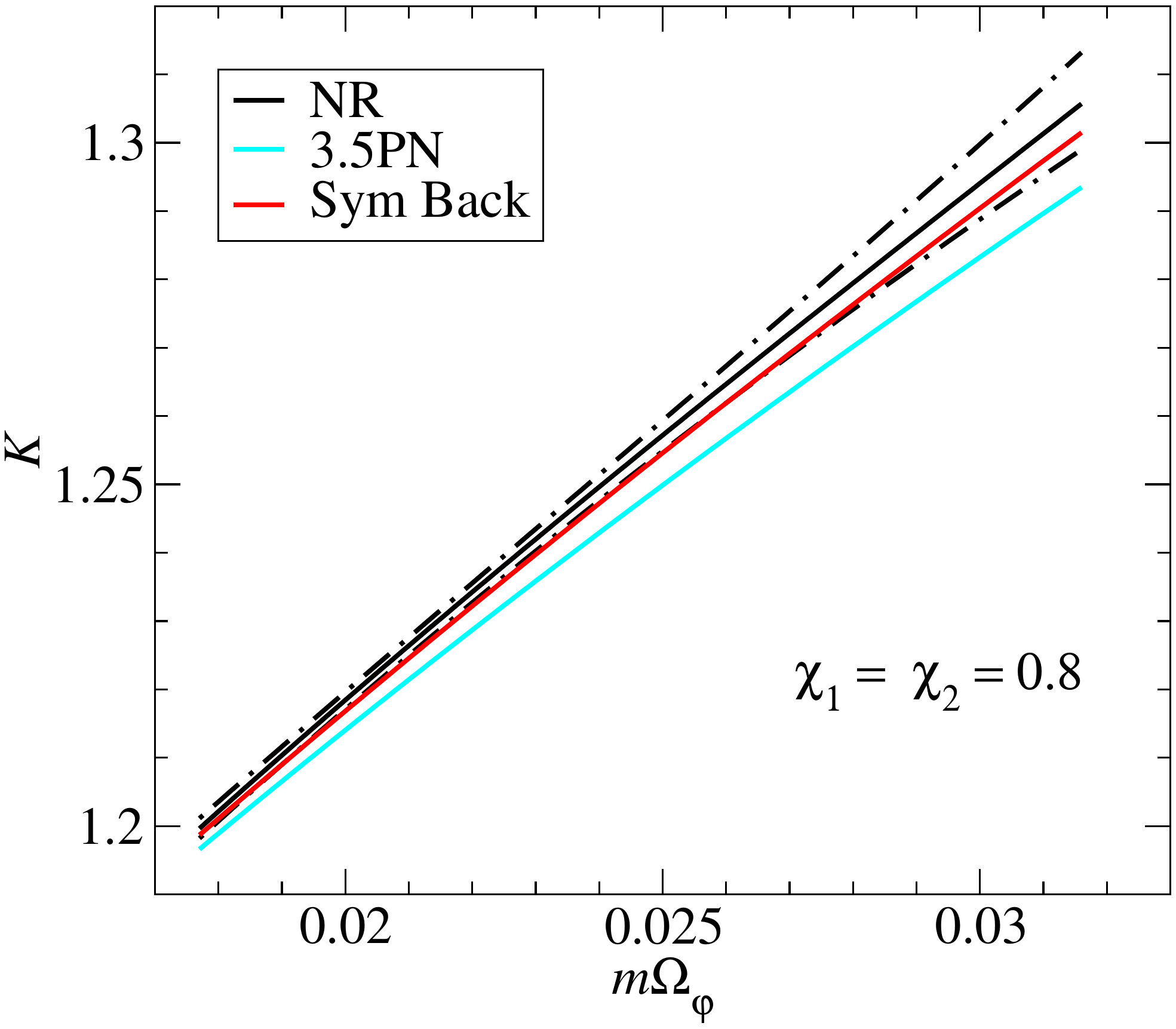}
	\hspace{1cm}
	\includegraphics[scale=0.32]{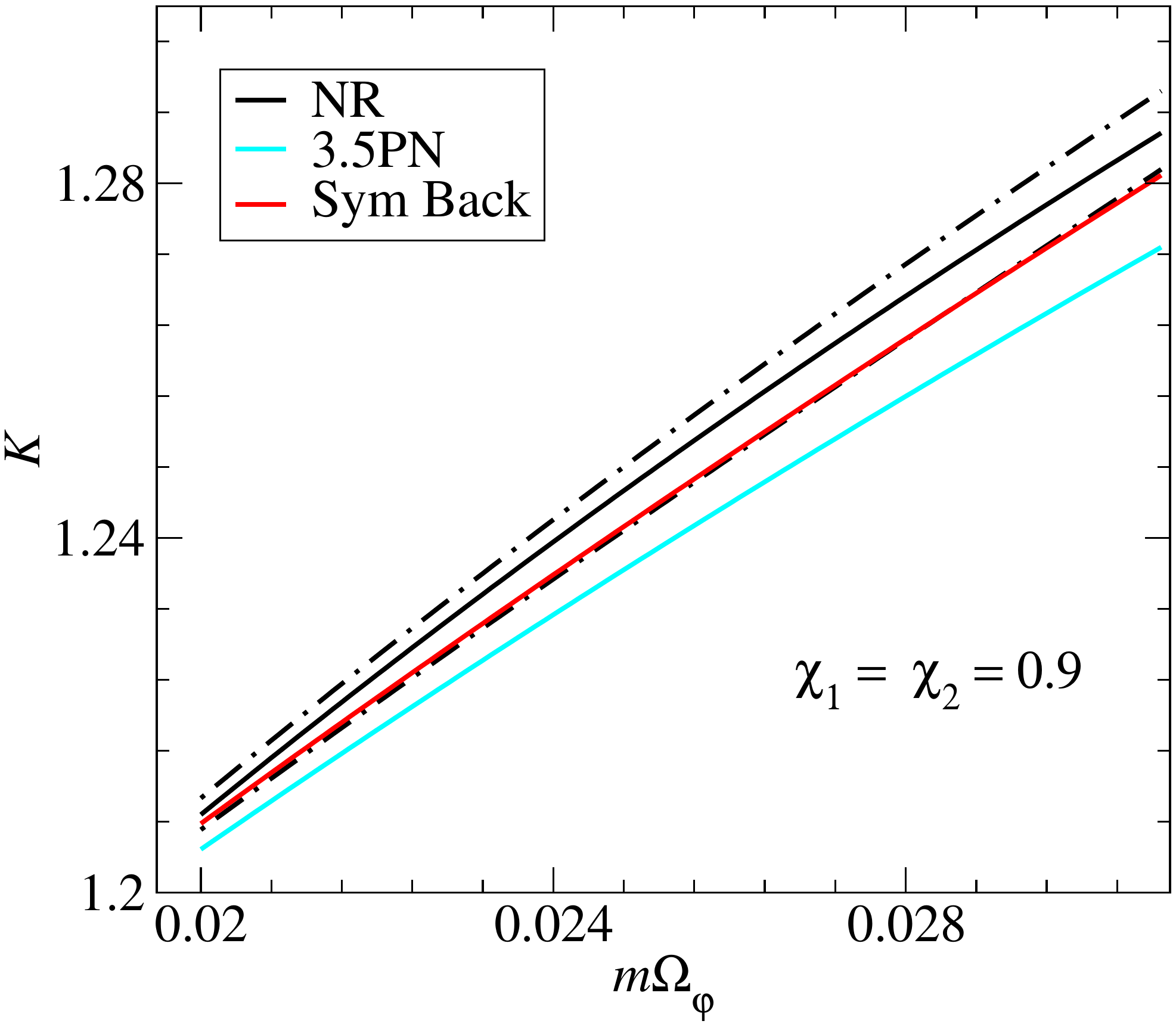}
	\hspace{1cm}
	\includegraphics[scale=0.32]{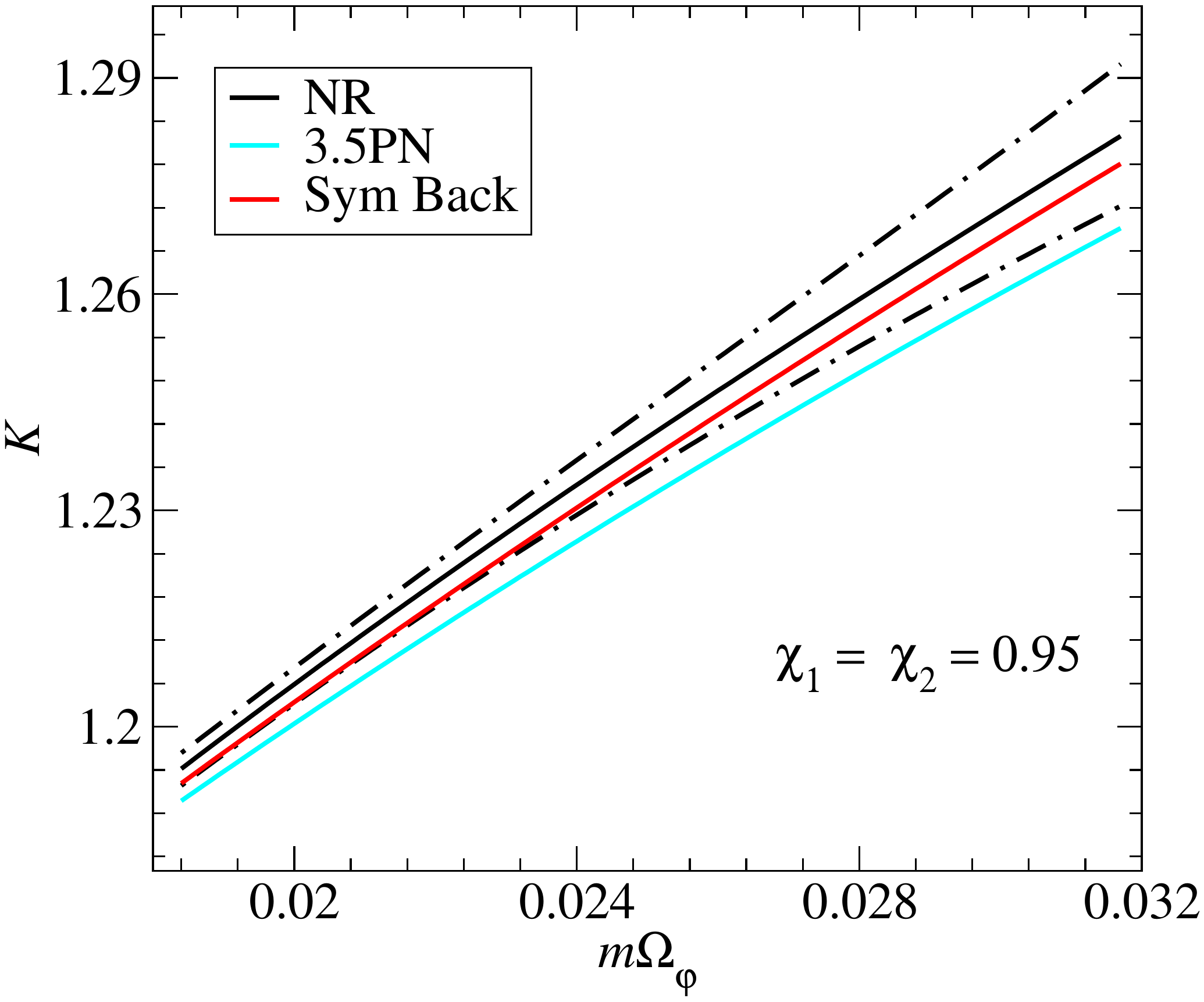}
	\hspace{1cm}
	\includegraphics[scale=0.32]{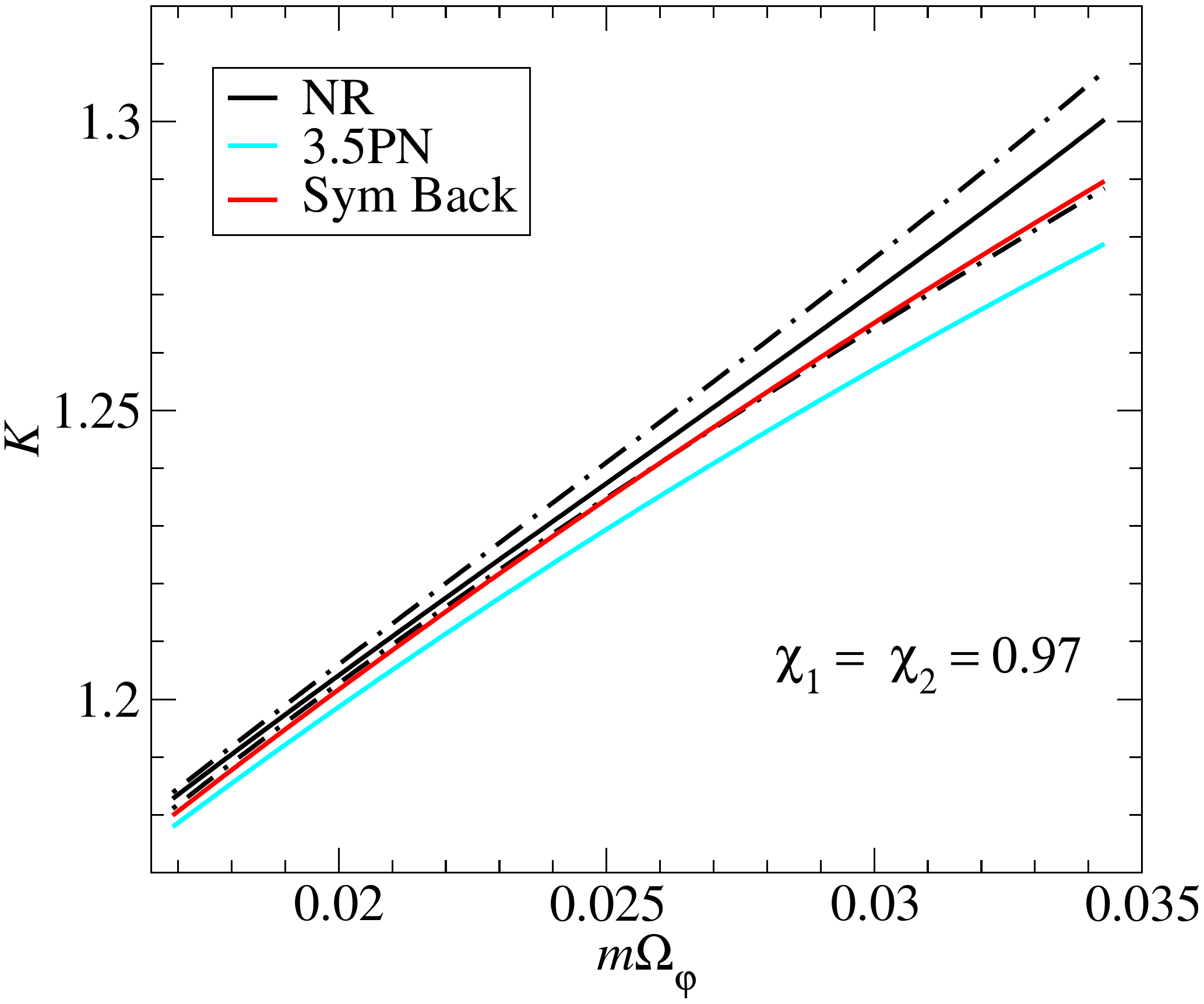}
	\caption{The periastron advance $K$ as a function of the circular-orbit frequency $m\Omega_\varphi$ for equal-mass, equal-spin configurations, as computed using NR simulations (black), post-Newtonian theory to 3.5PN order (cyan), and the improved perturbative expansion \eqref{Wexp} to first order (red).}
	\label{fig:Fig5}
\end{figure*}

\subsection{Comparison for equal-mass, equal-spin configurations}\label{subsec:equal-spin}

In the previous two subsections, we relied upon the input from NR simulations to measure conservative GSF effects on the periastron advance for non-spinning BBH and binaries with one non-zero spin. In this subsection we shall invert that logic, comparing the prediction of perturbation theory (symmetrized in the masses and spins) to those of NR simulations of equal-mass binaries with equal spins $\chi_1 = \chi_2=-0.95,-0.9,0,0.6,0.8,0.9,0.95,0.97$. Figure \ref{fig:Fig5} shows that the predictions of the improved perturbative expansion \eqref{Wexp} used to first order in $\nu$ (red curves), with $W_1$ given by the exact GSF result in a Schwarzschild background, are in very good agreement with the NR results (black curves), even for nearly extremal spins.

Importantly, the red curves in Fig.~\ref{fig:Fig5} were plotted using the inverse sum $(W_\text{SB} + \nu \, W_1)^{-1/2}$, without any further expansion in powers of the symmetric mass ratio, because $W_\text{SB}$ depends implicitly on $\nu$ through the spin variable $\chi_0 = \chi_s + \Delta \, \chi_a$. Note also that the improved perturbative expression does not include all the correct spin information. Indeed, as pointed out in Sec.~\ref{subsec:sym}, the symmetric background $W_\text{SB}$ does not capture the 3.5PN spin-orbit terms, nor the $\calO(\nu^2)$ contributions to the 2.5PN spin-orbit and 3PN spin-spin terms. A proper comparison between the expansion \eqref{Wexp} used to first order in $\nu$ and the NR results should make use of the (so far unknown) GSF correction to the periastron advance of a spinning particle in a Kerr background; here we merely made use of the GSF correction to the periastron advance of a non-spinning particle in a Schwarzschild background.

\subsection{Discussion of the results}

We conclude that, at least for the cases studied in this paper (see Refs.~\cite{Sm.79,FiDe.84,Fa.al.04,Sp.al2.11,Le.al2.12,Na.13} for other examples), the expansion \eqref{Wexp} in powers of $\nu$ used to \textit{first} order provides a better approximation to the exact NR results than the usual PN expansion \eqref{W_PN} used to \textit{third} order. Loosely speaking, this observation suggests that relativistic corrections dominate over finite mass-ratio corrections. This striking observation can be understood, at a heuristic level, as follows:
\begin{itemize}
	\item[(i)] In the formal expansion \eqref{Wexp}, the mass-ratio corrections $W_n$ ($n \geqslant 2$) are suppressed by factors of $\nu^n$ and $\nu^{n-1}$ relative to the leading-order contributions $W_\text{SB}$ and $W_1$, where the symmetric mass ratio ranges in $0 < \nu \leqslant 1/4$;
	\item[(ii)] The contribution $\calO(\nu^n)$ in Eq.~\eqref{Wexp} does not appear before the $n$PN order, i.e., higher mass-ratio corrections are further suppressed by increasingly high powers of the orbital velocity $0 < v \lesssim 0.3$.
\end{itemize}

For larger orbital frequencies (smaller separations), the NR results become much less accurate (see Sec.~\ref{sec:NR}), such that it becomes difficult to assess whether the additional corrections $\calO(\nu^2)$ and higher become significant, in which case the mass-ratio degeneracy observed in Figs.~\ref{fig:SF_non-spin} and \ref{fig:SF_spin-down} would be lifted. Furthermore, as the binary gets increasingly closer to the final plunge and merger, the adiabatic approximation must break down and purely conservative effects on the periastron advance can no longer be disentangled from the dissipative effects of radiation-reaction. A comparison to the conservative piece of the GSF correction to the geodesic periastron advance then becomes meaningless.

\section{Summary and Prospects}\label{sec:sum}

We have studied the periastron advance in binary systems of spinning black holes on quasi-circular orbits, for spins aligned or anti-aligned with the orbital angular momentum, by using numerical-relativity (NR) simulations, the post-Newtonian (PN) approximation and black-hole perturbation theory. For the range of orbital frequencies, mass ratios and spins considered, the 3.5PN approximation reproduces the NR results to within a few percent; this (dis)agreement deteriorates with increasing frequency and mass ratio (more unequal masses).

Motivated by the mathematical structure of the PN expansion, we then devised a simple method to impose the symmetry by exchange of the bodies' labels on the perturbative formula. The resulting ``symmetric background'' recovers most spin effects up to 3PN order. We then introduced a new type of expansion in powers of the symmetric mass ratio, using the symmetric background as a zeroth-order approximation. This allowed us, by comparison to the NR results, to measure the gravitational self-force (GSF) correction to the periastron advance of a non-spinning particle orbiting a black hole of mass $M$ and spin $S = -0.5 M^2$. This is one of the first results encoding the effect of the conservative GSF on the motion of a particle in a Kerr background; see \cite{Sh.al.12} for another example. That such a milestone was obtained by combining information from NR simulations, PN expansions, and black-hole perturbations illustrates the powerful interplay of these approximation methods and numerical techniques.

Numerical relativity simulations can thus be used to gain information regarding perturbative GSF effects on the dynamics of compact-object binaries. However, given the high computational cost and limited accuracy of such simulations, using NR data to develop accurate templates for extreme mass ratio inspirals is unpractical; clearly, standard perturbative methods \cite{SaTa.03,Ba.09,Po.al.11} are far better suited to model the dynamics and gravitational-wave emission of such systems.

However, this work supports the idea that by inverting the logic followed in Secs.~\ref{subsec:exp}--\ref{subsec:Kerr}, the results of perturbative GSF calculations may prove useful for the development of accurate waveforms for binary systems of \textit{spinning} compact objects with \textit{moderate mass ratios}; see Sec.~\ref{subsec:equal-spin}. The ``symmetrization'' introduced in Sec.~\ref{sec:sym} could in principle be applied to other coordinate-invariant diagnostics of the binary dynamics and wave emission, such as the binding energy, the total angular momentum, the fluxes of energy and angular momentum, and the gravitational-wave polarizations themselves. The addition of finite mass-ratio corrections coming from perturbative GSF calculations on top of such symmetric backgrounds, using perturbative expansions of the type \eqref{Wexp}, suggests a novel method to devise highly-accurate approximations to the exact results, even for comparable-mass binaries.

\begin{acknowledgments}
A.B. and A.L.T. acknowledge support from NSF through Grants PHY-0903631 and PHY-1208881. A.B. also acknowledges support from NASA through Grant NNX09AI81G and A.L.T. from the Maryland Center for Fundamental Physics. A.M. and H.P. acknowledge support from NSERC of Canada, from the Canada Research Chairs Program, and from the Canadian Institute for Advanced Research. D.H., L.K., G.L., and S.T. acknowledge support from the Sherman Fairchild Foundation and NSF Grants PHY-1306125 and PHYS-1005426 at Cornell. M.S., B.S., and N.T. gratefully acknowledge support from the Sherman Fairchild Foundation and NSF Grants PHY-1068881, PHY-1005655, and DMS-1065438 at Caltech. The numerical relativity simulations were performed at the GPC supercomputer at the SciNet HPC Consortium \cite{Lo.al4.10}; SciNet is funded by: the Canada Foundation for Innovation (CFI) under the auspices of Compute Canada; the Government of Ontario; Ontario Research Fund--Research Excellence; and the University of Toronto. Further computations were performed on the Caltech computer cluster Zwicky, which was funded by the Sherman Fairchild Foundation and the NSF MRI-R2 Grant PHY-0960291, on SHC at Caltech, which is supported by the Sherman Fairchild Foundation, and on the NSF XSEDE network under Grant TG-PHY990007N.
\end{acknowledgments}

\bibliography{./ListeRef}

\begin{thebibliography}{65}%
\makeatletter
\providecommand \@ifxundefined [1]{%
 \@ifx{#1\undefined}
}%
\providecommand \@ifnum [1]{%
 \ifnum #1\expandafter \@firstoftwo
 \else \expandafter \@secondoftwo
 \fi
}%
\providecommand \@ifx [1]{%
 \ifx #1\expandafter \@firstoftwo
 \else \expandafter \@secondoftwo
 \fi
}%
\providecommand \natexlab [1]{#1}%
\providecommand \enquote  [1]{``#1''}%
\providecommand \bibnamefont  [1]{#1}%
\providecommand \bibfnamefont [1]{#1}%
\providecommand \citenamefont [1]{#1}%
\providecommand \href@noop [0]{\@secondoftwo}%
\providecommand \href [0]{\begingroup \@sanitize@url \@href}%
\providecommand \@href[1]{\@@startlink{#1}\@@href}%
\providecommand \@@href[1]{\endgroup#1\@@endlink}%
\providecommand \@sanitize@url [0]{\catcode `\\12\catcode `\$12\catcode
  `\&12\catcode `\#12\catcode `\^12\catcode `\_12\catcode `\%12\relax}%
\providecommand \@@startlink[1]{}%
\providecommand \@@endlink[0]{}%
\providecommand \url  [0]{\begingroup\@sanitize@url \@url }%
\providecommand \@url [1]{\endgroup\@href {#1}{\urlprefix }}%
\providecommand \urlprefix  [0]{URL }%
\providecommand \Eprint [0]{\href }%
\providecommand \doibase [0]{http://dx.doi.org/}%
\providecommand \selectlanguage [0]{\@gobble}%
\providecommand \bibinfo  [0]{\@secondoftwo}%
\providecommand \bibfield  [0]{\@secondoftwo}%
\providecommand \translation [1]{[#1]}%
\providecommand \BibitemOpen [0]{}%
\providecommand \bibitemStop [0]{}%
\providecommand \bibitemNoStop [0]{.\EOS\space}%
\providecommand \EOS [0]{\spacefactor3000\relax}%
\providecommand \BibitemShut  [1]{\csname bibitem#1\endcsname}%
\let\auto@bib@innerbib\@empty
%</preamble>
\bibitem [{\citenamefont {Einstein}(1915)}]{Ei.15}%
  \BibitemOpen
  \bibfield  {author} {\bibinfo {author} {\bibfnamefont {A.}~\bibnamefont
  {Einstein}},\ }\href@noop {} {\bibfield  {journal} {\bibinfo  {journal}
  {Sitzber. Preuss. Akad. Wiss.}\ ,\ \bibinfo {pages} {831}} (\bibinfo {year}
  {1915})}\BibitemShut {NoStop}%
\bibitem [{\citenamefont {Stairs}(2003)}]{St.03}%
  \BibitemOpen
  \bibfield  {author} {\bibinfo {author} {\bibfnamefont {I.~H.}\ \bibnamefont
  {Stairs}},\ }\href@noop {} {\bibfield  {journal} {\bibinfo  {journal} {Living
  Rev. Rel.}\ }\textbf {\bibinfo {volume} {6}},\ \bibinfo {pages} {5} (\bibinfo
  {year} {2003})},\ \Eprint {http://arxiv.org/abs/arXiv:astro-ph/0307536}
  {arXiv:astro-ph/0307536} \BibitemShut {NoStop}%
\bibitem [{\citenamefont {Lorimer}(2008)}]{Lo.08}%
  \BibitemOpen
  \bibfield  {author} {\bibinfo {author} {\bibfnamefont {D.~R.}\ \bibnamefont
  {Lorimer}},\ }\href@noop {} {\bibfield  {journal} {\bibinfo  {journal}
  {Living Rev. Rel.}\ }\textbf {\bibinfo {volume} {11}},\ \bibinfo {pages} {8}
  (\bibinfo {year} {2008})},\ \Eprint {http://arxiv.org/abs/arXiv:0811.0762
  [astro-ph.CO]} {arXiv:0811.0762 [astro-ph.CO]} \BibitemShut {NoStop}%
\bibitem [{\citenamefont {Shoemaker}(2010)}]{Sh.10}%
  \BibitemOpen
  \bibfield  {author} {\bibinfo {author} {\bibfnamefont {D.}~\bibnamefont
  {Shoemaker}} (\bibinfo {collaboration} {{LIGO} Collaboration}),\ }\href
  {https://dcc.ligo.org/cgi-bin/DocDB/ShowDocument?docid=2974} {\  (\bibinfo
  {year} {2010})},\ \bibinfo {note} {{LIGO} Document T0900288-v3}\BibitemShut
  {NoStop}%
\bibitem [{\citenamefont {Acernese}\ \emph {et~al.}(2008)\citenamefont
  {Acernese} \emph {et~al.}}]{Ac.al.08}%
  \BibitemOpen
  \bibfield  {author} {\bibinfo {author} {\bibfnamefont {F.}~\bibnamefont
  {Acernese}} \emph {et~al.},\ }\href@noop {} {\bibfield  {journal} {\bibinfo
  {journal} {Class. Quant. Grav.}\ }\textbf {\bibinfo {volume} {25}},\ \bibinfo
  {pages} {114045} (\bibinfo {year} {2008})}\BibitemShut {NoStop}%
\bibitem [{eLISA()}]{eLISA}%
  \BibitemOpen
  eLISA,\ \href@noop {} {}\bibinfo {howpublished}
  {\href{http://www.elisascience.org}{http://www.elisascience.org}}\BibitemShut
  {NoStop}%
\bibitem [{\citenamefont {Blanchet}(2006)}]{Bl.06}%
  \BibitemOpen
  \bibfield  {author} {\bibinfo {author} {\bibfnamefont {L.}~\bibnamefont
  {Blanchet}},\ }\href@noop {} {\bibfield  {journal} {\bibinfo  {journal}
  {Living Rev. Rel.}\ }\textbf {\bibinfo {volume} {9}},\ \bibinfo {pages} {4}
  (\bibinfo {year} {2006})},\ \Eprint
  {http://arxiv.org/abs/arXiv:gr-qc/0202016} {arXiv:gr-qc/0202016} \BibitemShut
  {NoStop}%
\bibitem [{\citenamefont {Robertson}(1938)}]{Ro.38}%
  \BibitemOpen
  \bibfield  {author} {\bibinfo {author} {\bibfnamefont {H.~P.}\ \bibnamefont
  {Robertson}},\ }\href@noop {} {\bibfield  {journal} {\bibinfo  {journal}
  {Ann. Math.}\ }\textbf {\bibinfo {volume} {39}},\ \bibinfo {pages} {101}
  (\bibinfo {year} {1938})}\BibitemShut {NoStop}%
\bibitem [{\citenamefont {Damour}\ and\ \citenamefont
  {Sch{\"a}fer}(1988)}]{DaSc.88}%
  \BibitemOpen
  \bibfield  {author} {\bibinfo {author} {\bibfnamefont {T.}~\bibnamefont
  {Damour}}\ and\ \bibinfo {author} {\bibfnamefont {G.}~\bibnamefont
  {Sch{\"a}fer}},\ }\href@noop {} {\bibfield  {journal} {\bibinfo  {journal}
  {Nuovo Cim. B}\ }\textbf {\bibinfo {volume} {101}},\ \bibinfo {pages} {127}
  (\bibinfo {year} {1988})}\BibitemShut {NoStop}%
\bibitem [{\citenamefont {Damour}\ \emph
  {et~al.}(2000{\natexlab{a}})\citenamefont {Damour}, \citenamefont
  {Jaranowski},\ and\ \citenamefont {Sch{\"a}fer}}]{Da.al.00}%
  \BibitemOpen
  \bibfield  {author} {\bibinfo {author} {\bibfnamefont {T.}~\bibnamefont
  {Damour}}, \bibinfo {author} {\bibfnamefont {P.}~\bibnamefont {Jaranowski}},
  \ and\ \bibinfo {author} {\bibfnamefont {G.}~\bibnamefont {Sch{\"a}fer}},\
  }\href@noop {} {\bibfield  {journal} {\bibinfo  {journal} {Phys. Rev. D}\
  }\textbf {\bibinfo {volume} {62}},\ \bibinfo {pages} {044024} (\bibinfo
  {year} {2000}{\natexlab{a}})},\ \Eprint
  {http://arxiv.org/abs/arXiv:gr-qc/9912092} {arXiv:gr-qc/9912092} \BibitemShut
  {NoStop}%
\bibitem [{\citenamefont {Tessmer}\ \emph {et~al.}(2010)\citenamefont
  {Tessmer}, \citenamefont {Hartung},\ and\ \citenamefont
  {Sch{\"a}fer}}]{Te.al.10}%
  \BibitemOpen
  \bibfield  {author} {\bibinfo {author} {\bibfnamefont {M.}~\bibnamefont
  {Tessmer}}, \bibinfo {author} {\bibfnamefont {J.}~\bibnamefont {Hartung}}, \
  and\ \bibinfo {author} {\bibfnamefont {G.}~\bibnamefont {Sch{\"a}fer}},\
  }\href@noop {} {\bibfield  {journal} {\bibinfo  {journal} {Class. Quant.
  Grav.}\ }\textbf {\bibinfo {volume} {27}},\ \bibinfo {pages} {165005}
  (\bibinfo {year} {2010})},\ \Eprint {http://arxiv.org/abs/arXiv:1003.2735
  [gr-qc]} {arXiv:1003.2735 [gr-qc]} \BibitemShut {NoStop}%
\bibitem [{\citenamefont {Tessmer}\ \emph {et~al.}(2013)\citenamefont
  {Tessmer}, \citenamefont {Hartung},\ and\ \citenamefont
  {Sch{\"a}fer}}]{Te.al.13}%
  \BibitemOpen
  \bibfield  {author} {\bibinfo {author} {\bibfnamefont {M.}~\bibnamefont
  {Tessmer}}, \bibinfo {author} {\bibfnamefont {J.}~\bibnamefont {Hartung}}, \
  and\ \bibinfo {author} {\bibfnamefont {G.}~\bibnamefont {Sch{\"a}fer}},\
  }\href@noop {} {\bibfield  {journal} {\bibinfo  {journal} {Class. Quant.
  Grav.}\ }\textbf {\bibinfo {volume} {30}},\ \bibinfo {pages} {015007}
  (\bibinfo {year} {2013})},\ \Eprint {http://arxiv.org/abs/arXiv:1207.6961
  [gr-qc]} {arXiv:1207.6961 [gr-qc]} \BibitemShut {NoStop}%
\bibitem [{\citenamefont {K{\"o}nigsd{\"o}rffer}\ and\ \citenamefont
  {Gopakumar}(2005)}]{KoGo.05}%
  \BibitemOpen
  \bibfield  {author} {\bibinfo {author} {\bibfnamefont {C.}~\bibnamefont
  {K{\"o}nigsd{\"o}rffer}}\ and\ \bibinfo {author} {\bibfnamefont
  {A.}~\bibnamefont {Gopakumar}},\ }\href@noop {} {\bibfield  {journal}
  {\bibinfo  {journal} {Phys. Rev. D}\ }\textbf {\bibinfo {volume} {71}},\
  \bibinfo {pages} {024039} (\bibinfo {year} {2005})},\ \Eprint
  {http://arxiv.org/abs/arXiv:gr-qc/0501011} {arXiv:gr-qc/0501011} \BibitemShut
  {NoStop}%
\bibitem [{\citenamefont {Tessmer}(2009)}]{Te.09}%
  \BibitemOpen
  \bibfield  {author} {\bibinfo {author} {\bibfnamefont {M.}~\bibnamefont
  {Tessmer}},\ }\href@noop {} {\bibfield  {journal} {\bibinfo  {journal} {Phys.
  Rev. D}\ }\textbf {\bibinfo {volume} {80}},\ \bibinfo {pages} {124034}
  (\bibinfo {year} {2009})},\ \Eprint {http://arxiv.org/abs/arXiv:0910.5931
  [gr-qc]} {arXiv:0910.5931 [gr-qc]} \BibitemShut {NoStop}%
\bibitem [{\citenamefont {Sasaki}\ and\ \citenamefont
  {Tagoshi}(2003)}]{SaTa.03}%
  \BibitemOpen
  \bibfield  {author} {\bibinfo {author} {\bibfnamefont {M.}~\bibnamefont
  {Sasaki}}\ and\ \bibinfo {author} {\bibfnamefont {H.}~\bibnamefont
  {Tagoshi}},\ }\href@noop {} {\bibfield  {journal} {\bibinfo  {journal}
  {Living Rev. Rel.}\ }\textbf {\bibinfo {volume} {6}},\ \bibinfo {pages} {5}
  (\bibinfo {year} {2003})},\ \Eprint
  {http://arxiv.org/abs/arXiv:gr-qc/0306120} {arXiv:gr-qc/0306120} \BibitemShut
  {NoStop}%
\bibitem [{\citenamefont {Barack}(2009)}]{Ba.09}%
  \BibitemOpen
  \bibfield  {author} {\bibinfo {author} {\bibfnamefont {L.}~\bibnamefont
  {Barack}},\ }\href@noop {} {\bibfield  {journal} {\bibinfo  {journal} {Class.
  Quant. Grav.}\ }\textbf {\bibinfo {volume} {26}},\ \bibinfo {pages} {213001}
  (\bibinfo {year} {2009})},\ \Eprint {http://arxiv.org/abs/arXiv:0908.1664
  [gr-qc]} {arXiv:0908.1664 [gr-qc]} \BibitemShut {NoStop}%
\bibitem [{\citenamefont {Poisson}\ \emph {et~al.}(2011)\citenamefont
  {Poisson}, \citenamefont {Pound},\ and\ \citenamefont {Vega}}]{Po.al.11}%
  \BibitemOpen
  \bibfield  {author} {\bibinfo {author} {\bibfnamefont {E.}~\bibnamefont
  {Poisson}}, \bibinfo {author} {\bibfnamefont {A.}~\bibnamefont {Pound}}, \
  and\ \bibinfo {author} {\bibfnamefont {I.}~\bibnamefont {Vega}},\ }\href@noop
  {} {\bibfield  {journal} {\bibinfo  {journal} {Living Rev. Rel.}\ }\textbf
  {\bibinfo {volume} {14}},\ \bibinfo {pages} {7} (\bibinfo {year} {2011})},\
  \Eprint {http://arxiv.org/abs/arXiv:1102.0529 [gr-qc]} {arXiv:1102.0529
  [gr-qc]} \BibitemShut {NoStop}%
\bibitem [{\citenamefont {Cutler}\ \emph {et~al.}(1994)\citenamefont {Cutler},
  \citenamefont {Kennefick},\ and\ \citenamefont {Poisson}}]{Cu.al.94}%
  \BibitemOpen
  \bibfield  {author} {\bibinfo {author} {\bibfnamefont {C.}~\bibnamefont
  {Cutler}}, \bibinfo {author} {\bibfnamefont {D.}~\bibnamefont {Kennefick}}, \
  and\ \bibinfo {author} {\bibfnamefont {E.}~\bibnamefont {Poisson}},\
  }\href@noop {} {\bibfield  {journal} {\bibinfo  {journal} {Phys. Rev. D}\
  }\textbf {\bibinfo {volume} {50}},\ \bibinfo {pages} {3816} (\bibinfo {year}
  {1994})}\BibitemShut {NoStop}%
\bibitem [{\citenamefont {Schmidt}(2002)}]{Sc.02}%
  \BibitemOpen
  \bibfield  {author} {\bibinfo {author} {\bibfnamefont {W.}~\bibnamefont
  {Schmidt}},\ }\href@noop {} {\bibfield  {journal} {\bibinfo  {journal}
  {Class. Quant. Grav.}\ }\textbf {\bibinfo {volume} {19}},\ \bibinfo {pages}
  {2743} (\bibinfo {year} {2002})},\ \Eprint
  {http://arxiv.org/abs/arXiv:gr-qc/0202090} {arXiv:gr-qc/0202090} \BibitemShut
  {NoStop}%
\bibitem [{\citenamefont {Hinderer}\ \emph {et~al.}(2013)\citenamefont
  {Hinderer} \emph {et~al.}}]{Hi.al.13}%
  \BibitemOpen
  \bibfield  {author} {\bibinfo {author} {\bibfnamefont {T.}~\bibnamefont
  {Hinderer}} \emph {et~al.},\ }\href@noop {} {\bibfield  {journal} {\bibinfo
  {journal} {Phys. Rev. D}\ }\textbf {\bibinfo {volume} {88}},\ \bibinfo
  {pages} {084005} (\bibinfo {year} {2013})},\ \Eprint
  {http://arxiv.org/abs/arXiv:1309.0544 [gr-qc]} {arXiv:1309.0544 [gr-qc]}
  \BibitemShut {NoStop}%
\bibitem [{\citenamefont {Barack}\ and\ \citenamefont {Sago}(2011)}]{BaSa.11}%
  \BibitemOpen
  \bibfield  {author} {\bibinfo {author} {\bibfnamefont {L.}~\bibnamefont
  {Barack}}\ and\ \bibinfo {author} {\bibfnamefont {N.}~\bibnamefont {Sago}},\
  }\href@noop {} {\bibfield  {journal} {\bibinfo  {journal} {Phys. Rev. D}\
  }\textbf {\bibinfo {volume} {83}},\ \bibinfo {pages} {084023} (\bibinfo
  {year} {2011})},\ \Eprint {http://arxiv.org/abs/arXiv:1101.3331 [gr-qc]}
  {arXiv:1101.3331 [gr-qc]} \BibitemShut {NoStop}%
\bibitem [{\citenamefont {Buonanno}\ and\ \citenamefont
  {Damour}(1999)}]{BuDa.99}%
  \BibitemOpen
  \bibfield  {author} {\bibinfo {author} {\bibfnamefont {A.}~\bibnamefont
  {Buonanno}}\ and\ \bibinfo {author} {\bibfnamefont {T.}~\bibnamefont
  {Damour}},\ }\href@noop {} {\bibfield  {journal} {\bibinfo  {journal} {Phys.
  Rev. D}\ }\textbf {\bibinfo {volume} {59}},\ \bibinfo {pages} {084006}
  (\bibinfo {year} {1999})},\ \Eprint
  {http://arxiv.org/abs/arXiv:gr-qc/9811091} {arXiv:gr-qc/9811091} \BibitemShut
  {NoStop}%
\bibitem [{\citenamefont {Damour}\ \emph
  {et~al.}(2000{\natexlab{b}})\citenamefont {Damour}, \citenamefont
  {Jaranowski},\ and\ \citenamefont {Sch{\"a}fer}}]{Da.al3.00}%
  \BibitemOpen
  \bibfield  {author} {\bibinfo {author} {\bibfnamefont {T.}~\bibnamefont
  {Damour}}, \bibinfo {author} {\bibfnamefont {P.}~\bibnamefont {Jaranowski}},
  \ and\ \bibinfo {author} {\bibfnamefont {G.}~\bibnamefont {Sch{\"a}fer}},\
  }\href@noop {} {\bibfield  {journal} {\bibinfo  {journal} {Phys. Rev. D}\
  }\textbf {\bibinfo {volume} {62}},\ \bibinfo {pages} {084011} (\bibinfo
  {year} {2000}{\natexlab{b}})},\ \Eprint
  {http://arxiv.org/abs/arXiv:gr-qc/0005034} {arXiv:gr-qc/0005034} \BibitemShut
  {NoStop}%
\bibitem [{\citenamefont {Damour}\ \emph
  {et~al.}(2008{\natexlab{a}})\citenamefont {Damour}, \citenamefont
  {Jaranowski},\ and\ \citenamefont {Sch{\"a}fer}}]{Da.al2.08}%
  \BibitemOpen
  \bibfield  {author} {\bibinfo {author} {\bibfnamefont {T.}~\bibnamefont
  {Damour}}, \bibinfo {author} {\bibfnamefont {P.}~\bibnamefont {Jaranowski}},
  \ and\ \bibinfo {author} {\bibfnamefont {G.}~\bibnamefont {Sch{\"a}fer}},\
  }\href@noop {} {\bibfield  {journal} {\bibinfo  {journal} {Phys. Rev. D}\
  }\textbf {\bibinfo {volume} {78}},\ \bibinfo {pages} {024009} (\bibinfo
  {year} {2008}{\natexlab{a}})},\ \Eprint {http://arxiv.org/abs/arXiv:0803.0915
  [gr-qc]} {arXiv:0803.0915 [gr-qc]} \BibitemShut {NoStop}%
\bibitem [{\citenamefont {Barausse}\ and\ \citenamefont
  {Buonanno}(2010)}]{BaBu.10}%
  \BibitemOpen
  \bibfield  {author} {\bibinfo {author} {\bibfnamefont {E.}~\bibnamefont
  {Barausse}}\ and\ \bibinfo {author} {\bibfnamefont {A.}~\bibnamefont
  {Buonanno}},\ }\href@noop {} {\bibfield  {journal} {\bibinfo  {journal}
  {Phys. Rev. D}\ }\textbf {\bibinfo {volume} {81}},\ \bibinfo {pages} {084024}
  (\bibinfo {year} {2010})},\ \Eprint {http://arxiv.org/abs/arXiv:0912.3517
  [gr-qc]} {arXiv:0912.3517 [gr-qc]} \BibitemShut {NoStop}%
\bibitem [{\citenamefont {Damour}(2010)}]{Da.10}%
  \BibitemOpen
  \bibfield  {author} {\bibinfo {author} {\bibfnamefont {T.}~\bibnamefont
  {Damour}},\ }\href@noop {} {\bibfield  {journal} {\bibinfo  {journal} {Phys.
  Rev. D}\ }\textbf {\bibinfo {volume} {81}},\ \bibinfo {pages} {024017}
  (\bibinfo {year} {2010})},\ \Eprint {http://arxiv.org/abs/arXiv:0910.5533
  [gr-qc]} {arXiv:0910.5533 [gr-qc]} \BibitemShut {NoStop}%
\bibitem [{\citenamefont {Pretorius}(2005)}]{Pr.05}%
  \BibitemOpen
  \bibfield  {author} {\bibinfo {author} {\bibfnamefont {F.}~\bibnamefont
  {Pretorius}},\ }\href@noop {} {\bibfield  {journal} {\bibinfo  {journal}
  {Phys. Rev. Lett.}\ }\textbf {\bibinfo {volume} {95}},\ \bibinfo {pages}
  {121101} (\bibinfo {year} {2005})},\ \Eprint
  {http://arxiv.org/abs/arXiv:gr-qc/0507014} {arXiv:gr-qc/0507014} \BibitemShut
  {NoStop}%
\bibitem [{\citenamefont {Baker}\ \emph {et~al.}(2006)\citenamefont {Baker},
  \citenamefont {Centrella}, \citenamefont {Choi}, \citenamefont {Koppitz},\
  and\ \citenamefont {{van Meter}}}]{Ba.al.06}%
  \BibitemOpen
  \bibfield  {author} {\bibinfo {author} {\bibfnamefont {J.~G.}\ \bibnamefont
  {Baker}}, \bibinfo {author} {\bibfnamefont {J.}~\bibnamefont {Centrella}},
  \bibinfo {author} {\bibfnamefont {D.-I.}\ \bibnamefont {Choi}}, \bibinfo
  {author} {\bibfnamefont {M.}~\bibnamefont {Koppitz}}, \ and\ \bibinfo
  {author} {\bibfnamefont {J.}~\bibnamefont {{van Meter}}},\ }\href@noop {}
  {\bibfield  {journal} {\bibinfo  {journal} {Phys. Rev. Lett.}\ }\textbf
  {\bibinfo {volume} {96}},\ \bibinfo {pages} {111102} (\bibinfo {year}
  {2006})},\ \Eprint {http://arxiv.org/abs/arXiv:gr-qc/0511103}
  {arXiv:gr-qc/0511103} \BibitemShut {NoStop}%
\bibitem [{\citenamefont {Campanelli}\ \emph {et~al.}(2006)\citenamefont
  {Campanelli}, \citenamefont {Lousto}, \citenamefont {Marronetti},\ and\
  \citenamefont {Zlochower}}]{Ca.al.06}%
  \BibitemOpen
  \bibfield  {author} {\bibinfo {author} {\bibfnamefont {M.}~\bibnamefont
  {Campanelli}}, \bibinfo {author} {\bibfnamefont {C.~O.}\ \bibnamefont
  {Lousto}}, \bibinfo {author} {\bibfnamefont {P.}~\bibnamefont {Marronetti}},
  \ and\ \bibinfo {author} {\bibfnamefont {Y.}~\bibnamefont {Zlochower}},\
  }\href@noop {} {\bibfield  {journal} {\bibinfo  {journal} {Phys. Rev. Lett.}\
  }\textbf {\bibinfo {volume} {96}},\ \bibinfo {pages} {111101} (\bibinfo
  {year} {2006})},\ \Eprint {http://arxiv.org/abs/arXiv:gr-qc/0511048}
  {arXiv:gr-qc/0511048} \BibitemShut {NoStop}%
\bibitem [{\citenamefont {Pfeiffer}(2012)}]{Pf.12}%
  \BibitemOpen
  \bibfield  {author} {\bibinfo {author} {\bibfnamefont {H.~P.}\ \bibnamefont
  {Pfeiffer}},\ }\href@noop {} {\bibfield  {journal} {\bibinfo  {journal}
  {Class. Quant. Grav.}\ }\textbf {\bibinfo {volume} {29}},\ \bibinfo {pages}
  {124004} (\bibinfo {year} {2012})},\ \Eprint
  {http://arxiv.org/abs/arXiv:1203.5166 [gr-qc]} {arXiv:1203.5166 [gr-qc]}
  \BibitemShut {NoStop}%
\bibitem [{\citenamefont {Mrou{\'e}}\ \emph {et~al.}(2010)\citenamefont
  {Mrou{\'e}}, \citenamefont {Pfeiffer}, \citenamefont {Kidder},\ and\
  \citenamefont {Teukolsky}}]{Mr.al.10}%
  \BibitemOpen
  \bibfield  {author} {\bibinfo {author} {\bibfnamefont {A.~H.}\ \bibnamefont
  {Mrou{\'e}}}, \bibinfo {author} {\bibfnamefont {H.~P.}\ \bibnamefont
  {Pfeiffer}}, \bibinfo {author} {\bibfnamefont {L.~E.}\ \bibnamefont
  {Kidder}}, \ and\ \bibinfo {author} {\bibfnamefont {S.~A.}\ \bibnamefont
  {Teukolsky}},\ }\href@noop {} {\bibfield  {journal} {\bibinfo  {journal}
  {Phys. Rev. D}\ }\textbf {\bibinfo {volume} {82}},\ \bibinfo {pages} {124016}
  (\bibinfo {year} {2010})},\ \Eprint {http://arxiv.org/abs/arXiv:1004.4697
  [gr-qc]} {arXiv:1004.4697 [gr-qc]} \BibitemShut {NoStop}%
\bibitem [{\citenamefont {{Le Tiec}}\ \emph {et~al.}(2011)\citenamefont {{Le
  Tiec}} \emph {et~al.}}]{Le.al.11}%
  \BibitemOpen
  \bibfield  {author} {\bibinfo {author} {\bibfnamefont {A.}~\bibnamefont {{Le
  Tiec}}} \emph {et~al.},\ }\href@noop {} {\bibfield  {journal} {\bibinfo
  {journal} {Phys. Rev. Lett.}\ }\textbf {\bibinfo {volume} {107}},\ \bibinfo
  {pages} {141101} (\bibinfo {year} {2011})},\ \Eprint
  {http://arxiv.org/abs/arXiv:1106.3278 [gr-qc]} {arXiv:1106.3278 [gr-qc]}
  \BibitemShut {NoStop}%
\bibitem [{\citenamefont {Foucart}\ \emph {et~al.}(2013)\citenamefont {Foucart}
  \emph {et~al.}}]{Fo.al.13}%
  \BibitemOpen
  \bibfield  {author} {\bibinfo {author} {\bibfnamefont {F.}~\bibnamefont
  {Foucart}} \emph {et~al.},\ }\href@noop {} {\bibfield  {journal} {\bibinfo
  {journal} {Phys. Rev. D}\ }\textbf {\bibinfo {volume} {88}},\ \bibinfo
  {pages} {064017} (\bibinfo {year} {2013})},\ \Eprint
  {http://arxiv.org/abs/arXiv:1307.7685 [gr-qc]} {arXiv:1307.7685 [gr-qc]}
  \BibitemShut {NoStop}%
\bibitem [{\citenamefont {Hemberger}\ \emph {et~al.}(2013)\citenamefont
  {Hemberger} \emph {et~al.}}]{He.al.13}%
  \BibitemOpen
  \bibfield  {author} {\bibinfo {author} {\bibfnamefont {D.~A.}\ \bibnamefont
  {Hemberger}} \emph {et~al.},\ }\href@noop {} {\bibfield  {journal} {\bibinfo
  {journal} {Phys. Rev. D}\ }\textbf {\bibinfo {volume} {88}},\ \bibinfo
  {pages} {064014} (\bibinfo {year} {2013})},\ \Eprint
  {http://arxiv.org/abs/arXiv:1305.5991 [gr-qc]} {arXiv:1305.5991 [gr-qc]}
  \BibitemShut {NoStop}%
\bibitem [{\citenamefont {Lovelace}\ \emph {et~al.}(2012)\citenamefont
  {Lovelace}, \citenamefont {Boyle}, \citenamefont {Scheel},\ and\
  \citenamefont {Szil{\'a}gyi}}]{Lo.al.12}%
  \BibitemOpen
  \bibfield  {author} {\bibinfo {author} {\bibfnamefont {G.}~\bibnamefont
  {Lovelace}}, \bibinfo {author} {\bibfnamefont {M.}~\bibnamefont {Boyle}},
  \bibinfo {author} {\bibfnamefont {M.~A.}\ \bibnamefont {Scheel}}, \ and\
  \bibinfo {author} {\bibfnamefont {B.}~\bibnamefont {Szil{\'a}gyi}},\
  }\href@noop {} {\bibfield  {journal} {\bibinfo  {journal} {Class. Quant.
  Grav.}\ }\textbf {\bibinfo {volume} {29}},\ \bibinfo {pages} {045003}
  (\bibinfo {year} {2012})},\ \Eprint {http://arxiv.org/abs/arXiv:1110.2229
  [gr-qc]} {arXiv:1110.2229 [gr-qc]} \BibitemShut {NoStop}%
\bibitem [{\citenamefont {Mrou{\'e}}\ and\ \citenamefont
  {Pfeiffer}(2012)}]{MrPf.12}%
  \BibitemOpen
  \bibfield  {author} {\bibinfo {author} {\bibfnamefont {A.~H.}\ \bibnamefont
  {Mrou{\'e}}}\ and\ \bibinfo {author} {\bibfnamefont {H.~P.}\ \bibnamefont
  {Pfeiffer}},\ }\href@noop {} {\  (\bibinfo {year} {2012})},\ \Eprint
  {http://arxiv.org/abs/arXiv:1210.2958 [gr-qc]} {arXiv:1210.2958 [gr-qc]}
  \BibitemShut {NoStop}%
\bibitem [{\citenamefont {Smarr}(1979)}]{Sm.79}%
  \BibitemOpen
  \bibfield  {author} {\bibinfo {author} {\bibfnamefont {L.}~\bibnamefont
  {Smarr}},\ }in\ \href@noop {} {\emph {\bibinfo {booktitle} {Sources of
  gravitational radiation}}},\ \bibinfo {editor} {edited by\ \bibinfo {editor}
  {\bibfnamefont {L.}~\bibnamefont {Smarr}}}\ (\bibinfo  {publisher} {Cambridge
  University Press},\ \bibinfo {address} {Cambridge},\ \bibinfo {year} {1979})\
  p.\ \bibinfo {pages} {245}\BibitemShut {NoStop}%
\bibitem [{\citenamefont {Fitchett}\ and\ \citenamefont
  {Detweiler}(1984)}]{FiDe.84}%
  \BibitemOpen
  \bibfield  {author} {\bibinfo {author} {\bibfnamefont {M.~J.}\ \bibnamefont
  {Fitchett}}\ and\ \bibinfo {author} {\bibfnamefont {S.}~\bibnamefont
  {Detweiler}},\ }\href@noop {} {\bibfield  {journal} {\bibinfo  {journal}
  {Mon. Not. Roy. Astron. Soc.}\ }\textbf {\bibinfo {volume} {211}},\ \bibinfo
  {pages} {933} (\bibinfo {year} {1984})}\BibitemShut {NoStop}%
\bibitem [{\citenamefont {Favata}\ \emph {et~al.}(2004)\citenamefont {Favata},
  \citenamefont {Hughes},\ and\ \citenamefont {Holz}}]{Fa.al.04}%
  \BibitemOpen
  \bibfield  {author} {\bibinfo {author} {\bibfnamefont {M.}~\bibnamefont
  {Favata}}, \bibinfo {author} {\bibfnamefont {S.~A.}\ \bibnamefont {Hughes}},
  \ and\ \bibinfo {author} {\bibfnamefont {D.~E.}\ \bibnamefont {Holz}},\
  }\href@noop {} {\bibfield  {journal} {\bibinfo  {journal} {Astrophys. J.}\
  }\textbf {\bibinfo {volume} {607}},\ \bibinfo {pages} {L5} (\bibinfo {year}
  {2004})},\ \Eprint {http://arxiv.org/abs/arXiv:astro-ph/0402056}
  {arXiv:astro-ph/0402056} \BibitemShut {NoStop}%
\bibitem [{\citenamefont {Sperhake}\ \emph {et~al.}(2011)\citenamefont
  {Sperhake}, \citenamefont {Cardoso}, \citenamefont {Ott}, \citenamefont
  {Schnetter},\ and\ \citenamefont {Witek}}]{Sp.al2.11}%
  \BibitemOpen
  \bibfield  {author} {\bibinfo {author} {\bibfnamefont {U.}~\bibnamefont
  {Sperhake}}, \bibinfo {author} {\bibfnamefont {V.}~\bibnamefont {Cardoso}},
  \bibinfo {author} {\bibfnamefont {C.~D.}\ \bibnamefont {Ott}}, \bibinfo
  {author} {\bibfnamefont {E.}~\bibnamefont {Schnetter}}, \ and\ \bibinfo
  {author} {\bibfnamefont {H.}~\bibnamefont {Witek}},\ }\href@noop {}
  {\bibfield  {journal} {\bibinfo  {journal} {Phys. Rev. D}\ }\textbf {\bibinfo
  {volume} {84}},\ \bibinfo {pages} {084038} (\bibinfo {year} {2011})},\
  \Eprint {http://arxiv.org/abs/arXiv:1105.5391 [gr-qc]} {arXiv:1105.5391
  [gr-qc]} \BibitemShut {NoStop}%
\bibitem [{\citenamefont {{Le Tiec}}\ \emph {et~al.}(2012)\citenamefont {{Le
  Tiec}}, \citenamefont {Barausse},\ and\ \citenamefont
  {Buonanno}}]{Le.al2.12}%
  \BibitemOpen
  \bibfield  {author} {\bibinfo {author} {\bibfnamefont {A.}~\bibnamefont {{Le
  Tiec}}}, \bibinfo {author} {\bibfnamefont {E.}~\bibnamefont {Barausse}}, \
  and\ \bibinfo {author} {\bibfnamefont {A.}~\bibnamefont {Buonanno}},\
  }\href@noop {} {\bibfield  {journal} {\bibinfo  {journal} {Phys. Rev. Lett.}\
  }\textbf {\bibinfo {volume} {108}},\ \bibinfo {pages} {131103} (\bibinfo
  {year} {2012})},\ \Eprint {http://arxiv.org/abs/arXiv:1111.5609 [gr-qc]}
  {arXiv:1111.5609 [gr-qc]} \BibitemShut {NoStop}%
\bibitem [{\citenamefont {Nagar}(2013)}]{Na.13}%
  \BibitemOpen
  \bibfield  {author} {\bibinfo {author} {\bibfnamefont {A.}~\bibnamefont
  {Nagar}},\ }\href@noop {} {\  (\bibinfo {year} {2013})},\ \Eprint
  {http://arxiv.org/abs/arXiv:1306.6299 [gr-qc]} {arXiv:1306.6299 [gr-qc]}
  \BibitemShut {NoStop}%
\bibitem [{\citenamefont {Mrou{\'e}}\ \emph {et~al.}(2013)\citenamefont
  {Mrou{\'e}} \emph {et~al.}}]{Mr.al.13}%
  \BibitemOpen
  \bibfield  {author} {\bibinfo {author} {\bibfnamefont {A.~H.}\ \bibnamefont
  {Mrou{\'e}}} \emph {et~al.},\ }\href@noop {} {\  (\bibinfo {year} {2013})},\
  \Eprint {http://arxiv.org/abs/arXiv:1304.6077 [gr-qc]} {arXiv:1304.6077
  [gr-qc]} \BibitemShut {NoStop}%
\bibitem [{\citenamefont {Buchman}\ \emph {et~al.}(2012)\citenamefont
  {Buchman}, \citenamefont {Pfeiffer}, \citenamefont {Scheel},\ and\
  \citenamefont {Szil{\'a}gyi}}]{Bu.al.12}%
  \BibitemOpen
  \bibfield  {author} {\bibinfo {author} {\bibfnamefont {L.~T.}\ \bibnamefont
  {Buchman}}, \bibinfo {author} {\bibfnamefont {H.~P.}\ \bibnamefont
  {Pfeiffer}}, \bibinfo {author} {\bibfnamefont {M.~A.}\ \bibnamefont
  {Scheel}}, \ and\ \bibinfo {author} {\bibfnamefont {B.}~\bibnamefont
  {Szil{\'a}gyi}},\ }\href@noop {} {\bibfield  {journal} {\bibinfo  {journal}
  {Phys. Rev. D}\ }\textbf {\bibinfo {volume} {86}},\ \bibinfo {pages} {084033}
  (\bibinfo {year} {2012})},\ \Eprint {http://arxiv.org/abs/arXiv:1206.3015
  [gr-qc]} {arXiv:1206.3015 [gr-qc]} \BibitemShut {NoStop}%
\bibitem [{\citenamefont {MacDonald}\ \emph {et~al.}(2013)\citenamefont
  {MacDonald} \emph {et~al.}}]{Ma.al2.13}%
  \BibitemOpen
  \bibfield  {author} {\bibinfo {author} {\bibfnamefont {I.}~\bibnamefont
  {MacDonald}} \emph {et~al.},\ }\href@noop {} {\bibfield  {journal} {\bibinfo
  {journal} {Phys. Rev. D}\ }\textbf {\bibinfo {volume} {87}},\ \bibinfo
  {pages} {024009} (\bibinfo {year} {2013})},\ \Eprint
  {http://arxiv.org/abs/arXiv:1210.3007 [gr-qc]} {arXiv:1210.3007 [gr-qc]}
  \BibitemShut {NoStop}%
\bibitem [{SpEC()}]{SpECwebsite}%
  \BibitemOpen
  SpEC,\ \href@noop {} {}\bibinfo {howpublished}
  {\href{http://www.black-holes.org/SpEC.html}{http://www.black-holes.org/SpEC.html}}\BibitemShut
  {NoStop}%
\bibitem [{\citenamefont {Hergt}\ and\ \citenamefont
  {Sch{\"a}fer}(2008{\natexlab{a}})}]{HeSc.08}%
  \BibitemOpen
  \bibfield  {author} {\bibinfo {author} {\bibfnamefont {S.}~\bibnamefont
  {Hergt}}\ and\ \bibinfo {author} {\bibfnamefont {G.}~\bibnamefont
  {Sch{\"a}fer}},\ }\href@noop {} {\bibfield  {journal} {\bibinfo  {journal}
  {Phys. Rev. D}\ }\textbf {\bibinfo {volume} {77}},\ \bibinfo {pages} {104001}
  (\bibinfo {year} {2008}{\natexlab{a}})},\ \Eprint
  {http://arxiv.org/abs/arXiv:0712.1515 [gr-qc]} {arXiv:0712.1515 [gr-qc]}
  \BibitemShut {NoStop}%
\bibitem [{\citenamefont {Hergt}\ and\ \citenamefont
  {Sch{\"a}fer}(2008{\natexlab{b}})}]{HeSc2.08}%
  \BibitemOpen
  \bibfield  {author} {\bibinfo {author} {\bibfnamefont {S.}~\bibnamefont
  {Hergt}}\ and\ \bibinfo {author} {\bibfnamefont {G.}~\bibnamefont
  {Sch{\"a}fer}},\ }\href@noop {} {\bibfield  {journal} {\bibinfo  {journal}
  {Phys. Rev. D}\ }\textbf {\bibinfo {volume} {78}},\ \bibinfo {pages} {124004}
  (\bibinfo {year} {2008}{\natexlab{b}})},\ \Eprint
  {http://arxiv.org/abs/arXiv:0809.2208 [gr-qc]} {arXiv:0809.2208 [gr-qc]}
  \BibitemShut {NoStop}%
\bibitem [{\citenamefont {Blanchet}\ \emph {et~al.}(2006)\citenamefont
  {Blanchet}, \citenamefont {Buonanno},\ and\ \citenamefont {Faye}}]{Bl.al.06}%
  \BibitemOpen
  \bibfield  {author} {\bibinfo {author} {\bibfnamefont {L.}~\bibnamefont
  {Blanchet}}, \bibinfo {author} {\bibfnamefont {A.}~\bibnamefont {Buonanno}},
  \ and\ \bibinfo {author} {\bibfnamefont {G.}~\bibnamefont {Faye}},\
  }\href@noop {} {\bibfield  {journal} {\bibinfo  {journal} {Phys. Rev. D}\
  }\textbf {\bibinfo {volume} {74}},\ \bibinfo {pages} {104034} (\bibinfo
  {year} {2006})},\ \bibinfo {note} {\textit{{E}rrata:} Phys. Rev. D
  \textbf{75}, 049903(E) (2007) \& Phys. Rev. D \textbf{81}, 089901(E)
  (2010)},\ \Eprint {http://arxiv.org/abs/arXiv:gr-qc/0605140}
  {arXiv:gr-qc/0605140} \BibitemShut {NoStop}%
\bibitem [{\citenamefont {Damour}\ \emph
  {et~al.}(2008{\natexlab{b}})\citenamefont {Damour}, \citenamefont
  {Jaranowski},\ and\ \citenamefont {Sch{\"a}fer}}]{Da.al.08}%
  \BibitemOpen
  \bibfield  {author} {\bibinfo {author} {\bibfnamefont {T.}~\bibnamefont
  {Damour}}, \bibinfo {author} {\bibfnamefont {P.}~\bibnamefont {Jaranowski}},
  \ and\ \bibinfo {author} {\bibfnamefont {G.}~\bibnamefont {Sch{\"a}fer}},\
  }\href@noop {} {\bibfield  {journal} {\bibinfo  {journal} {Phys. Rev. D}\
  }\textbf {\bibinfo {volume} {77}},\ \bibinfo {pages} {064032} (\bibinfo
  {year} {2008}{\natexlab{b}})},\ \Eprint {http://arxiv.org/abs/arXiv:0711.1048
  [gr-qc]} {arXiv:0711.1048 [gr-qc]} \BibitemShut {NoStop}%
\bibitem [{\citenamefont {Barack}\ \emph {et~al.}(2010)\citenamefont {Barack},
  \citenamefont {Damour},\ and\ \citenamefont {Sago}}]{Ba.al.10}%
  \BibitemOpen
  \bibfield  {author} {\bibinfo {author} {\bibfnamefont {L.}~\bibnamefont
  {Barack}}, \bibinfo {author} {\bibfnamefont {T.}~\bibnamefont {Damour}}, \
  and\ \bibinfo {author} {\bibfnamefont {N.}~\bibnamefont {Sago}},\ }\href@noop
  {} {\bibfield  {journal} {\bibinfo  {journal} {Phys. Rev. D}\ }\textbf
  {\bibinfo {volume} {82}},\ \bibinfo {pages} {084036} (\bibinfo {year}
  {2010})},\ \Eprint {http://arxiv.org/abs/arXiv:1008.0935 [gr-qc]}
  {arXiv:1008.0935 [gr-qc]} \BibitemShut {NoStop}%
\bibitem [{\citenamefont {Barausse}\ \emph {et~al.}(2012)\citenamefont
  {Barausse}, \citenamefont {Buonanno},\ and\ \citenamefont {{Le
  Tiec}}}]{Ba.al.12}%
  \BibitemOpen
  \bibfield  {author} {\bibinfo {author} {\bibfnamefont {E.}~\bibnamefont
  {Barausse}}, \bibinfo {author} {\bibfnamefont {A.}~\bibnamefont {Buonanno}},
  \ and\ \bibinfo {author} {\bibfnamefont {A.}~\bibnamefont {{Le Tiec}}},\
  }\href@noop {} {\bibfield  {journal} {\bibinfo  {journal} {Phys. Rev. D}\
  }\textbf {\bibinfo {volume} {85}},\ \bibinfo {pages} {064010} (\bibinfo
  {year} {2012})},\ \Eprint {http://arxiv.org/abs/arXiv:1111.5610 [gr-qc]}
  {arXiv:1111.5610 [gr-qc]} \BibitemShut {NoStop}%
\bibitem [{\citenamefont {Abramowicz}\ and\ \citenamefont
  {Fragile}(2013)}]{AbFr.13}%
  \BibitemOpen
  \bibfield  {author} {\bibinfo {author} {\bibfnamefont {M.~A.}\ \bibnamefont
  {Abramowicz}}\ and\ \bibinfo {author} {\bibfnamefont {P.~C.}\ \bibnamefont
  {Fragile}},\ }\href@noop {} {\bibfield  {journal} {\bibinfo  {journal}
  {Living Rev. Rel.}\ }\textbf {\bibinfo {volume} {16}},\ \bibinfo {pages} {1}
  (\bibinfo {year} {2013})},\ \Eprint {http://arxiv.org/abs/arXiv:1104.5499
  [astro-ph.HE]} {arXiv:1104.5499 [astro-ph.HE]} \BibitemShut {NoStop}%
\bibitem [{\citenamefont {Bardeen}\ \emph {et~al.}(1972)\citenamefont
  {Bardeen}, \citenamefont {Press},\ and\ \citenamefont
  {Teukolsky}}]{Ba.al.72}%
  \BibitemOpen
  \bibfield  {author} {\bibinfo {author} {\bibfnamefont {J.~M.}\ \bibnamefont
  {Bardeen}}, \bibinfo {author} {\bibfnamefont {W.~H.}\ \bibnamefont {Press}},
  \ and\ \bibinfo {author} {\bibfnamefont {S.~A.}\ \bibnamefont {Teukolsky}},\
  }\href@noop {} {\bibfield  {journal} {\bibinfo  {journal} {Astrophys. J.}\
  }\textbf {\bibinfo {volume} {178}},\ \bibinfo {pages} {347} (\bibinfo {year}
  {1972})}\BibitemShut {NoStop}%
\bibitem [{\citenamefont {Mino}(2003)}]{Mi.03}%
  \BibitemOpen
  \bibfield  {author} {\bibinfo {author} {\bibfnamefont {Y.}~\bibnamefont
  {Mino}},\ }\href@noop {} {\bibfield  {journal} {\bibinfo  {journal} {Phys.
  Rev. D}\ }\textbf {\bibinfo {volume} {67}},\ \bibinfo {pages} {084027}
  (\bibinfo {year} {2003})},\ \Eprint
  {http://arxiv.org/abs/arXiv:gr-qc/0302075} {arXiv:gr-qc/0302075} \BibitemShut
  {NoStop}%
\bibitem [{\citenamefont {Favata}(2011)}]{Fa2.11}%
  \BibitemOpen
  \bibfield  {author} {\bibinfo {author} {\bibfnamefont {M.}~\bibnamefont
  {Favata}},\ }\href@noop {} {\bibfield  {journal} {\bibinfo  {journal} {Phys.
  Rev. D}\ }\textbf {\bibinfo {volume} {83}},\ \bibinfo {pages} {024028}
  (\bibinfo {year} {2011})},\ \Eprint {http://arxiv.org/abs/arXiv:1010.2553
  [gr-qc]} {arXiv:1010.2553 [gr-qc]} \BibitemShut {NoStop}%
\bibitem [{\citenamefont {Boh{\'e}}\ \emph {et~al.}(2013)\citenamefont
  {Boh{\'e}}, \citenamefont {Marsat},\ and\ \citenamefont
  {Blanchet}}]{Bo.al.13}%
  \BibitemOpen
  \bibfield  {author} {\bibinfo {author} {\bibfnamefont {A.}~\bibnamefont
  {Boh{\'e}}}, \bibinfo {author} {\bibfnamefont {S.}~\bibnamefont {Marsat}}, \
  and\ \bibinfo {author} {\bibfnamefont {L.}~\bibnamefont {Blanchet}},\
  }\href@noop {} {\bibfield  {journal} {\bibinfo  {journal} {Class. Quant.
  Grav.}\ }\textbf {\bibinfo {volume} {30}},\ \bibinfo {pages} {135009}
  (\bibinfo {year} {2013})},\ \Eprint {http://arxiv.org/abs/arXiv:1303.7412
  [gr-qc]} {arXiv:1303.7412 [gr-qc]} \BibitemShut {NoStop}%
\bibitem [{\citenamefont {Boyle}\ \emph {et~al.}(2008)\citenamefont {Boyle},
  \citenamefont {Kesden},\ and\ \citenamefont {Nissanke}}]{Bo.al2.08}%
  \BibitemOpen
  \bibfield  {author} {\bibinfo {author} {\bibfnamefont {L.}~\bibnamefont
  {Boyle}}, \bibinfo {author} {\bibfnamefont {M.}~\bibnamefont {Kesden}}, \
  and\ \bibinfo {author} {\bibfnamefont {S.}~\bibnamefont {Nissanke}},\
  }\href@noop {} {\bibfield  {journal} {\bibinfo  {journal} {Phys. Rev. Lett.}\
  }\textbf {\bibinfo {volume} {100}},\ \bibinfo {pages} {151101} (\bibinfo
  {year} {2008})},\ \Eprint {http://arxiv.org/abs/arXiv:0709.0299 [gr-qc]}
  {arXiv:0709.0299 [gr-qc]} \BibitemShut {NoStop}%
\bibitem [{\citenamefont {Boyle}\ and\ \citenamefont {Kesden}(2008)}]{BoKe.08}%
  \BibitemOpen
  \bibfield  {author} {\bibinfo {author} {\bibfnamefont {L.}~\bibnamefont
  {Boyle}}\ and\ \bibinfo {author} {\bibfnamefont {M.}~\bibnamefont {Kesden}},\
  }\href@noop {} {\bibfield  {journal} {\bibinfo  {journal} {Phys. Rev. D}\
  }\textbf {\bibinfo {volume} {78}},\ \bibinfo {pages} {024017} (\bibinfo
  {year} {2008})},\ \Eprint {http://arxiv.org/abs/arXiv:0712.2819 [gr-qc]}
  {arXiv:0712.2819 [gr-qc]} \BibitemShut {NoStop}%
\bibitem [{\citenamefont {Blanchet}\ and\ \citenamefont
  {Damour}(1988)}]{BlDa.88}%
  \BibitemOpen
  \bibfield  {author} {\bibinfo {author} {\bibfnamefont {L.}~\bibnamefont
  {Blanchet}}\ and\ \bibinfo {author} {\bibfnamefont {T.}~\bibnamefont
  {Damour}},\ }\href@noop {} {\bibfield  {journal} {\bibinfo  {journal} {Phys.
  Rev. D}\ }\textbf {\bibinfo {volume} {37}},\ \bibinfo {pages} {1410}
  (\bibinfo {year} {1988})}\BibitemShut {NoStop}%
\bibitem [{\citenamefont {Blanchet}\ \emph {et~al.}(2010)\citenamefont
  {Blanchet}, \citenamefont {Detweiler}, \citenamefont {{Le Tiec}},\ and\
  \citenamefont {Whiting}}]{Bl.al2.10}%
  \BibitemOpen
  \bibfield  {author} {\bibinfo {author} {\bibfnamefont {L.}~\bibnamefont
  {Blanchet}}, \bibinfo {author} {\bibfnamefont {S.}~\bibnamefont {Detweiler}},
  \bibinfo {author} {\bibfnamefont {A.}~\bibnamefont {{Le Tiec}}}, \ and\
  \bibinfo {author} {\bibfnamefont {B.~F.}\ \bibnamefont {Whiting}},\
  }\href@noop {} {\bibfield  {journal} {\bibinfo  {journal} {Phys. Rev. D}\
  }\textbf {\bibinfo {volume} {81}},\ \bibinfo {pages} {084033} (\bibinfo
  {year} {2010})},\ \Eprint {http://arxiv.org/abs/arXiv:1002.0726 [gr-qc]}
  {arXiv:1002.0726 [gr-qc]} \BibitemShut {NoStop}%
\bibitem [{\citenamefont {Akcay}\ \emph {et~al.}(2012)\citenamefont {Akcay},
  \citenamefont {Barack}, \citenamefont {Damour},\ and\ \citenamefont
  {Sago}}]{Ak.al.12}%
  \BibitemOpen
  \bibfield  {author} {\bibinfo {author} {\bibfnamefont {S.}~\bibnamefont
  {Akcay}}, \bibinfo {author} {\bibfnamefont {L.}~\bibnamefont {Barack}},
  \bibinfo {author} {\bibfnamefont {T.}~\bibnamefont {Damour}}, \ and\ \bibinfo
  {author} {\bibfnamefont {N.}~\bibnamefont {Sago}},\ }\href@noop {} {\bibfield
   {journal} {\bibinfo  {journal} {Phys. Rev. D}\ }\textbf {\bibinfo {volume}
  {86}},\ \bibinfo {pages} {104041} (\bibinfo {year} {2012})},\ \Eprint
  {http://arxiv.org/abs/arXiv:1209.0964 [gr-qc]} {arXiv:1209.0964 [gr-qc]}
  \BibitemShut {NoStop}%
\bibitem [{\citenamefont {Damour}(2001)}]{Da.01}%
  \BibitemOpen
  \bibfield  {author} {\bibinfo {author} {\bibfnamefont {T.}~\bibnamefont
  {Damour}},\ }\href@noop {} {\bibfield  {journal} {\bibinfo  {journal} {Phys.
  Rev. D}\ }\textbf {\bibinfo {volume} {64}},\ \bibinfo {pages} {124013}
  (\bibinfo {year} {2001})},\ \Eprint
  {http://arxiv.org/abs/arXiv:gr-qc/0103018} {arXiv:gr-qc/0103018} \BibitemShut
  {NoStop}%
\bibitem [{\citenamefont {Shah}\ \emph {et~al.}(2012)\citenamefont {Shah},
  \citenamefont {Friedman},\ and\ \citenamefont {Keidl}}]{Sh.al.12}%
  \BibitemOpen
  \bibfield  {author} {\bibinfo {author} {\bibfnamefont {A.~G.}\ \bibnamefont
  {Shah}}, \bibinfo {author} {\bibfnamefont {J.~L.}\ \bibnamefont {Friedman}},
  \ and\ \bibinfo {author} {\bibfnamefont {T.~S.}\ \bibnamefont {Keidl}},\
  }\href@noop {} {\bibfield  {journal} {\bibinfo  {journal} {Phys. Rev. D}\
  }\textbf {\bibinfo {volume} {86}},\ \bibinfo {pages} {084059} (\bibinfo
  {year} {2012})},\ \Eprint {http://arxiv.org/abs/arXiv:1207.5595 [gr-qc]}
  {arXiv:1207.5595 [gr-qc]} \BibitemShut {NoStop}%
\bibitem [{\citenamefont {Loken}\ \emph {et~al.}(2010)\citenamefont {Loken}
  \emph {et~al.}}]{Lo.al4.10}%
  \BibitemOpen
  \bibfield  {author} {\bibinfo {author} {\bibfnamefont {C.}~\bibnamefont
  {Loken}} \emph {et~al.},\ }\href@noop {} {\bibfield  {journal} {\bibinfo
  {journal} {J. Phys.: Conf. Ser.}\ }\textbf {\bibinfo {volume} {256}},\
  \bibinfo {pages} {012026} (\bibinfo {year} {2010})}\BibitemShut {NoStop}%
\end{thebibliography}%

\end{document}